\documentclass[ejs,noshowframe]{imsart}
\RequirePackage{amsthm,amsmath,amsfonts,amssymb}
\RequirePackage[authoryear]{natbib}
\RequirePackage[colorlinks,citecolor=blue,urlcolor=blue]{hyperref}
\RequirePackage{graphicx}
\RequirePackage{tabularx,multirow,booktabs,graphicx}
\graphicspath{{images/}}
\arxiv{2010.00000}
\startlocaldefs
\numberwithin{equation}{section}

\theoremstyle{plain}

\newtheorem{thm1}{Theorem}[section]
\newtheorem{lemma}{Lemma}[section]
\newtheorem{as1}{Assumption}
\newtheorem{rmk1}{Remark}[section]
\newtheorem{eg1}{Example}[section]

\newtheorem{pr1}{Proposition}
\theoremstyle{remark}

\endlocaldefs

\begin{document}
\allowdisplaybreaks
\begin{frontmatter}
\title{Large-dimensional Robust Factor Analysis with Group Structure\thanks{The author would like  to express sincere appreciation to Professor Yundong Tu from Peking University for his invaluable assistance with the coding aspects of this research.} \hspace{.2cm}}
\runtitle{Large-dimensional Robust Factor Analysis with Group Structure}

\begin{aug}
\author[A]{\fnms{Yong}~\snm{He}\ead[label=e1]{heyong@sdu.edu.cn}},
\author[A]{\fnms{Xiaoyang}~\snm{Ma}\ead[label=e2]{}},
\author[A]{\fnms{Xingheng}~\snm{Wang}\ead[label=e3]{}}
\and
\author[A]{\fnms{Yalin}~\snm{Wang}\ead[label=e4]{}}
\address[A]{Shandong University\printead[presep={\\ }]{e1}}
\runauthor{Y. He et al.}
\end{aug}

\begin{abstract}
In this paper, we focus on exploiting the group structure for  large-dimensional  factor models, which captures the homogeneous effects of common factors on individuals within the same group.  In view of the fact that datasets in macroeconomics and finance are typically heavy-tailed, we propose to identify the unknown group structure using the agglomerative hierarchical clustering algorithm and an information criterion with the robust two-step (RTS) estimates as  initial values. The loadings and factors are then re-estimated conditional on the identified groups. Theoretically, we demonstrate the consistency of the estimators for both group membership and the number of groups determined by the information criterion. Under finite second moment condition, we provide the convergence rate for the newly estimated factor loadings with group information, which are shown to achieve efficiency gains compared to those obtained without group structure information. Numerical simulations and real data analysis demonstrate the nice finite sample performance of our proposed approach in the presence of both group structure and heavy-tailedness.
\end{abstract}


\end{frontmatter}

\section{Introduction}
		Factor model, serving as a powerful tool for data simplification and dimensionality reduction, is drawing growing attention in the ``big data" era. The core idea is to explain the correlation structure of  observed variables through a few latent common factors, and study the internal dependence between variables, so as to explore the basic structure of observed data. With the rapid development of information technology, massive datasets are now ubiquitous and  large-dimensional approximate factor model becomes more and more important and has broad applications in various research areas such as finance, economics, psychology, sociology, genomics, and computational biology, among others.
	
	The concept of factor models dates back to the early 1900s, when psychologist Spearman proposed it to do statistical analysis of intelligence test scores \citep{spearman1904general}. In the mid-20th century, the work of Harold Hotelling \citep{hotelling1933analysis} further developed the method of factor analysis.
	The past two decades have witnessed an emerging use of factor analysis that makes it possible to summarize information from a large number of time series while keeping the empirical framework at a manageable magnitude \citep{forni1998let, stock2002a, stock2002b, stock2003}. The benefits of factor analysis in economics and finance have been demonstrated in a range of applications, including \cite{forni1997money,forni1998let,stock2002b,stock2003,bernanke2003zero,giannone2002using}, etc. One of the most important ways to do factor analysis is by principal component analysis (PCA) \citep{pearson1901lines, hotelling1933analysis}.  PCA  is easy to implement and provides consistent estimators for factors and factor loadings when both the cross section dimension $N$ and the time dimension $T$ goes to infinity. Representative works include, but are not limited to,  \cite{bai2002determining, bai2003inferential, stock2002a, stock2002b, onatski2009testing, ahn2013estimation, trapani2018randomized}.
	
	However, all of the aforementioned works assume that the fourth moments or even higher moments of factors and idiosyncratic errors are bounded, which is actually an idealization of the complex random real world. As the modern scientific research moves along, robust methods to deal with heavy-tailed data has drawn more and more attention. Such heavy-tailed data is often encountered in scientific fields, including financial engineering \citep{cont2001}, neurobiology \citep{roberts2015}, and biomedical imaging, and often cannot be effectively modeled by probabilistic models with high order moments. Therefore, PCA as a way to do factor analysis  has certain limitations in the presence of heavy-tailed data \citep{fan2018,he2022large,he2024matrix}, as it is equivalent to least squares which is well known to be sensitive to outliers. The elliptical distribution family provides a more reasonable characterization of  heavy-tailed data, which include a large class of heavy-tail distributions such as $t$-distribution. The factor model under the elliptical distribution framework, known as elliptical factor model, is also well-developed in the literature, which assumes that the factors and idiosyncratic errors  jointly follow an elliptical distribution.  For example, \cite{fan2018} considered large-scale covariance matrix estimation through the framework of Elliptical Distribution (FED); \cite{yu2019} proposed a robust estimation method for the number of factors in a large-dimensional factor model under the FED condition. Spatial Kendall's tau statistic, as an effective tool to measure the ``correlation" between ellptical variables, has drawn growing attention  since it was first proposed in 1998 \citep{choi1998}. The study of \cite{marden1999} shows that the population spatial Kendall's tau matrix and the covariance matrix share the same eigenspace, see also \cite{Han2018} for high-dimensional robust principal component analysis.
	 \cite{he2022large} further proposed a robust two-step (RTS) procedure to estimate the factor loadings, scores and common components under the FED without requiring any moment constraints. The convergence rates of the estimators are the same as those by  PCA method \citep{bai2003inferential}.
	
	Additionally, the large-dimensional approximate factor model increases the number of unknown parameters in the system as each cross-sectional unit is equipped with a heterogeneous loading vector to the common factors. In order to reduce the parameter dimension, the factor loading matrix is expected to have some sparse structure. For example, \cite{Tsai2010} studied the monthly U.S. excess stock returns using a constrained factor model, where it is assumed that stocks within the same sector of industry have the same loadings on risk factors. In an analysis of China real estate market, \cite{xiang2023determining} divided cities into four groups based on their level of economic development, assuming that market and government factors have the same impact on house price growth within each group. However, the grouping assumptions in both studies were pre-specified but not supported by any economic theory. It is often difficult to obtain prior information about how individuals should be grouped.
	
	Inspired by recent research of panel data model considering the group structure of the regression coefficient \citep{wang2018homogeneity,vogt2017classification,chen2019estimating,chen2021nonparametric,guo2022homogeneity},
	\cite{tu2023} firstly proposed a data-driven unsupervised grouping method, which considers the grouping structure of factor loadings in the large-dimentional approximate factor model. 
	In particular, \cite{tu2023} investigated how such loading homogeneity affect the factor analysis, and how to recover the underlying grouping structure.  \cite{tu2023} identified the grouping structure of factor loading matrix by the agglomerative hierarchical clustering (AHC) algorithm and an information criterion, and re-estimated the factors and factor loading matrix by least-squares optimization under the condition that the loading vectors in same group are equal. It is also demonstrated that when data exhibit a grouping structure, the estimates obtained from the grouped data significantly outperform the PCA estimates obtained prior to grouping.
	
	The goal of this paper is to propose robust estimators of the factors and factor loadings in the presence of both group structure and heavy-tailedness. The contributions of this paper lie in the following aspects. Firstly, for  heavy-tailed data, we use the robust two-step (RTS) method  \citep{he2022large} to give more robust initial estimators of the factors and factor loadings before grouping.
 We establish the asymptotic properties of the proposed estimators, which shows that our proposed method can correctly identify the group membership and estimate the group number with probability approaching 1 under finite second moment condition.  Secondly, we investigate the asymptotic properties of the refined loading estimators derived from the inferred group structure, and find that in the presence of group homogeneity, the squared deviations between the oracle loading estimator (when the group structure is known as a priori) and those lying in the true loading space vanish at a rate faster than $\delta_{N T}^2=\min \{N, T\}$ derived in Proposition \ref{pr1} by \cite{he2022large}, which ignores such homogeneity information.
 Additionally, the convergence rate of the loading estimator  is consistent with that of the estimator using PCA as initial estimate by \cite{tu2023}, while we only require finite second moment condition in contrast with the finite eighth moment condition in \cite{tu2023}.
 We also give a uniform convergence rate for the individual factor loading estimators, which is crucial to derive the clustering consistency and acts as an important complement for the theoretical results in \cite{he2022large}.
Finally, numerical simulations and empirical analysis clearly demonstrate the improved performance of the estimators after grouping. 
	
	The rest of this paper is organized as follows. In Section \ref{sec2}, we introduce the model setup, i.e., the high dimensional elliptical approximate factor model with grouping structure. We then propose the clustering method and an information criterion for identifying the latent group structure and estimating the unknown common components. In Section \ref{sec3}, we investigate the theoretical properties of the estimators we proposed. Simulation results are given in Section \ref{sec4}. We also applied the grouped algorithm to analyze a large-scale real macroeconomic dataset in Section \ref{sec5}. Further discussions and future research directions are left in Section \ref{sec6}. Proofs of the main
theorems and auxiliary lemmas appear in Appendix \ref{appA}-\ref{appC}.
	
	We introduce the notations adopted throughout the paper as an ending of this section.
	For any vector $\boldsymbol{a} $,  $\|\boldsymbol{a}\|_q$ denotes its $L_q$-norm,  $q\ge 1$, $\|\boldsymbol{a}\|$ also denotes $L_2$-norm. $\boldsymbol{a}\stackrel{d}{=}\boldsymbol{b}$ means distributions of $\boldsymbol{a}$ and $\boldsymbol{b}$ are the same. For a (random) matrix $\mathbf{A} $ of dimension \( p \times q \), $\mathbf{A}^\top$ denotes the transpose of $\mathbf{A} $,  both $\|\mathbf{A}\|_2$ and $\|\mathbf{A}\|$ denote the spectral norm of  $\mathbf{A} $, and $\|\mathbf{A}\|_F$ denotes the Frobenius norm of $\mathbf{A}$. Let $tr(\mathbf{A})$ be the trace of $\mathbf{A}$, and $\mathbf{P}_A$ be the projection matrix on the real linear space spanned by the column vectors of the matrix $\mathbf{A}$. Denote $\mathbf{A}_{i \cdot}$ as the $i$-th row of $\mathbf{A}$, which is a \( p \times 1 \) vector, and $\mathbf{A}_{\cdot j}$ as the $j$-th column of $\mathbf{A}$. $\mathbf{I}_q$ is denoted as the identity matrix of order $q$. Let $\lambda_i(\mathbf{A})$ be the $i$-th eigenvalue of the non-negative definite matrix $\mathbf{A}$ in descending order. The cardinality of the set $\mathcal{G}$ is denoted as $|\mathcal{G}|$.
	For two series $\left\{a_n\right\}_{n \geq 1}$ and $\left\{b_n\right\}_{n \geq 1}$, if there is a constant $C$ such that $a_n \geq C b_n, \forall n \geq 0$, we denote it as $a_n \gtrsim b_n$. If there is a constant $C$ such that $a_n \leq C b_n, \forall n \leq 0$, it is written as $a_n \lesssim b_n$. If both are true, it is written as $a_n \asymp b_n$.
	For two series of random variables $\left\{X_n\right\}_{n \geq 1}$ and $\left\{Y_n\right\}_{n \geq 1}$,  $X_n \gtrsim Y_n$ means $Y_n = O_p (X_n) $, $X_n \lesssim Y_n$ means $X_n =O_p(Y_n)$, and $X_n \asymp Y_n$ means $Y_n = O_p (X_n) $ and $X_n =O_p(Y_n)$. The constant $C$ and $M$ in different lines can be nonidentical.
	\section{Models and Methodology}\label{sec2}
	\subsection{Model Setup}
	Suppose that the panel dataset $\{y_{it}\}_{i\le N, t\le T}$ has a factor model representation:
	\begin{equation}\label{factor model}
		y_{it}=\boldsymbol{\lambda}_i^\top \boldsymbol{f}_t +\epsilon_{it},\quad i\le N,\quad t\le T,\quad \text{or in vector form},\quad \boldsymbol{y}_t=\mathbf{\Lambda}\boldsymbol{f}_t+\boldsymbol{\epsilon}_t,
	\end{equation}
	where $\boldsymbol{y}_t=(y_{1t},\ldots,y_{Nt})^\top \in \mathbb{R}^N$, $\boldsymbol{f}_t=(f_{1t},\ldots,f_{mt})^\top \in \mathbb{R}^m$ are the unobserved common factors, $\mathbf{\Lambda}=(\boldsymbol{\lambda}_1,\ldots,\boldsymbol{\lambda}_N)^\top$ is the $N\times m$ factor loading matrix with the $i$-th row being the individual specific loading vector $\boldsymbol{\lambda}_i=(\lambda_{i1},
	\ldots,\lambda_{im})^\top$, and $\boldsymbol{\epsilon}_t=(\epsilon_{1t},\ldots,\epsilon_{Nt})^\top$ is the $N\times 1$ vector of the idiosyncratic errors. The cross-sectional dimension $N$ and the time length $T$ are both allowed to increase to infinity, while the number $m$ is usually unknown but assumed to be fixed.
	
	Let $\mathbf{F}=(\boldsymbol{f}_1,\ldots,\boldsymbol{f}_T)^\top$, $\mathbf{Y}=(\boldsymbol{y}_1,\ldots,\boldsymbol{y}_T)^\top$ and $\boldsymbol{\epsilon}=(\boldsymbol{\epsilon}_1,\ldots,\boldsymbol{\epsilon}_T)^\top$. Then we have the following matrix representation form $$
	\mathbf{Y}=\mathbf{F}\mathbf{\Lambda}^\top+\boldsymbol{\epsilon}.
	$$
	
	As mentioned in the introduction and precisely stated in Assumption \ref{as1} below, we assume that $(\boldsymbol{f}_t^{\top}, \boldsymbol{\epsilon}_t^{\top})^{\top}$ is a series of temporally independent and identically distributed random vectors generated from an elliptical distribution.
	
	For a random vector $\boldsymbol{Z}=\left(Z_1, \ldots, Z_p\right)^{\top}$ following an elliptical distribution, denoted by $\boldsymbol{Z} \sim$ $E D(\boldsymbol{\mu}, \boldsymbol{\Sigma}, \xi)$, we mean that
	$$
	\boldsymbol{Z} \stackrel{d}{=} \boldsymbol{\mu}+\xi \mathbf{A} \boldsymbol{U},
	$$
	where $\boldsymbol{\mu} \in \mathbb{R}^p, \boldsymbol{U}$ is a random vector uniformly distributed on the unit sphere $S^{q-1}$ in $\mathbb{R}^q$, $ \xi \geq 0$ is a scalar random variable independent of $\boldsymbol{U}$, $ \mathbf{A} \in \mathbb{R}^{p \times q}$ is a deterministic matrix satisfying $\mathbf{A A}^{\top}=\boldsymbol{\Sigma}$ with $\boldsymbol{\Sigma}$ called scatter matrix whose rank is $q$. Elliptical distribution family has some nice properties, e.g., the marginal distributions, conditional distributions and distributions of linear combinations of elliptical vectors are also elliptical.
	
	Let the scatter matrices of $\boldsymbol{f}_t$ and $\boldsymbol{\epsilon}_t$ be $\boldsymbol{\Sigma}_f$ and $\boldsymbol{\Sigma}_\epsilon$, respectively. For reason of identifiability, we impose the following constraints:
	$$
	\boldsymbol{\Sigma}_f=\mathbf{I}_m \text { and }\left\|\operatorname{diag}\left(\boldsymbol{\Sigma}_\epsilon\right)\right\|_{\infty}=1,
	$$
	which is borrowed from  \cite{han2014scale}, \cite{yu2019} and \cite{he2022large}. Thus, for the factor model (\ref{factor model}) under the FED condition, the scatter matrix of $\boldsymbol{y}_t$, $ \boldsymbol{\Sigma}_y$, is composed of a low-rank part $\mathbf{\Lambda} \mathbf{\Lambda}^{\top}$ and a sparse part $\boldsymbol{\Sigma}_\epsilon$, i.e, $\boldsymbol{\Sigma}_y=\mathbf{\Lambda \Lambda}^{\top}+\boldsymbol{\Sigma}_\epsilon$.
	
	For model (\ref{factor model}), the factor loadings and scores can be estimated by a robust two-step (RTS)  procedure as advocated by \cite{he2022large}. If $m$ is unknown, we can estimate the number of factors $m$ consistently as in \cite{yu2019}. In the following, we briefly review the RTS procedure in \cite{he2022large}.
First we introduce the population spatial Kendall's tau matrix.
Let $\boldsymbol{Y}\sim ED(\boldsymbol{\mu},\boldsymbol{\Sigma},\xi)$ and $\widetilde{\boldsymbol{Y}}$ be an independent copy of $\boldsymbol{Y}$. Then the population spatial Kendall's tau matrix is defined as
$$
\mathbf{K}=E\left\{\frac{(\boldsymbol{Y}-\widetilde{\boldsymbol{Y}})(\boldsymbol{Y}-\widetilde{\boldsymbol{Y}})^{\top}}{\left\|\boldsymbol{Y}-\widetilde{\boldsymbol{Y}}\right\|^2}\right\}.
$$
    We can estimate the spatial Kendall's tau matrix $\mathbf{K}$ by a second-order U-statistic. Specifically, assume $\{\boldsymbol{Y}_1,\dots,\boldsymbol{Y}_T\}$ is a series of $T$ independent data points following the distribution $\boldsymbol{Y}\sim ED(\boldsymbol{\mu},\boldsymbol{\Sigma},\xi)$. The sample version spacial Kendall's tau matrix is
	$$\widehat{\mathbf{K}}=\frac{2}{T(T-1)} \sum_{t<t^{\prime}} \frac{\left(\boldsymbol{Y}_t-\boldsymbol{Y}_{t^{\prime}}\right)\left(\boldsymbol{Y}_t-\boldsymbol{Y}_{t^{\prime}}\right)^{\top}}{\left\|\boldsymbol{Y}_t-\boldsymbol{Y}_{t^{\prime}}\right\|_2^2} .$$

 Thus in the first step, we estimate the spatial Kendall's tau matrix of $\boldsymbol{y}_t$ by \begin{equation}\label{Kendall estimator}
		\widehat{\mathbf{K}}_y=\frac{2}{T(T-1)} \sum_{t<t^{\prime}} \frac{\left(\boldsymbol{y}_t-\boldsymbol{y}_{t^{\prime}}\right)\left(\boldsymbol{y}_t-\boldsymbol{y}_{t^{\prime}}\right)^{\top}}{\left\|\boldsymbol{y}_t-\boldsymbol{y}_{t^{\prime}}\right\|_2^2} .
	\end{equation}

 \cite{marden1999} reveal that all the eigenvectors of the spatial Kendall's tau matrix are identical to the eigenvectors of the scatter matrix with the same descending order. We estimate $\mathbf{\Lambda}$ by the eigenvectors of $\widehat{\mathbf{K}}_y$ as in \cite{he2022large}.
	Thus we estimate the factor loading matrix $\mathbf{\Lambda}$ by $\widetilde{\mathbf{\Lambda}}=(\tilde{\boldsymbol{\lambda}}_1,\ldots,\tilde{\boldsymbol{\lambda}}_N)^\top$, which are $\sqrt{N}$ times the leading $m$ eigenvectors of $\widehat{\mathbf{K}}_y$. Now in the second step, the estimated factors are $\widetilde{\mathbf{F}}=(\tilde{\boldsymbol{f}}_1,\ldots,\tilde{\boldsymbol{f}}_T)^\top=\mathbf{Y}\widetilde{\mathbf{\Lambda}}/N$ by the least square optimization, using the normalization that $\widetilde{\mathbf{\Lambda}}^\top\widetilde{\mathbf{\Lambda}}/N=\mathbf{I}_{m}$.
	
	Although the dimension of the loading vector $\boldsymbol{\lambda}_i$, i.e. the number of factors $m$ is often relatively small, the individual specific nature of the loading vector makes the total number of loading vectors to be estimated equal to $N$. As a consequent, the convergence rate of the loading estimator is rather slow, as shown by \cite{he2022large}.  To address the issue of dealing with an excessive number of parameters, it is common to exert specific forms of structural sparsity, which leads to a reduction in the dimension of the relevant parameters.
	
	In this study, our attention is directed towards the previously mentioned approximate linear factor model characterized by a latent group homogeneity structure, wherein factor loadings remain consistent across units categorized within the same group. More precisely, we posit the existence of an index set partition $\{1, \ldots, N\}$, denoted by $\left\{\mathcal{G}_1, \ldots, \mathcal{G}_{K_0}\right\}$ such that
	\begin{equation}\label{homo}
		\boldsymbol{\lambda}_i=\sum_{k=1}^{K_0} \boldsymbol{\lambda}_{\mathcal{G}, k} \cdot \mathbf{1} \left\{i \in \mathcal{G}_k\right\} \quad \text { and } \quad \mathcal{G}_k \cap \mathcal{G}_j=\emptyset \quad \text { for } \quad k \neq j,
	\end{equation}
	where $\emptyset$ denotes the empty set, $K_0$ is the true value of group number, $\mathbf{1}\{\cdot\}$ denotes the indicator function and $\boldsymbol{\lambda}_{\mathcal{G}, k}$ is the common value shared by all loading vectors whose index is in $\mathcal{G}_k$. This structural group homogeneity dictates that individuals within a given group prossess a shared group-specific loading vector, reducing the dimension of the loading vectors from $N$ to $K_0$. We assume that the number of groups $K_0$ is finite but unknown. To explore this unknown homogeneity structure, we will estimate the unknown group number $K_0$, identify the membership of the groups $\mathcal{G}_1, \ldots, \mathcal{G}_{K_0}$ and obtain the group-specific loading vector estimates in the following subsections.
	\subsection{A clustering algorithm}
	
	In this subsection, we apply the classical AHC algorithm to explore the latent homogeneity structure in the loading vectors $\boldsymbol{\lambda}_i$'s. The AHC method has been widely used for cluster analysis over the past few decades, see for example, \cite{everitt2011cluster, vogt2017classification, chen2019estimating,chen2021nonparametric}, among others.
	
	To apply the AHC algorithm, we start by defining a distance measure for the estimated loading vectors. To be specific, for any $i, j \in\{1, \ldots, N\}$, define the $L_1$-distance between $\tilde{\boldsymbol{\lambda}}_i$ and $\tilde{\boldsymbol{\lambda}}_j$ as:
	\begin{equation}\label{de0}
		\hat{\Delta}_{i j}=\frac{1}{m}\left\|\tilde{\boldsymbol{\lambda}}_i-\tilde{\boldsymbol{\lambda}}_j\right\|_1=\frac{1}{m} \sum_{k=1}^{m}\left|\tilde{\lambda}_{i k}-\tilde{\lambda}_{j k}\right| .
	\end{equation}
	
	In fact, there exists an $m \times m$ rotation matrix $\widehat{\mathbf{H}}$, with $\operatorname{rank}(\widehat{\mathbf{H}})=m$ such that $\hat{\Delta}_{i j}$ can be viewed as a natural estimate of $\Delta_{i j}$, where
	\begin{equation}\label{delta}
		\Delta_{i j}=\frac{1}{m}\left\|\widehat{\mathbf{H}}^{\top}\left(\boldsymbol{\lambda}_i-\boldsymbol{\lambda}_j\right)\right\|_1.
	\end{equation}
	
	From (\ref{delta}), it is clear that $\Delta_{i j}=0$ for $i, j \in \mathcal{G}_k$ and $\Delta_{i j} \neq 0$ for $i \in \mathcal{G}_{k_1}$ and $j \in \mathcal{G}_{k_2}$, whenever $k_1 \neq k_2$. Then denote $\boldsymbol{\Delta}=\left(\Delta_{i j}\right)$ as a distance matrix among the loading vectors, with the $(i, j)$-entry being $\Delta_{i j}$ defined in (\ref{delta}). Correspondingly, denote the estimate of $\boldsymbol{\Delta}$ as $\hat{\boldsymbol{\Delta}}_N$, whose $(i, j)$-entry is $\hat{\Delta}_{i j}$ defined in (\ref{de0}).
	
	For a given number of groups $K$, with $1 \leq K \leq N$, the Agglomerative Hierarchical Clustering (AHC) algorithm proceeds in four steps.
	Firstly, an initialization is performed, where each cross-sectional unit is considered as a separate group, resulting in a total of $N$ groups. At this point, the initial distance matrix is defined as $$\hat{\boldsymbol{\Delta}}:=\hat{\boldsymbol{\Delta}}_N.$$ Subsequently, the closest groups are merged by searching for the smallest non-diagonal element in the distance matrix $\hat{\boldsymbol{\Delta}}$, and the two groups corresponding to this element are combined into a new group. After the merge, it is necessary to recalculate the distances between the remaining groups. At this stage, the complete distance is used to define the distance between the new groups $\mathcal{A}$ and $\mathcal{B}$: $$\mathcal{D}(\mathcal{A}, \mathcal{B}):=\max _{i \in \mathcal{A}, j \in \mathcal{B}} \hat{\Delta}_{i j},$$ thereby obtaining the updated $(N-1)\times (N-1)$ distance matrix $\hat{\boldsymbol{\Delta}}$.
	Finally, the above steps of merging the closest groups and updating the distance matrix are repeated until the number of groups is reduced to the predetermined constant $K$. The set of indices obtained after $N-K$ merges is represented as $\widetilde{\mathcal{G}}_{1 \mid K}, \widetilde{\mathcal{G}}_{2 \mid K}, \ldots, \widetilde{\mathcal{G}}_{K \mid K}$, and thus the estimation of grouping under a given number of groups $K$ is: $$\widetilde{\mathcal{G}}(K)=\left\{\widetilde{\mathcal{G}}_{1 \mid K}, \widetilde{\mathcal{G}}_{2 \mid K}, \ldots, \widetilde{\mathcal{G}}_{K \mid K}\right\}.$$
	The steps of the AHC algorithm are summarized as follows:
	\begin{table}[htbp]
		\centering
		\begin{tabular}{p{\linewidth}}
			\toprule 
			$\textbf{The AHC algorithm}$ \\
			\midrule
			$\textbf{STEP1}$ :\\
			Start with $N$ groups, each containing one cross-sectional unit, and set $\hat{\boldsymbol{\Delta}}=\hat{\boldsymbol{\Delta}}_N$.\\
			$\textbf{STEP2}:$ \\
			Search for the smallest distance among the off-diagonal elements of $\hat{\boldsymbol{\Delta}}$ and merge the corresponding two groups.\\
			$\textbf{STEP3}:$ \\
			Re-calculate the distance among the remaining groups and update the distance matrix $\hat{\boldsymbol{\Delta}}$. For the distance between groups $\mathcal{A}$ and $\mathcal{B}$, use complete linkage: $\mathcal{D}(\mathcal{A}, \mathcal{B})=\max _{i \in \mathcal{A}, j \in \mathcal{B}} \hat{\Delta}_{i j}$.\\
			$\textbf{STEP4}:$ \\
			Repeat Steps 2-3 until the number of groups reduces to $K$. Denote the current index sets as $\widetilde{\mathcal{G}}_{1 \mid K}, \widetilde{\mathcal{G}}_{2 \mid K}, \ldots, \widetilde{\mathcal{G}}_{K \mid K}$. Then $\widetilde{\mathcal{G}}(K)=\left\{\widetilde{\mathcal{G}}_{1 \mid K}, \widetilde{\mathcal{G}}_{2 \mid K}, \ldots, \widetilde{\mathcal{G}}_{K \mid K}\right\}$ is the estimated group membership for the given group number $K$.\\
			\bottomrule 
		\end{tabular}
	\end{table}
	
	\subsection{Estimation Procedure}
	In this subsection, we discuss how to select the true group number $K_0$ using an information criterion when it is unknown.
	
	For the clusters $\widetilde{\mathcal{G}}(K)=\left\{\widetilde{\mathcal{G}}_{1 \mid K}, \widetilde{\mathcal{G}}_{2 \mid K}, \ldots, \widetilde{\mathcal{G}}_{K \mid K}\right\}$ identified in Step 4 of the AHC algorithm for a given $K$, we construct a goodness of fit measure making use of the homogeneity structure. In order to demonstrate the group homogeneity, denote the factor loadings in the $K$ groups as $\boldsymbol{\lambda}_{1 \mid K}, \ldots, \boldsymbol{\lambda}_{K \mid K}$.
	Since $\tilde{\boldsymbol{f}}_t$ is a consistent estimator of $\boldsymbol{f}_t$ up to some rotation from RTS in \cite{he2022large}, we estimate the factor loadings with group structure by minimizing the following loss function
	\begin{equation}\label{opti}
		(N T)^{-1} \sum_{i=1}^N \sum_{t=1}^T\left(y_{i t}-\boldsymbol{\lambda}_i \tilde{\boldsymbol{f}}_t\right)^2,
	\end{equation}
	subject to the group structure restriction
	$$
	\boldsymbol{\lambda}_i=\boldsymbol{\lambda}_{k \mid K} \quad \text { for } \quad i \in \widetilde{\mathcal{G}}_{k \mid K},\quad i=1, \ldots, N,\quad k=1, \ldots, K .
	$$
	
	Solving the above optimization problem with constraints and obtain
	\begin{equation}\label{groupedlambda}
		\hat{\boldsymbol{\lambda}}_{i \mid \widetilde{\mathcal{G}}(K)}=\frac{1}{|\widetilde{\mathcal{G}}_{k \mid K}|}(\widetilde{\mathbf{F}}^\top \widetilde{\mathbf{F}})^{-1} \widetilde{\mathbf{F}}^{\top} \sum_{i \in \widetilde{\mathcal{G}}_{k \mid K}} \underline{\boldsymbol{y}}_i, \text { if } i \in \widetilde{\mathcal{G}}_{k \mid K},
	\end{equation}
	where $\widetilde{\mathbf{F}}$ is estimated by RTS in Section 2.1, and $\underline{\boldsymbol{y}}_i=\left(y_{i 1}, \ldots, y_{i T}\right)^{\top}$. The estimator of $\mathbf{\Lambda}$ is $\widehat{\mathbf{\Lambda}}_{\widetilde{\mathcal{G}}(K)}=\left(\hat{\boldsymbol{\lambda}}_{1 \mid \widetilde{\mathcal{G}}_{(K)}}, \ldots, \hat{\boldsymbol{\lambda}}_{N \mid \tilde{\mathcal{G}}_{(K)}}\right)^{\top}$.
	With the above quantities, consider the following information criterion
	\begin{equation}\label{info}
		\mathbb{I C}(K)=\log [S(K)]+K \cdot \rho,
	\end{equation}
	where the tuning parameter $\rho \in(0,1)$ may rely on $N, T$. In simulation studies, we give the choice of $\rho$ in the case of finite samples. The goodness-of-fit measure is
	\begin{equation}\label{SK}
		S(K):=(N T)^{-1} \sum_{i=1}^N \sum_{t=1}^T \left(y_{i t}-\hat{\boldsymbol{\lambda}}_{i \mid \tilde{\mathcal{G}}(K)}^{\top} \tilde{\boldsymbol{f}}_t\right)^2 .
	\end{equation}
	
	The group number is then selected as
	\begin{equation}\label{groupnumber}
		\hat{K}=\arg \min _{1 \leq K \leq \bar{K}} \mathbb{I} \mathbb{C}(K),
	\end{equation}
	where $\bar{K}$ is a pre-specified large positive integer.
	In practice, the $\mathrm{AHC}$ algorithm provides a path of clustering when the number of groups $K$ reduces from $N$ to 1 , along which $\hat{K}$ is selected by (\ref{groupnumber}). Upon this choice of $\hat{K}$, obtain the estimated clustering as $\widehat{\mathcal{G}}=\left\{\widetilde{\mathcal{G}}_{1 \mid \hat{K}}, \widetilde{\mathcal{G}}_{2 \mid \hat{K}}, \ldots, \widetilde{\mathcal{G}}_{\hat{K} \mid \hat{K}}\right\}$, where $\hat{K}$ replaces $K$ in Step 4 of the AHC algorithm. Finally, the group-specific loading matrix $\mathbf{\Lambda}$ is estimated by $\widehat{\mathbf{\Lambda}}=\widehat{\mathbf{\Lambda}}_{\widehat{\mathcal{G}}}$, where $\widehat{\mathcal{G}}$ replaces $\widetilde{\mathcal{G}}(K)$ used in (\ref{groupedlambda}).
	
	Based on the group-specific loading matrix estimate $\widehat{\mathbf{\Lambda}}=\left(\hat{\boldsymbol{\lambda}}_1, \ldots, \hat{\boldsymbol{\lambda}}_N\right)^{\top}$, we can re-estimate the factors by solving the following least squares problem
	\begin{equation}
		\min _{\boldsymbol{f}_t} \sum_{i=1}^N\left(y_{i t}-\hat{\boldsymbol{\lambda}}_i^{\top} \boldsymbol{f}_t\right)^2, \quad t=1, \ldots, T .
	\end{equation}
	This leads to the estimator for $\boldsymbol{f}_t$ by $\hat{\boldsymbol{f}}_t=\left(\sum_{i=1}^N \hat{\boldsymbol{\lambda}}_i \hat{\boldsymbol{\lambda}}_i^{\top}\right)^{-1}\left(\sum_{i=1}^N \hat{\boldsymbol{\lambda}}_i y_{i t}\right),\quad t=1, \ldots, T$.

	We summarize the algorithmic steps of our robust factor analysis (RFA) model as follow:
\begin{table}[htbp]
	\centering
	\label{Algorithm2}
	\begin{tabular}{p{\linewidth}}
		\toprule 
		$\textbf{The RFA algorithm with group structure}$\\
		\midrule
		$\textbf{STEP1}$ :\\
		Estimate the spatial Kendall’s tau matrix of \(\boldsymbol{y}_t\) by \(\widehat{\textbf{K}}_y=\frac{2}{T(T-1)}\sum_{t<t'}\frac{(\boldsymbol{y}_t-\boldsymbol{y}_{t'})(\boldsymbol{y}_t-\boldsymbol{y}_{t'})^{\top}}{||\boldsymbol{y}_t-\boldsymbol{y}_{t'}||^2_2}\). Then estimate $\boldsymbol{\Lambda}$ by the eigenvectors of the spatial Kendall’s tau
		matrix, denoted as \(\widetilde{\mathbf{\Lambda}}\), and estimate the factor score \(\mathbf{F}\) by a cross-sectional least square regression based on the estimated factor loadings, i.e., \(\widetilde{\mathbf{F}}=\mathbf{Y}\widetilde{\mathbf{\Lambda}}/N\).\\
		$\textbf{STEP2}:$ \\
		Apply the information criterion to select the group number \(\widehat{K}\) and use the AHC algorithm to classify the loading vectors into \(\widehat{K}\) groups.\\
		$\textbf{STEP3}:$ \\
		Re-estimate the factors loadings by minimize the loss function
		\[{(NT)}^{-1}\sum_{i=1}^{N}\sum_{t=1}^{T}(y_{it}-\boldsymbol{\lambda}_{i}^\top\tilde{\boldsymbol{f}}_t)^2,\]
		subject to the group structure restriction \(\boldsymbol{\lambda}_{i}=\boldsymbol{\lambda}_{k|\widehat{K}}\) for  \(i\in\widetilde{\mathcal{G}}_{k|\widehat{K}},k=1,\dots,\hat{K}\). Based on the group-specific loading matrix estimate \(\widehat{\boldsymbol{\Lambda}}\), re-estimate the factors by \(\hat{\boldsymbol{f}}_t=\left(\sum_{i=1}^{N}\hat{\boldsymbol{\lambda}}_i\hat{\boldsymbol{\lambda}}_i^{\top}\right)^{-1}
		\left(\sum_{i=1}^{N}\hat{\boldsymbol{\lambda}}_i y_{it}\right) \) for \(t=1,\dots,T\).
		\\
		\bottomrule 
	\end{tabular}
\end{table}
	
	\section{Asymptotic theory}\label{sec3}
	To investigate the asymptotic properties of the estimators, the following technical assumptions are needed.

	\begin{as1}\label{as1}
		$$
		\left(\begin{array}{c}
			\boldsymbol{f}_t \\
			\boldsymbol{\epsilon}_t
		\end{array}\right)=\xi_t\left(\begin{array}{cc}
			\mathbf{I}_m & \mathbf{0} \\
			\mathbf{0} & \mathbf{A}
		\end{array}\right) \frac{\boldsymbol{g}_t}{\left\|\boldsymbol{g}_t\right\|},
		$$
		where $\xi_t$'s are independent samples of a scalar random variable $\xi$, and $\boldsymbol{g}_t$'s are independent Gaussian samples from $\boldsymbol{g} \sim \mathcal{N}\left(\mathbf{0}, \mathbf{I}_{m+N}\right)$. $m$ is fixed. Further, $\xi$ and $\boldsymbol{g}$ are independent and $\xi / \sqrt{N}=O_p(1)$ as $N \rightarrow \infty$. Therefore, $\left(\boldsymbol{f}_t^{\top}, \boldsymbol{\epsilon}_t^{\top}\right)^{\top}$ are independent samples from $E D\left(\mathbf{0}, \boldsymbol{\Sigma}_0, \xi\right)$ for $t=1, \ldots, T$ where $\boldsymbol{\Sigma}_0=\left(\begin{array}{cc}\mathbf{I}_m & \mathbf{0} \\ \mathbf{0} & \boldsymbol{\Sigma}_\epsilon\end{array}\right)$, and $\boldsymbol{\Sigma}_\epsilon=\mathbf{A A}^{\top}$. To make the model identifiable, we further assume that $\left\|\operatorname{diag}\left(\boldsymbol{\Sigma}_0\right)\right\|_{\infty}=1$.
	\end{as1}
	
	\begin{as1}\label{as2}
		$\left\|\boldsymbol{\lambda}_i\right\| \leq \overline{\boldsymbol{\lambda}}<\infty$, and $\mathbf{\Lambda}^{\top} \mathbf{\Lambda} / N \rightarrow \mathbf{V}$ as $N \rightarrow \infty$, where $\mathbf{V}$ is a positive definite matrix. There exist positive constants $c_1, c_2$ such that $c_2 \leq \lambda_m(\mathbf{V})<\cdots<\lambda_1(\mathbf{V}) \leq c_1$.
	\end{as1}
	
	\begin{as1}\label{as3}
		There exist positive constants $c_1$ and $c_2$, such that
		$c_2 \leq \lambda_{\min }\left(\boldsymbol{\Sigma}_\epsilon\right)\\ \leq \lambda_{\max }\left(\boldsymbol{\Sigma}_\epsilon\right) \leq c_1$.
	\end{as1}
	
	Assumptions \ref{as1}-\ref{as3} are common in the literature such as in \cite{yu2019,he2022large}. Assumption \ref{as1} asserts that the random vectors $(\boldsymbol{f}_t^{\top},\boldsymbol{\epsilon}_t^{\top})^{\top}$ are i.i.d. and follow an elliptical distribution. Consequently, it implies that the vectors $\boldsymbol{y}_t$ are also i.i.d. and follow an elliptical distribution with scatter matrix $\boldsymbol{\Sigma}_y = \boldsymbol{\Lambda\Lambda}^{\top} + \boldsymbol{\Sigma}_{\epsilon}$. Assumption \ref{as2} assume $\left\|\boldsymbol{\lambda}_i\right\|, i=1,\ldots,N$ are uniformly bounded and $\boldsymbol{\Lambda}^{\top}\boldsymbol{\Lambda}/N$ converges to a positive definite matrix with bounded maximum and minimum eigenvalues, which guarantees that the number of groups should be larger than the number of factors.
 We additionally stipulate that the eigenvalues $\lambda_j(\mathbf{V})$ of matrix $\mathbf{V}$ are distinct to ensure the corresponding eigenvectors are identifiable. Assumption \ref{as3} requires that the eigenvalues of $\boldsymbol{\Sigma}_{\epsilon}$ are bounded from below and above, which essentially makes the idiosyncratic error negligible with respect to the common components. And it is a typical way to control the cross-sectional correlations of the idiosyncratic errors.
	
	\begin{as1}\label{as4}
		$ T^{-1}\sum_{t=1}^{T}\boldsymbol{f}_t\boldsymbol{f}_t^\top \xrightarrow{p} \boldsymbol{\Sigma}_F $ as $T\rightarrow \infty$, where $\boldsymbol{\Sigma}_F$ is a positive definite matrix.
	\end{as1}
	
	\begin{as1}\label{as4-1}
		There exists a finite positive constant $M$ such that, for all $N$ and $T$, \newline
		(a) $\mathbb{E}\left[\epsilon_{i t} \epsilon_{i s}\right]=\tau_{i, t s}$ with $\left|\tau_{i, t s}\right| \leq\left|\tau_{t s}\right|$ for some $\tau_{t s}$ and for all $i$, and $T^{-1} \sum_{t=1}^T$\\$ \sum_{s=1}^T\left|\tau_{t s}\right| \leq$ $M$;\newline
		(b) $\mathbb{E}\left[T^{-1} \sum_{t=1}^T \epsilon_{i t} \epsilon_{j t}\right]=\sigma_{i j},\left|\sigma_{i i}\right| \leq M$ for all $i$, and $N^{-1} \sum_{i=1}^N \sum_{j=1}^N\left|\sigma_{i j}\right| \leq M$.
		
	\end{as1}

    \begin{as1}\label{as8}
		There is weak dependence between factors and idiosyncratic errors:$$
		\mathbb{E}\left[\frac{1}{N}\sum_{i=1}^{N}\left\|\frac{1}{\sqrt{T}}\sum_{t=1}^{T}\boldsymbol{f}_t\epsilon_{it}\right\|^2\right]\le M_1,
		$$
		where $M_1$ is a finite positive constant.
	\end{as1}
	
	Assumption \ref{as4} states that the second moment of the factors exist, allowing for a series of heavy-tailed distributions including the $t_3$ distribution. While Assumption \ref{as4-1} permits limited time series and cross-sectional dependence in the idiosyncratic component, which indicates that the model has an approximate factor structure rather than a strict one. Moreover, Assumption \ref{as8} implies weak dependence between factors and idiosyncratic errors. These assumptions were also introduced in \cite{bai2002determining,bai2003inferential}. Assumption \ref{as4} and Assumption \ref{as4-1} are relaxed versions of Assumption A and Assumption C in \cite{bai2002determining}. These assumptions relax the fourth-order moment restrictions on the factors and the eighth-order moment restrictions on the idiosyncratic errors relative to \cite{bai2002determining}, making them more widely applicable.
	
	\begin{as1}\label{as5}
		(a) There exists a constant $\tau_1 \in(0,1)$ such that
		$$
		\min _{1 \leq k \leq K_0}\left|\mathcal{G}_k\right| \geq \tau_1 \cdot N
		$$
		
		(b) The tuning parameter $\rho$ used in the information criterion satisfies: (i) $C_{N T} \cdot \rho \rightarrow$ $\infty$, and (ii) $\rho / \zeta^2 \rightarrow 0$ and $\max \left\{\sqrt{N / T}, N^{-\frac{1}{4}}\right\}=o(\zeta)$, as $N, T \rightarrow \infty$, where $\zeta$ is defined in Theorem \ref{thm1} and $C_{N T}=\min \left\{\sqrt{N_{K_0}}, \sqrt{T}\right\}$, with $N_{K_0}=\min \left\{\left|\mathcal{G}_k\right|, k=1, \ldots, K_0\right\}$.
	\end{as1}
	

	Assumption \ref{as5} is essential for establishing the consistency of $\hat{K}$ derived from the information criterion in (\ref{info}). According to Assumption \ref{as5}(a), the group sizes are of the same order of magnitude $N$. This condition is also exerted by \cite{tu2023}, and is a bit stronger than Assumption 1 in \cite{wang2018homogeneity}, which allows the fixed group size for some groups in a panel regression setup. Assumption \ref{as5}(b) imposes mild constraints on $\rho$, $C_{NT}$, and $\zeta$, indicating that the tuning parameter $\rho$ should converge to zero at a appropriate rate.
	
    Under Assumptions \ref{as1}-\ref{as3}, the conditions of Theorems 3.1 in \cite{he2022large} are easily satisfied, and we can similarly prove the following vector-version proposition.
	
	\begin{pr1}\label{pr1}
		Suppose that Assumptions \ref{as1}-\ref{as3} hold, there exist a series of matrices $\widehat{\mathbf{H}}$ (dependent on $T, N$ and $\widetilde{\mathbf{\Lambda}})$ so that $\widehat{\mathbf{H}}^{\top} \mathbf{V} \widehat{\mathbf{H}} \stackrel{p}{\rightarrow} \mathbf{I}_m$, such that
		$$
		\delta_{N T}^2\left\|\tilde{\boldsymbol{\lambda}}_i-\widehat{\mathbf{H}}^{\top} \boldsymbol{\lambda}_i\right\|_2^2=O_p(1),
		$$
		for each $i$, as $N, T \rightarrow \infty$, where $\delta_{N T}=\min\{\sqrt{N},\sqrt{T}\}$ and
		$$
		\widehat{\mathbf{H}}=\mathbf{M}_1 \mathbf{\Lambda}^{\top} \widetilde{\mathbf{\Lambda}} \widehat{\mathbf{V}}^{-1}.
		$$
		
	\end{pr1}
	
	\begin{rmk1}
		Proposition \ref{pr1} gives the vector version of Theorem 3.1 in \cite{he2022large}. The improvement of the corresponding convergence rate after giving the group structure to the loading matrix is embodied in the subsequent Theorem \ref{thm2}.
	\end{rmk1}
	
	\begin{pr1}\label{pr2}
		Suppose that Assumptions \ref{as1}-\ref{as3} hold, for the series of matrices $\widehat{\mathbf{H}}$ given in Proposition \ref{pr1} (dependent on $T, N$ and $\widetilde{\mathbf{\Lambda}})$, we have
		$$
		\max_{1\le i\le N}\left\|\tilde{\boldsymbol{\lambda}}_i-\widehat{\mathbf{H}}^{\top} \boldsymbol{\lambda}_i\right\|_2=O_p(N^{-\frac{1}{2}})+O_p(\left(N/T\right)^{\frac{1}{2}}),
		$$
		as $N, T \rightarrow \infty$.
	\end{pr1}
	\begin{rmk1}
		Proposition \ref{pr2} gives for the first time a uniform convergence rate for the individual specific factor loading estimator. It is crucial in the proof of the clustering consistency stated below in Theorem \ref{thm1}.
	\end{rmk1}
	
	%
	%
	
	\begin{thm1}\label{thm1}
		Denote $\zeta=\min _{1 \leq k \neq l<K_0} \min _{i \in \mathcal{G}_k, j \in \mathcal{G}_l} \Delta_{i j}$. Let Assumptions \ref{as1}-\ref{as3} hold and suppose that $\zeta$ fulfills the condition $\max \left\{\sqrt{N / T}, N^{-\frac{1}{2}}\right\}=o(\zeta)$. If the number of latent groups $K_0$ is known a priori, then
		$$
		P\left(\left\{\widehat{\mathcal{G}}_1, \widehat{\mathcal{G}}_2, \ldots, \widehat{\mathcal{G}}_{K_0}\right\}=\left\{\mathcal{G}_1, \mathcal{G}_2, \ldots, \mathcal{G}_{K_0}\right\}\right) \rightarrow 1,
		$$
		as $N, T \rightarrow \infty$.
	\end{thm1}

		Theorem \ref{thm1} shows that when the number of latent groups is known, the group membership $\left\{\mathcal{G}_1, \ldots, \mathcal{G}_{K_0}\right\}$ can be consistently estimated via the AHC algorithm. The result of Theorem \ref{thm1} does not require any restriction on moment conditions and therefore generalizes well to heavy-tailed distributions. Theorem \ref{thm1} implicitly imposes a constraint on the relative growth rates of the two dimensions $N$ and $T$. The constraint is necessary to ensure the uniform consistency of the estimated loadings. Additionally, the minimum signal strength $\zeta$ should either remain constant or diminish towards zero at a rate greater than $\max \left\{\sqrt{N / T}, N^{-\frac{1}{2}}\right\}$. We observe similar consistency results in time-varying and functional coefficient models, such as \cite{chen2019estimating} and \cite{chen2021nonparametric}, where regressors are observed. Since the latent factors are unobserved in our factor setup and the estimation error exists, it brings additional technical challenges.
	
	To analyze the asymptotic properties of the final  loading estimator $\hat{\boldsymbol{\lambda}}_i$, we first investigate the asymptotic properties of the ``oracle" loading estimators when the true group structure $\left\{\mathcal{G}_1, \ldots, \mathcal{G}_{K_0}\right\}$ is known as a priori.  Under the assumption of homogeneity (\ref{homo}), the constraints in the optimization problem (\ref{opti}) can be rewritten as
	$$
	\boldsymbol{\lambda}_i=\boldsymbol{\lambda}_{\mathcal{G}, k}, \quad \text { for }  i \in \mathcal{G}_k, \ \  k=1, \ldots, K_0 .
	$$
	
	The constrained optimization problem (\ref{opti}) yields the oracle loading estimator $\hat{\boldsymbol{\lambda}}_i^*$ as
	$$
	\hat{\boldsymbol{\lambda}}_i^* =\frac{1}{\left|\mathcal{G}_k\right|}\left(\widetilde{\mathbf{F}}^{\top} \widetilde{\mathbf{F}}\right)^{-1} \widetilde{\mathbf{F}}^{\top} \sum_{i\in\mathcal{G}_k} \underline{\boldsymbol{y}}_{i}, \quad \text { for } i \in \mathcal{G}_k, \ \ k=1, \ldots, K_0,
	$$
	and the following theorem demonstrates the consistency of the oracle estimator $\hat{\boldsymbol{\lambda}}_i^*$.
	
	\begin{thm1}\label{thm2}
		Suppose that Assumptions \ref{as1}-\ref{as4} and Assunption \ref{as5}(a) are satisfied. Then given the true group structure $\mathcal{G}=\left\{\mathcal{G}_1, \ldots, \mathcal{G}_{K_0}\right\}$,
		$$
		B_{N T}(k)^2\left\|\hat{\boldsymbol{\lambda}}_i^*-\widehat{\mathbf{H}}^{\top} \boldsymbol{\lambda}_{\mathcal{G}, k}\right\|_2^2=O_p(1),
		$$
		for each $i \in \mathcal{G}_k, k=1, \ldots, K_0$, where $B_{N T}(k)=\min \left\{\sqrt{\left|\mathcal{G}_k\right| N}, \sqrt{ T}\right\}$.
	\end{thm1}
		Theorem \ref{thm2} indicates that, under the assumption of group homogeneity, the squared deviations between the oracle loading estimator and those lying in the true loading space vanish at a rate of $B_{N T}(k)^2$. This rate is faster than $\delta_{N T}^2=\min \{N, T\}$ given in Proposition \ref{pr1} for the loading estimator that does not exploit the homogeneity structure.
	
	The clustering consistency established in Theorem \ref{thm1} relies on the prior knowledge of the true group number $K_0$. We then demonstrate in the next theorem that the information criterion (\ref{info}) can asymptotically determine the number of groups accurately.
	\begin{thm1}\label{thm3}
		Suppose that Assumptions \ref{as1}-\ref{as8} are satisfied. Then we have,
		
		(a)
		$$
		P\left(\hat{K}=K_0\right) \rightarrow 1 ;
		$$
		
		(b)
		$$
		P\left(\widehat{\mathbf{\Lambda}}=\widehat{\mathbf{\Lambda}}^*\right) \rightarrow 1,
		$$
		
		as $ N, T \rightarrow \infty $.
	\end{thm1}

	As a consequence, the final group-specific loading matrix estimator $\widehat{\boldsymbol{\Lambda}}$ equals the oracle estimator $\widehat{\mathbf{\Lambda}}^*=\left(\hat{\boldsymbol{\lambda}}_1^*, \ldots, \hat{\boldsymbol{\lambda}}_N^*\right)^{\top}$ with probability approaching 1.

\section{Simulation Study}\label{sec4}
	In this section, we illustrate the finite-sample properties of the proposed methodology by two simulated examples. The first example shows that in the case of heavy-tailed distributions, the RTS initial estimator yields a significant advantage for the estimator with group structure and further demonstrates the effect of group distance on the performance of the clustering algorithm, with $K_{0}=4$ and a set of other parameter specifications. The mainly purpose of the second example is to evaluate the information criterion used to select the group number under several different homogeneity structures, where we set the true group number as $K_{0}=1,2,3$, respectively. In the following, for ease of presentation, we denote PCA as the clustering method by \cite{tu2023} and RTS as the proposed grouping method in this paper, as these methods differs significantly in terms of initial estimates.

 Next, we consider the metric for  estimation accuracy. In both of two examples, we chose $S=500$ replications. We then provide the mean squared error (MSE) as a measure to gauge the properties of the common component estimates. Specifically, in order to measure the accuracy of estimates of the common component $c_{i t}=\boldsymbol{\lambda}_{i}^{\top } \boldsymbol{f}_{t}$ in one replication, we follow the evaluation criteria in \cite{Massacci2016} to compute the mean squared error (MSE), which, for the pre-clustering common component estimates, $\tilde{c}_{i t}=\tilde{\boldsymbol{\lambda}}_{i}^{\top} \tilde{\boldsymbol{f}}_{t}$, is defined as

$$
\text{PreC-MSE}=(N T)^{-1} \sum_{i=1}^{N} \sum_{t=1}^{T}\left(\tilde{c}_{i t}-c_{i t}\right)^{2}
$$
{\setlength{\parindent}{0cm}
	and for the post-clustering common component estimates, $\hat{c}_{i t}=\hat{\boldsymbol{\lambda}}_{i}^{\top} \hat{\boldsymbol{f}}_{t}$,}

$$
\text {PostC-MSE}=(N T)^{-1} \sum_{i=1}^{N} \sum_{t=1}^{T}\left(\hat{c}_{i t}-c_{i t}\right)^{2}.
$$

To assess the selection performance by the information criterion and evaluate the clustering performance, we adopt two measures from \cite{Chen2018}, that is, the normalised mutual information (NMI) and the Purity, both of which examine how close the estimated set of group indexes is to the true set of group indexes. Specifically, denote $\mathcal{G}=\left\{\mathcal{G}_{1}, \mathcal{G}_{2}, \ldots, \mathcal{G}_{K_{0}}\right\}$ and $\widehat{\mathcal{G}}=\left\{\widehat{\mathcal{G}}_{1}, \widehat{\mathcal{G}}_{2}, \ldots, \widehat{\mathcal{G}}_{\hat{K}}\right\}$ as the true set of group indexes and the estimated set of group indexes of $\{1,2, \ldots, N\}$, respectively. Then the NMI between $\mathcal{G}$ and $\widehat{\mathcal{G}}$ is defined as

$$
\text{NMI}(\mathcal{G}, \widehat{\mathcal{G}})=\frac{I(\mathcal{G}, \widehat{\mathcal{G}})}{[H(\mathcal{G})+H(\widehat{\mathcal{G}})] / 2}
$$
{\setlength{\parindent}{0cm}
	where $H(\mathcal{G})$ and $H(\widehat{\mathcal{G}})$ are the entropies of $\mathcal{G}$ and $\widehat{\mathcal{G}}$, respectively, i.e.,}

$$
H(\mathcal{G})=-\sum_{i=1}^{K_{0}} \frac{|\mathcal{G}_{i}|}{N} \log _{2}\left(\frac{|\mathcal{G}_{i}|}{N}\right),\quad H(\widehat{\mathcal{G}})=-\sum_{i=1}^{\hat{K}} \frac{|\widehat{\mathcal{G}}_{i}|}{N} \log _{2}\left(\frac{|\widehat{\mathcal{G}}_{i}|}{N}\right)
$$
{\setlength{\parindent}{0cm}
	and $I(\mathcal{G}, \widehat{\mathcal{G}})$ is the mutual information between $\mathcal{G}$ and $\widehat{\mathcal{G}}$ defined as}

$$
I(\mathcal{G}, \widehat{\mathcal{G}})=\sum_{i=1}^{K_{0}} \sum_{j=1}^{\hat{K}} \frac{|\mathcal{G}_{i} \cap \widehat{\mathcal{G}}_{j}|}{N} \cdot \log _{2}\left(\frac{N|\mathcal{G}_{i} \cap \widehat{\mathcal{G}}_{j}|}{|\mathcal{G}_{i}||\widehat{\mathcal{G}}_{j}|}\right).
$$

The Purity is defined as

$$
\text{Purity}(\mathcal{G}, \widehat{\mathcal{G}})=\frac{1}{N} \sum_{j=1}^{\hat{K}} \max _{1 \leq i \leq K_{0}}|\mathcal{G}_{i} \cap \widehat{\mathcal{G}}_{j}|.
$$

The NMI and the Purity are both ranging from 0 to 1, with a larger value indicating that the two index sets are closer. We chose the tuning parameter $\rho$ to be $log(min\{N_K,T\})/{min\{N_K,T\}}$ in our finite sample simulations, where $N_K=min\{|\widehat{\mathcal{G}}_{k|K}|,k=1,\dots,K\}$.

		\begin{eg1}\label{ex2}
			We first consider a two-factor model with 4 latent groups of homogeneous loading vectors, \textsl{i.e.},
		$$
	y_{i t}=\lambda_{i 1} f_{1 t}+\lambda_{i 2} f_{2 t}+e_{i t}, \quad i=1, \ldots, N, \quad t=1, \ldots, T
$$
{\setlength{\parindent}{0cm}
where \(\left(
{\boldsymbol{f}}_{t}^{\top},
{\boldsymbol{\epsilon}}_{t}^{\top}
\right)^{\top} \)
are $\textsl{i.i.d.}$ jointly elliptical random samples from multivariate skewed-$t_3$ distribution. We set $K_{0}=4$ and the loading vectors specified as}
$$
\lambda_{i 1}=\left\{\begin{array}{ll}
2, & \text { if } i \in \mathcal{G}_{1} ; \\
0, & \text { if } i \in \mathcal{G}_{2} ; \\
1, & \text { if } i \in \mathcal{G}_{3} ; \\
2+\delta, & \text { if } i \in \mathcal{G}_{4},
\end{array}\right. \qquad
\lambda_{i 2}=\left\{\begin{array}{ll}
0, & \text { if } i \in \mathcal{G}_{1} ; \\
2, & \text { if } i \in \mathcal{G}_{2} ; \\
2+\delta, & \text { if } i \in \mathcal{G}_{3} ; \\
1, & \text { if } i \in \mathcal{G}_{4},
\end{array}\right.
$$
where the cardinalities of the four groups are $N_{l}=0.25 N,$ $l=1, \ldots, 4$, and the sample size is set as $N=120,160,200,$ $T=100,200$. The parameter $\delta$ adjusts the distance between the loading vectors from any two different groups, except for the distance between group 1 and 2. We set $\delta=0.4,0.6$, with a larger value indicating stronger heterogeneity among the groups.
		\end{eg1}

\begin{table}[t]
\centering
\caption{Estimation results for Example \ref{ex2}.}
\label{Table42}
\resizebox{12cm}{4cm}{
	\begin{tabular}{ccccccccccccc} 
		\toprule 
		\multirow[b]{2}{*}{T} & \multirow[b]{2}{*}{$\delta$} & \multirow[b]{2}{*}{$N$} &\multirow[b]{2}{*}{Method}& \multirow[b]{2}{*}{$\bar{m}$} &  \multicolumn{4}{c}{Frequency} & \multicolumn{2}{c}{MSE$(\times 10)$}  & &\\
		\cmidrule(lr){6-9}
		\cmidrule(lr){10-11}
		&  &  &  & & $\hat{K}=2$ & $\hat{K}=3$ & $\hat{K}=4$ & $\hat{K}=5$ & PreC-MSE & PostC-MSE &NMI &Purity \\
		\hline
		100&0.4&120&PCA & 2(0) &153 &94 &241 &10 &4.89&3.02&0.82&0.77\\
		&  &  & RTS & 2(0) &143 &59   &298  &0&1.28&0.52&0.89&0.83
		\\
		& &160&PCA&2(0)&59&60&364&12&3.16&1.69&0.88&0.88\\
		& &   &RTS&2(0)&46&29&425&0 &1.18&0.42&0.96&0.94\\
		& &200&PCA&2(0)&39&37&396&18&2.76&1.09&0.90&0.91\\
		& &   &PTS&2(0)&16&10&474&0 &1.28&0.33&0.99&0.98\\
		&0.6&120&PCA&2(0)&91&39&352&14&4.30&1.36&0.89&0.87\\
		&   &   &RTS&2(0)&70&19&411&0 &2.55&0.52&0.95&0.92\\
		&   &160&PCA&2(0)&32&31&415&17&3.32&1.16&0.92&0.92\\
		&   &   &RTS&2(0)&18&6 &476&0 &1.84&0.40&0.99&0.98\\
		&   &200&PCA&2(0)&27&24&421&19&3.12&1.08&0.92&0.93\\
		&   &   &RTS&2(0)&9 &3 &488&0 &1.63&0.32&0.99&0.99\\
		\hline
		200&0.4&120&PCA & 2(0)&138&54&300&4 &2.38&0.98&0.86&0.82\\
		&  &  & RTS & 2(0) &138&35&327&0 &1.09&0.51&0.90&0.84
		\\
		& &160&PCA&2(0)&45&34&412&7 &2.21&0.74&0.92&0.91\\
		& &   &RTS&2(0)&34&14&452&0 &1.06&0.37&0.97&0.96\\
		& &200&PCA&2(0)&11&29&439&16&2.52&0.69&0.93&0.95\\
		& &   &PTS&2(0)&7&8&485&0 &1.41&0.31&0.99&0.99\\
		&0.6&120&PCA&2(0)&61&21&413&1&2.23&0.86&0.94&0.92\\
		&   &   &RTS&2(0)&54& 9&437&0 &0.96&0.50&0.96&0.94\\
		&   &160&PCA&2(0)&16&13&413&1 &2.23&0.86&0.94&0.92\\
		&   &   &RTS&2(0)&9&3 &488&0 &1.01&0.39&0.99&0.99\\
		&   &200&PCA&2(0)&14&16&450&14&2.02&0.68&0.95&0.96\\
		&   &   &RTS&2(0)&3 &2 &495&0 &0.85&0.31&0.99&0.99\\
		\bottomrule 
	\end{tabular}
 }
\end{table}
Table \ref{Table42} presents the average value $(\bar{m})$ and the standard deviation (in parentheses) of the number of factors selected by the information criterion of \cite{bai2002determining}. For all of the specifications, the estimated number of groups is at least 2 (i.e. $\hat{K}\ge 2$), and the estimated group number is never overestimated (i.e. $\hat{K}> 5$). Then, the frequency (over 500 replications) at which a number between 2 and 5 is selected as the group number by the information criterion (\ref{info}).
We can see that our proposed information criterion performs well across all values of $\delta$, $N$ and $T$. The simulation results show that our proposed criterion is capable of identifying the true group structures under each setting, conditional on the correct estimation of the number of groups. The results presented in Table \ref{Table42} demonstrate that in all these scenarios, the number of groups can almost always be estimated correctly in all replications, and the NMI and Purity values are all close to 1 when the sample size $N$ ranges from 120 to 200. On the other hand, increasing the sample size improves the estimation precision of both methods. The findings are also consistent with those of \cite{Tsai2010} that a large noise-to-signal ratio or a small sample size may bring substantial uncertainty in estimating both constrained and unconstrained factor models. Furthermore, Table \ref{Table42} reports the average values of MSE $(\times 10)$. We note that the average MSE values of the common component estimation (PostC-MSE) are slightly smaller than those of the pre-cluster estimation (PreC-MSE).
 Besides,
We can see that in different settings of \(\delta\), the bigger the \(\delta\) is, i.e., the further the distance is between these loading vectors from different groups, the easier it is to determine the true group number.

Finally, both the RTS and PCA methods demonstrate improved performance with larger sample sizes or longer time series lengths.
In addition to the findings presented in Example \ref{ex1} below, it is noteworthy that in heavy-tailed distribution scenarios, the RTS-based estimators outperform the PCA-based estimators by a large margin, in terms of PreC-MSE, PostC-MSE, NMI and Purity.

	\begin{eg1}\label{ex1}
		We next consider the two-factor model \citep{Boivin2006},
$$
y_{i t}=\lambda_{i 1} f_{1 t}+\lambda_{i 2} f_{2 t}+\sqrt{\theta_{i}} \epsilon_{i t}, i=1, \ldots, N, \quad t=1, \ldots, T
$$
{\setlength{\parindent}{0cm}
with}

$$
f_{m t}=0.5 f_{m, t-1}+u_{m t}, \quad u_{m t} \sim N(0,1), \quad m=1,2 .
$$
We choose the model dimensions to be $T=200, N=50,90,100,120,150$, and the error terms $\epsilon_{i t}$ are independently drawn from normal distribution $N(0, \kappa)$. The parameters $\theta_{i}$ and $\kappa$ together control the relative importance of the information contained in common factors $f_{1 t}$, $f_{2 t}$ and idiosyncratic errors $\epsilon_{i t}$. Following \cite{Choi2021CanonicalCM}, we set $\theta_{i}=4\left(\lambda_{i 1}^{2}+\lambda_{i 2}^{2}\right) / 3$ so that the variances contributed by common factors and idiosyncratic errors are equal when $\kappa=1$. To allow for different degrees of noise-to-signal ratio, we set $\kappa=0.5,1,2$. In particular, consider the case where three groups of observations are obtained, with sample sizes $N_{1}, N_{2}, N_{3}$, respectively, such that $N=N_{1}+N_{2}+N_{3}$. That is,
\begin{enumerate}
	\item[(a)] $N_{1}: y_{i t}=2 f_{1 t}+\sqrt{16 / 3} \epsilon_{i t}$;
	\item[(b)] $N_{2}: y_{i t}=2 f_{2 t}+\sqrt{16 / 3} \epsilon_{i t}$;
	\item[(c)] $N_{3}: y_{i t}=2.4 f_{1 t}+3.2 f_{2 t}+\sqrt{64 / 3} \epsilon_{i t}$.
\end{enumerate}
	\end{eg1}

	We consider  4 different scenarios, in which the combinations of $N, N_{1}, N_{2}, N_{3}$ are specified in Table \ref{Table41}.
	The results for Example \ref{ex1} are also reported in Table \ref{Table41}.
 We can see that the clustering method identifies the group membership accurately as indicated by the high NMI and Purity values. In addition, Table \ref{Table41} further reports the average values of MSE $(\times 10)$, NMI and Purity over the replications for which the true number of groups $K_{0}$ is correctly identified.
 \begin{table}[t]
 	\centering
 	\caption{Estimation results for Example \ref{ex1}.}
 	\label{Table41}
 	\resizebox{12cm}{4.5cm}{
 		\begin{tabular}{cccccccccccccccc}
 			\toprule 
 			\multirow[b]{2}{*}{Scenario} &\multirow[b]{2}{*}{Method}& \multirow[b]{2}{*}{$N_{1}$} & \multirow[b]{2}{*}{$N_{2}$} & \multirow[b]{2}{*}{$N_{3}$} & \multirow[b]{2}{*}{$N$} & \multirow[b]{2}{*}{$\bar{m}$} & \multicolumn{5}{c}{Frequency} & \multicolumn{2}{c}{MSE$(\times 10)$}  & &\\
 			\cmidrule(lr){8-12}
 			\cmidrule(lr){13-14}
 			&  &  &  &  &  &  & $\hat{K}=1$ & $\hat{K}=2$ & $\hat{K}=3$ & $\hat{K}=4$ & $\hat{K}=5$  & PreC & PostC &NMI &Purity \\
 			\hline
 			\multicolumn{16}{l}{Panel A: $\kappa=0.5$} \\
 			\multirow[c]{2}{*}{1} & PCA & 50 & 0 & 0 & 50 & 1(0) & 500 &0 &0 &0 &0   &0.08  &0.05  &-&1.00\\
 			& RTS & 50 & 0 & 0 & 50 & 1(0) & 500 &0 &0 &0 &0 & 0.08 &0.05 &-&1.00
 			\\
 			\multirow[c]{2}{*}{2} & PCA & 50 & 50 & 0 & 100 & 2(0) & 0 &500 &0 &0 &0   & 0.80 & 0.53&1.00&1.00\\
 			& RTS & 50 & 50 & 0 & 100 & 2(0) & 0 &500 &0 &0 &0 & 0.82 & 0.53 &1.00&1.00
 			\\
 			\multirow[c]{2}{*}{3} & PCA & 50 & 0 & 50 & 100 & 2(0) & 0 &500 &0 &0 &0 &2.07  &1.34   &1.00&1.00\\
 			& RTS & 50 & 0 & 50 & 100 & 2(0) & 0 &500 &0 &0 &0&2.12  &1.34   &1.00&1.00 \\
 			\multirow[c]{2}{*}{4} & PCA & 30 & 30 & 30 & 90 & 2(0) & 0 &0 &500 &0 &0  &1.88  & 1.31 &1.00&1.00\\
 			& RTS & 30 & 30 & 30 & 90 & 2(0) & 0 &0 &500 &0 &0 & 1.91 & 1.31 &1.00&1.00\\
 			\multirow[c]{2}{*}{5} & PCA & 50 & 50 & 50 & 150 & 2(0) & 0 &0 &500 &0 &0 & 1.34&0.79 &1.00&1.00 \\
 			& RTS & 50 & 50 & 50 & 150 & 2(0) & 0 &0 &500 &0 &0 & 1.37&
 			0.78
 			&1.00&1.00
 			\\
 			\hline
 			\multicolumn{16}{l}{Panel B: $\kappa=1$} \\
 			\multirow[c]{2}{*}{1} & PCA & 50 & 0 & 0 & 50 & 1(0) & 500 &0 &0 &0 &0 &  0.15 & 0.10&-&1.00\\
 			& RTS & 50 & 0 & 0 & 50 & 1(0) & 500 &0 &0 &0 &0 &0.16 &0.10 &-&1.00
 			\\
 			\multirow[c]{2}{*}{2} & PCA & 50 & 50 & 0 & 100 & 2(0) & 0 &500 &0 &0 &0 & 1.62  &1.07 &1.00&1.00 \\
 			& RTS & 50 & 50 & 0 & 100 & 2(0) & 0 &500 &0 &0 &0  & 1.65 & 1.07 &1.00&1.00
 			\\
 			\multirow[c]{2}{*}{3} & PCA & 50 & 0 & 50 & 100 & 2(0) & 0 &500 &0 &0 &0 &4.28  &2.67    &1.00&1.00\\
 			& RTS & 50 & 0 & 50 & 100 & 2(0) & 0 &500 &0 &0 &0&4.27  &2.67   &1.00&1.00 \\
 			\multirow[c]{2}{*}{4} & PCA & 30 & 30 & 30 & 90 & 2(0) & 0 &0 &500 &0 &0 & 3.90 & 2.66 &1.00&1.00\\
 			& RTS & 30 & 30 & 30 & 90 & 2(0) & 0 &0 &500 &0 &0  & 3.97 & 2.67 &1.00&1.00\\
 			\multirow[c]{2}{*}{5} & PCA & 50 & 50 & 50 & 150 & 2(0) & 0 &0 &500 &0 &0 &2.74 &1.59 &1.00&1.00 \\
 			& RTS & 50 & 50 & 50 & 150 & 2(0) & 0 &0 &500 &0 &0 &2.80 &1.59
 			&1.00&1.00
 			\\
 			\hline
 			\multicolumn{16}{l}{Panel C: $\kappa=2$} \\
 			\multirow[c]{2}{*}{1} & PCA & 50 & 0 & 0 & 50 & 1(0) & 500 &0 &0 &0 &0 &0.30  &0.20  &-&1.00\\
 			& RTS & 50 & 0 & 0 & 50 & 1(0) & 500 &0 &0 &0 &0 &0.31 &0.20 &-&1.00
 			\\
 			\multirow[c]{2}{*}{2} & PCA & 50 & 50 & 0 & 100 & 2(0) & 0 &500 &0 &0 &0 &3.31  & 2.13&1.00&1.00 \\
 			& RTS & 50 & 50 & 0 & 100 & 2(0) & 0 &500 &0 &0 &0 &3.35  & 2.13 &1.00&1.00
 			\\
 			\multirow[c]{2}{*}{3} & PCA & 50 & 0 & 50 & 100 & 2(0) & 0 &474 &26 &0 &0 & 8.89 & 5.16   &0.94&0.95\\
 			& RTS & 50 & 0 & 50 & 100 & 2(0) & 0 &468 &32 &0 &0& 8.92 &  5.11 &0.93&0.94 \\
 			\multirow[c]{2}{*}{4} & PCA & 30 & 30 & 30 & 90 & 2(0) & 0 &0 &477 &23 &0  &6.64  &5.38  &0.76&0.78\\
 			& RTS & 30 & 30 & 30 & 90 & 2(0) & 0 &0 &479 &21 &0  & 6.79 & 5.53 &0.76&0.78\\
 			\multirow[c]{2}{*}{5} & PCA & 50 & 50 & 50 & 150 & 2(0) & 0 &0 &480 &20 &0   & 5.57&3.66 &0.94&0.95 \\
 			& RTS & 50 & 50 & 50 & 150 & 2(0) & 0 &0 &475 &25 &0 &5.61 &3.73
 			&0.93&0.94
 			\\
 			\bottomrule
 		\end{tabular}
 	}
 \end{table}

 Overall,
as the data become noisier $(\kappa=2)$, the accuracy of clustering and the precision of estimating the number of groups deteriorate, especially if $N$ is small. Note that the NMI can not be calculated in scenario 1, as this measure requires that the true number of groups should be larger than one.  In addition, as expected, both estimation methods are adversely affected by the increase in the noise-to-signal ratio, as indicated by the rising average MSE values. When we increase the sample size,   the estimation precision of both methods is improved. In situations with underlying group structures, the estimates of all methods improve after grouping. Moreover, we observe that both the RTS and PCA methods exhibit similar estimation performance in all scenarios.
To conclude, our proposed method exhibits satisfactory performance in finite samples, particularly in scenarios involving four latent groups in the approximate factor model.
	
	\section{Real Data Analysis}\label{sec5}
	In this section, we apply the proposed method for large-dimensional approximate factor model to study an macroeconomic real dataset, i.e., U.S. monthly economic data which are available at U.S. Research Returns Data (\url{http://research.stlouisfed.org/econ/mccracken/fred-md/}). Our dataset consists of 41 variables ($N$ = 41) and 700 samples, which starts from January 1st, 1959, to May 1st, 2017. This dataset has been extensively analysed in existing studies under the factor model framework, while the homogeneity
	structure among the loading vectors has been fully ignored so far. To exclude structural instabilities, we focus on the subsample starting from August 1st, 1987, to July 1st, 2012, so we obtain a total of $T$ = 300 monthly observations for each time series.

\begin{figure}[htbp]
		\centering		\includegraphics[width=1\linewidth]{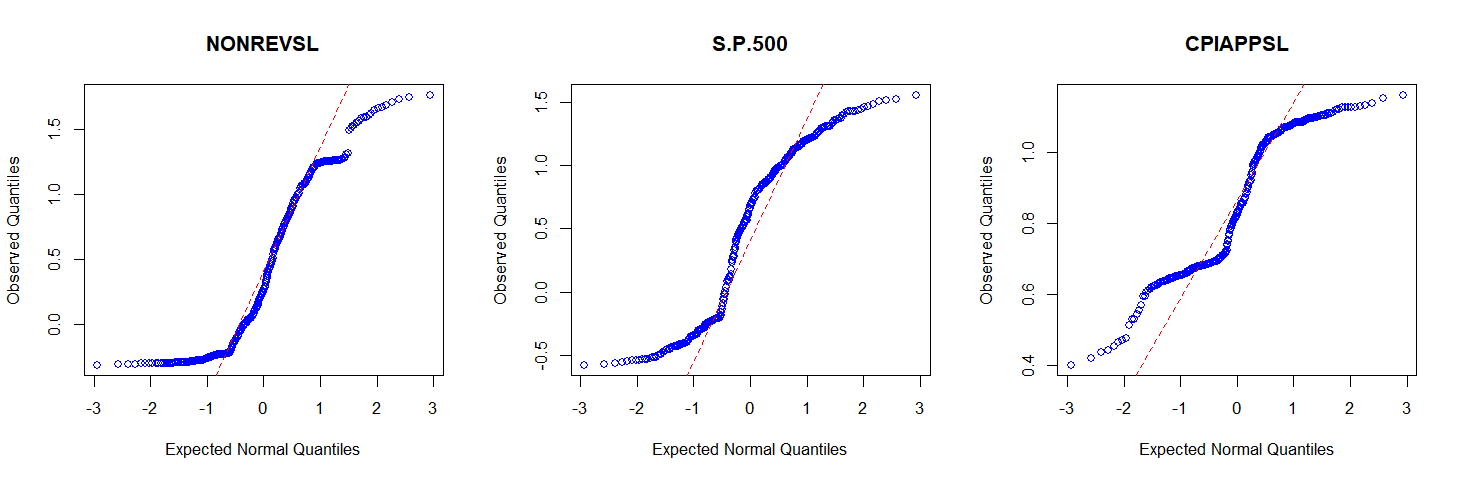}
		\caption{QQ plots of three variables.}
		\label{Figure4}
	\end{figure}

 First, due to space constraints, we only plot the QQ-plots of three variables in Figure \ref{Figure4}, from which we can conclude that the data points deviate significantly from the line in the tail region and the dataset exhibits heavy-tailed distribution characteristics.
	
\begin{figure}[htbp]
		\centering
  \includegraphics[width=0.7\linewidth]{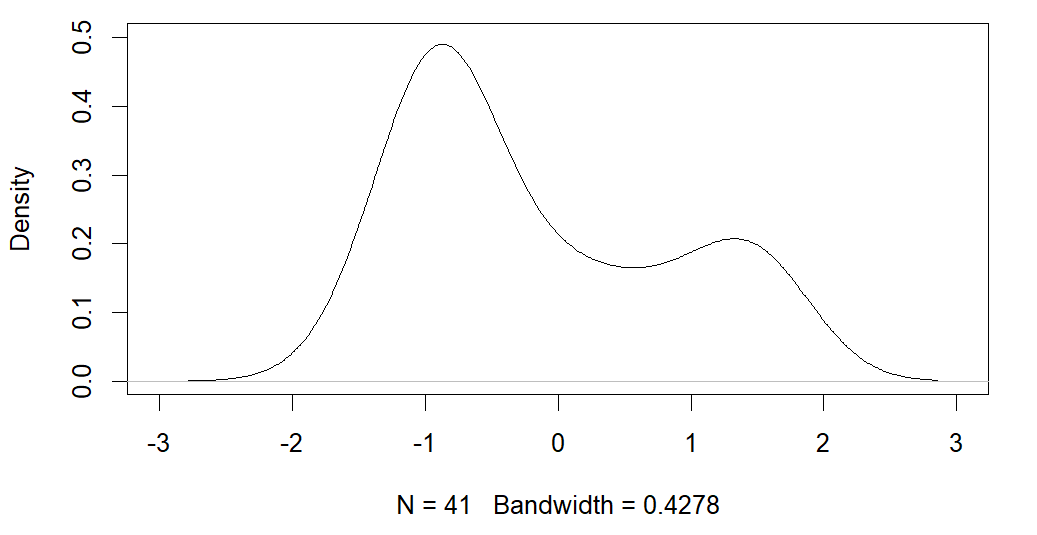}
		\caption{The kernel density function of factor loadings estimated by the RTS method.
		}
		\label{Figure3}
	\end{figure}

	Then we standardize each of the series and then select 4 factors by the information criterion \(IC_2\) of \cite{bai2002determining} in the training sample. We analyze whether the factor loading matrix corresponding to the data selected in this paper exhibits a grouping structure. We plot the kernel density function of the factor loading vectors estimated by the RTS method in Figure \ref{Figure3} and it shows that  the kernel density function exhibits a multi-modal structure, with two prominent peaks, reflecting the presence of a grouping structure in the dataset. This indicates the suitability of the proposed model based on grouping structure.
	
	The information criterion introduced in (\ref{groupnumber}) identifies 6 distinct groups in the loading vectors that initially estimated by the RTS method. We get the same result when using the PCA method. The grouping memberships for the two methods are shown in Figure \ref{Figure1} and Figure \ref{Figure2} below. Additionally, we display the factor loadings within each group after applying the RTS grouping method, as illustrated in Table \ref{Table63} below. From Table \ref{Table63}, it can be observed that Group 1 and Group 4 contain relatively more series, which correspond to the two most prominent peaks in the kernel density  shown in Figure \ref{Figure3}.
	\begin{table}[t]
		\centering
		\caption{RTS initial estimated loadings in each group.}
		\label{Table63}
		\begin{tabular}{p{1.5cm} >{\raggedright\arraybackslash}p{8cm} >{\centering\arraybackslash}p{1cm}}
			\hline
			Groups & Series & Numbers\\
			\hline
			Group 1&REALLN, NONREVSL, CONSPI, S\&P 500, S\&P:indust, PPICMM, CPITRNSL, CPIMEDSL, CUSR0000SAC, CUSR0000SAS, CPIULFSL, CUSR0000SA0L2, CUSR0000SA0L5, PCEPI, DNDGRG3M086SBEA, DSERRG3M086SBEA & 16\\
			\hline
			Group 2&S\&P div yield&1\\
			\hline
			Group 3&S\&P PE ratio, EXCAUSx&2\\
			\hline
			Group 4&FEDFUNDS, CP3Mx, TB3MS, TB6MS, GS1, GS5, GS10, AAA, BAA, EXJPUSx, EXUSUKx, CPIAPPSL, CUSR0000SAD, DDURRG3M086SBEA&14\\
			\hline
			Group 5&COMPAPFFx, TB3SMFFM, TB6SMFFM, T1YFFM&4\\
			\hline
			Group 6&T5YFFM, T10YFFM, AAAFFM, BAAFFM&4\\
			\hline
		\end{tabular}
	\end{table}
	
 \cite{McCracken2015} divides 134 monthly U.S. indicators into eight categories: (i) Money and Credit, (ii) Interest rate and Exchange Rates, (iii) Prices, and (iv) Stock Market, (v) Output and income, (vi) Labor market, (vii) Housing, (viii) Consumption, orders, and inventories.
 Referring to the detailed classification provided by \cite{McCracken2015}, it can be observed from Table \ref{Table63}, Figure \ref{Figure1} and Figure \ref{Figure2} that the AHC clustering algorithm proposed in this paper tends to group series belonging to the same category mentioned above into one cluster.
Observing the grouping results of the RTS method in Table \ref{Table63}, it can be found that all series in Group 5 and Group 6 belong to category (ii): Interest rate and Exchange Rates. Moreover, Group 5 mainly contains short-term fund rate-related data of less than one year, while Group 6 mainly contains long-term fund rate-related data of more than one year. Moreover, the series in Group 1 mainly include index data affecting GDP, such as the consumer price index, stock price index, and consumer confidence index, all of which have strong practical significance in their clustering. We plot the identified memberships under two initial estimators in Figure \ref{Figure1} and Figure \ref{Figure2}, it can also be observed that series with stronger correlations mostly appear in adjacent branches. For example, AAAFFM and BAAFFM, AAA and BAA, GS5 and GS10, T5YFFM and T10YFFM, etc.

	\begin{figure}[htbp]
		\centering
		\includegraphics[width=1\linewidth]{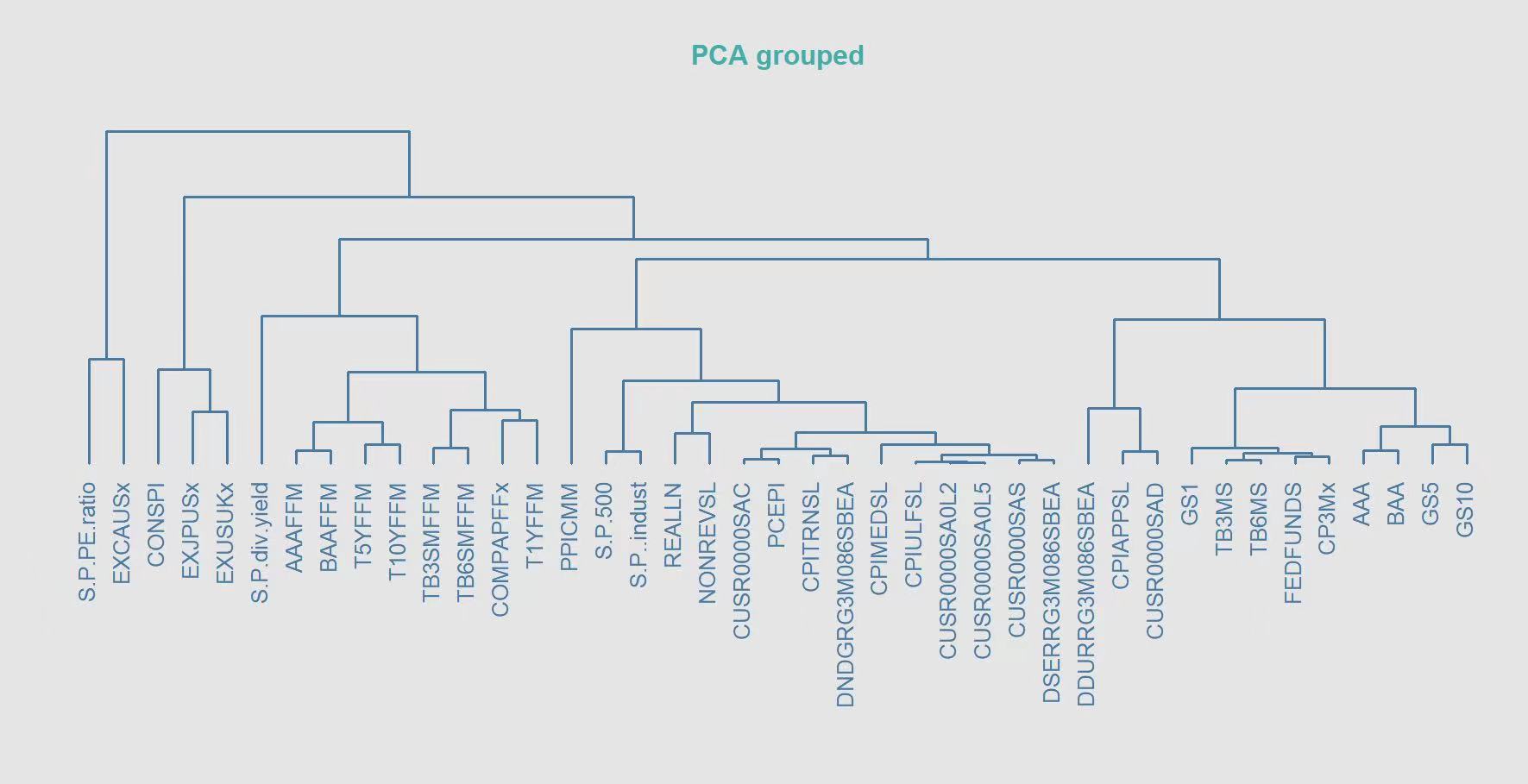}
		\caption{Identified membership under PCA initial estimator}
		\label{Figure1}
	\end{figure}
	\begin{figure}[htbp]
		\centering
		\includegraphics[width=1\linewidth]{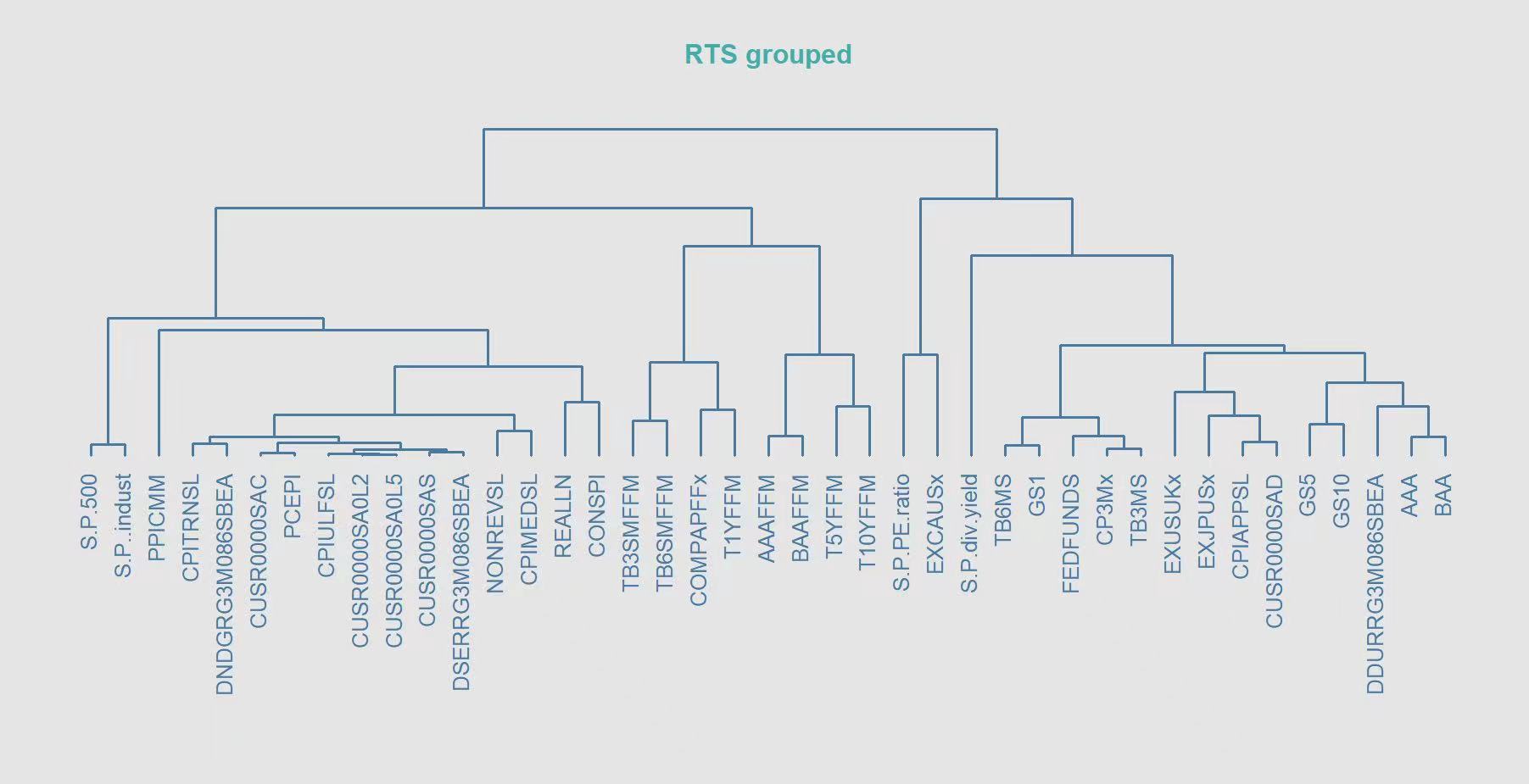}
		\caption{Identified membership under RTS initial estimator}
		\label{Figure2}
	\end{figure}

	Then we design a rolling scheme to evaluate the PCA and RTS methods based on their performances in forecasting. Now we take the RTS-based estimator as an example to illustrate, the process of the PCA method is similar.
	We obtain the RTS-based estimator $\widetilde{\boldsymbol{\Lambda}}$ and the post-clusting estimator $\widehat{\boldsymbol{\Lambda}}$ based on our proposed method. Then we establish a vector autoregressive (VAR) model with factor matrix \(\widetilde{\mathbf{F}}\)  estimated by RTS method in full samples. Specifically, we obtain a third-order vector autoregressive model, or VAR(3),  for the predicted values of factor. To measure the forecasting performance of the estimated factor loadings, we establish the $k$-step ahead prediction model ($k=1,2,3,4$), where $k$ is the lag order.
	
	\begin{table}[htbp]
		\centering
		\caption{The mean squared error of PCA and RTS estimations.}
		\label{Table62}
		\begin{center}
				\begin{tabular}{ccccc} 
					\toprule 
					k &1&2&3&4\\
					\midrule
					\(MSE-PCA_0\) &4.090&4.141&4.107&4.057\\
					\(MSE-PCA_1\) &3.781&3.823&3.797&3.752\\
					\(MSE-RTS_0\) &2.809&2.796&2.756&2.725\\
					\(MSE-RTS_1\) &2.374&2.394&2.373&2.336\\
					\bottomrule 
				\end{tabular}     
		\end{center}
	\end{table}

Finally, we compare the out-of-sample prediction performance. The performance is evaluated by the mean squared error (MSE) of the estimated value value and its real value. Specifically, we define
$$\text{Pre-}\boldsymbol{y}_k= \widetilde{\mathbf{\Lambda}}\boldsymbol{f}_{k},\ \ \text{Post-}\boldsymbol{y}_k= \widehat{\mathbf{\Lambda}}\boldsymbol{f}_{k},\ k=1,2,3,4.$$
where the factor \(\boldsymbol{f}_{k}\) is obtained from the VAR(3) model. Then the MSEs are given by
$$MSE-RTS_0(k)=\frac{\left\|\text{Pre-}\boldsymbol{y}_k-\text{Rel-}\boldsymbol{y}_{k}\right\|^2}{N},\ k=1,2,3,4.$$
$$MSE-RTS_1(k)=\frac{\left\|\text{Post-}\boldsymbol{y}_k-\text{Rel-}\boldsymbol{y}_{k}\right\|^2}{N},\ k=1,2,3,4.$$
 where $\text{Rel-}\boldsymbol{y}_{k}$ above denotes the real sample value. Table \ref{Table62} displays the MSE of the PCA and RTS method before and after grouping. The subscript 0 in Table \ref{Table62} denotes MSE before grouping while subscript 1 denotes after grouping, and we do 1-step to 4-step ahead forecast. We note that the errors of loading estimators obtained after homogeneity pursuit are considerably smaller than those of initial estimators. Furthermore, the errors of loading estimators obtained by RTS method is smaller than PCA method for  this heavy-tailed dataset. These results highlight the enhanced efficiency of the loading estimators after implementing the homogeneity pursuit procedure. Additionally, from Table \ref{Table62}, it can be found that as the lag order $k$ increases, the predicted MSE does not significantly increase. In fact, except for a slight increase in lag 2, it generally shows a decreasing trend. This indicates that the prediction method will maintain its accuracy over a longer time horizon. These results demonstrate the superiority of the proposed approach in practical applications.
The model proposed in this paper can accurately predict changes in US monthly macroeconomic data over several months, demonstrating high practical value.
	
	\section{Discussion}\label{sec6}
	In this paper, we propose a robust two-step estimation procedure for large-dimensional factor model to estimate the latent homogeneity structure in the loadings. Then we further use the constrained least squares method to obtain the group-specific loading vectors. The asymptotic properties of the estimators are established. Simulation studies demonstrate that our proposed approach outperforms the those which ignore the homogeneity structure of loadings in the model, especially for heavy-tailed datasets.  The empirical application to the U.S. macroeconomic datasets illustrates the practical merits of our methodology.
	
	In the future, we are also interested in several extended research directions. For example, we can extend the clustering problem of factor loadings to the matrix factor models \citep{ wang2019factor, chen2023statistical,he2023one,he2024matrix}, and the factors can be further separated into global factors and local factors to distinguish their respective roles in group homogeneity. The theoretical analysis is more challenging and we leave this as a future work.

\begin{appendix}
 \section{Auxiliary lemmas}\label{appA}
	\begin{lemma}\label{lm1}
		Under Assumptions \ref{as1}-\ref{as3}, as $\min \{T, N\} \rightarrow \infty$, we have
		$$
		\left\|\mathbf{M}_{2,i\cdot}\right\|_2^2=O_p\left(\frac{1}{T N^2}+\frac{1}{N^3}\right).
		$$
	\end{lemma}
	
	\begin{lemma}\label{lm2}
		Under Assumptions \ref{as1}-\ref{as3}, as $\min \{T, N\} \rightarrow \infty$, we have
		$$
		\left\|\mathbf{M}_{4,i\cdot} \widetilde{\mathbf{\Lambda}}\right\|_2^2=O_p\left(\frac{1}{T}+\frac{1}{N}\right)+O_p(1) \times \frac{1}{N}\|\widetilde{\mathbf{\Lambda}}-\mathbf{\Lambda} \widehat{\mathbf{H}}\|^2_F .
		$$
	\end{lemma}
	
	\begin{lemma}\label{lmA2}
		Let $\mathcal{M}(\mathcal{G})$ denote the event that $\left\{\widehat{\mathcal{G}}_1, \ldots, \widehat{\mathcal{G}}_{K_0}\right\}=\left\{\mathcal{G}_1, \ldots, \mathcal{G}_{K_0}\right\}$. Suppose that conditions of Theorem \ref{thm3} are satisfied. Then we have
		$$
		P\left[\mathbb{I} \mathbb{C}\left(K_0\right)<\mathbb{I} \mathbb{C}(K), 1 \leq K \leq K_0-1\right] \rightarrow 1,
		$$
		conditional on the event $\mathcal{M}(\mathcal{G})$.
	\end{lemma}

	\begin{lemma}\label{lmA3}
		Suppose that conditions of Lemma \ref{lmA2} are satisfied. Then we have
		$$
		P\left[\mathbb{I} \mathbb{C}\left(K_0\right)<\mathbb{I} \mathbb{C}(K), K_0+1 \leq K \leq \bar{K}\right] \rightarrow 1,
		$$
		conditional on the event $\mathcal{M}(\mathcal{G})$.
	\end{lemma}
	
	\begin{lemma}\label{bai2002lm4}
		Suppose that conditions of Lemma \ref{lmA3} are satisfied. Let $\mathbf{M}_{F}=\mathbf{I}-\mathbf{P}_{F}$ denote the idempotent matrix spanned by null space of $\mathbf{F}$ and define $
		V(\mathbf{F})=(N T)^{-1} \sum_{i =1}^{T} \underline{\boldsymbol{y}}_{i}^{\top} \mathbf{M}_{F} \underline{\boldsymbol{y}}_{i} $ ,
		$$
		V(\widetilde{\mathbf{F}})-V(\mathbf{F})=O_p(C_{N T}^{-2}).$$
	\end{lemma}
	\section{Proof of the main results}\label{appB}
	\subsection{Proof of Proposition \ref{pr1}}
	Define $\widehat{\mathbf{V}}$ as the diagonal matrix composed of the leading $m$ eigenvalues of $\widehat{\mathbf{K}}_y$. \textbf{Lemma A.3} in \cite{he2022large} implies that $\widehat{\mathbf{V}}$ is asymptotically invertible, $\|\widehat{\mathbf{V}}\|_F=O_p(1)$ and $\|\widehat{\mathbf{V}}^{-1}\|_F=O_p(1)$. Because $\widetilde{\mathbf{\Lambda}}=\sqrt{N} \widehat{\boldsymbol{\Gamma}}$ and $\widehat{\boldsymbol{\Gamma}}$ is composed of the leading eigenvectors of $\widehat{\mathbf{K}}_y$, we have
	$$
	\widehat{\mathbf{K}}_y \widetilde{\mathbf{\Lambda}}=\widetilde{\mathbf{\Lambda}} \widehat{\mathbf{V}}.
	$$
	Expand $\widehat{\mathbf{K}}_y$ by its definition, then
	$$
	\begin{aligned}
		\widetilde{\mathbf{\Lambda}} \widehat{\mathbf{V}}=
  & \frac{2}{T(T-1)} \sum_{1 \leq t<s \leq T} \frac{\left(\boldsymbol{y}_t-\boldsymbol{y}_s\right)\left(\boldsymbol{y}_t-\boldsymbol{y}_s\right)^{\top}}{\left\|\boldsymbol{y}_t-\boldsymbol{y}_s\right\|^2} \widetilde{\mathbf{\Lambda}} \\
		= & \frac{2}{T(T-1)} \sum_{1 \leq t<s \leq T} \frac{\left[\mathbf{\Lambda}\left(\boldsymbol{f}_t-\boldsymbol{f}_s\right)+\left(\boldsymbol{\epsilon}_t-\boldsymbol{\epsilon}_s\right)\right]\left[\mathbf{\Lambda}\left(\boldsymbol{f}_t-\boldsymbol{f}_s\right)+\left(\boldsymbol{\epsilon}_t-\boldsymbol{\epsilon}_s\right)\right]^{\top}}{\left\|\boldsymbol{y}_t-\boldsymbol{y}_s\right\|^2} \widetilde{\mathbf{\Lambda}} \\
		= & \frac{2}{T(T-1)} \sum_{1 \leq t<s \leq T} \frac{\mathbf{\Lambda}\left(\boldsymbol{f}_t-\boldsymbol{f}_s\right)\left(\boldsymbol{f}_t-\boldsymbol{f}_s\right)^{\top} \mathbf{\Lambda}^{\top}}{\left\|\boldsymbol{y}_t-\boldsymbol{y}_s\right\|^2} \widetilde{\mathbf{\Lambda}}\\
  &+\frac{2}{T(T-1)} \sum_{1 \leq t<s \leq T} \frac{\left(\boldsymbol{\epsilon}_t-\boldsymbol{\epsilon}_s\right)\left(\boldsymbol{f}_t-\boldsymbol{f}_s\right)^{\top} \mathbf{\Lambda}^{\top}}{\left\|\boldsymbol{y}_t-\boldsymbol{y}_s\right\|^2} \widetilde{\mathbf{\Lambda}} \\
		& +\frac{2}{T(T-1)} \sum_{1 \leq t<s \leq T} \frac{\mathbf{\Lambda}\left(\boldsymbol{f}_t-\boldsymbol{f}_s\right)\left(\boldsymbol{\epsilon}_t-\boldsymbol{\epsilon}_s\right)^{\top}}{\left\|\boldsymbol{y}_t-\boldsymbol{y}_s\right\|^2} \widetilde{\mathbf{\Lambda}}\\
  &+\frac{2}{T(T-1)} \sum_{1 \leq t<s \leq T} \frac{\left(\boldsymbol{\epsilon}_t-\boldsymbol{\epsilon}_s\right)\left(\boldsymbol{\epsilon}_t-\boldsymbol{\epsilon}_s\right)^{\top}}{\left\|\boldsymbol{y}_t-\boldsymbol{y}_s\right\|^2} \widetilde{\mathbf{\Lambda}}
	\end{aligned}
	$$
	For the ease of notations, we denote
	$$
 \begin{aligned}
		&\mathbf{M}_1=\frac{2}{T(T-1)} \sum_{1 \leq t<s \leq T} \frac{\left(\boldsymbol{f}_t-\boldsymbol{f}_s\right)\left(\boldsymbol{f}_t-\boldsymbol{f}_s\right)^{\top}}{\left\|\boldsymbol{y}_t-\boldsymbol{y}_s\right\|^2}, \\
  &\mathbf{M}_2=\frac{2}{T(T-1)} \sum_{1 \leq t<s \leq T} \frac{\left(\boldsymbol{\epsilon}_t-\boldsymbol{\epsilon}_s\right)\left(\boldsymbol{f}_t-\boldsymbol{f}_s\right)^{\top}}{\left\|\boldsymbol{y}_t-\boldsymbol{y}_s\right\|^2} \\
		& \mathbf{M}_3=\frac{2}{T(T-1)} \sum_{1 \leq t<s \leq T} \frac{\left(\boldsymbol{f}_t-\boldsymbol{f}_s\right)\left(\boldsymbol{\epsilon}_t-\boldsymbol{\epsilon}_s\right)^{\top}}{\left\|\boldsymbol{y}_t-\boldsymbol{y}_s\right\|^2}, \\
  &\mathbf{M}_4=\frac{2}{T(T-1)} \sum_{1 \leq t<s \leq T} \frac{\left(\boldsymbol{\epsilon}_t-\boldsymbol{\epsilon}_s\right)\left(\boldsymbol{\epsilon}_t-\boldsymbol{\epsilon}_s\right)^{\top}}{\left\|\boldsymbol{y}_t-\boldsymbol{y}_s\right\|^2}
	\end{aligned}
 $$
	and let $\widehat{\mathbf{H}}=\mathbf{M}_1 \mathbf{\Lambda}^{\top} \widetilde{\mathbf{\Lambda}} \widehat{\mathbf{V}}^{-1}$, then
	$$
	\widetilde{\mathbf{\Lambda}}-\mathbf{\Lambda} \widehat{\mathbf{H}}=\left(\mathbf{M}_2 \mathbf{\Lambda}^{\top} \widetilde{\mathbf{\Lambda}}+\mathbf{\Lambda} \mathbf{M}_3 \widetilde{\mathbf{\Lambda}}+\mathbf{M}_4 \widetilde{\mathbf{\Lambda}}\right) \widehat{\mathbf{V}}^{-1}.
	$$
	Since $\mathbf{\Lambda}^\top=(\boldsymbol{\lambda}_1,\ldots,\boldsymbol{\lambda}_N)$ and $\widetilde{\mathbf{\Lambda}}^\top=(\tilde{\boldsymbol{\lambda}}_1,\ldots,\tilde{\boldsymbol{\lambda}}_N)$, consider the columns of $\mathbf{\Lambda}^\top$ and $\widetilde{\mathbf{\Lambda}}^\top$ separately. Then through the expansion of $\widehat{\mathbf{K}}_y$, we get
	$$
	\tilde{\boldsymbol{\lambda}}_i -\widehat{\mathbf{H}}^{\top} \boldsymbol{\lambda}_i = \widehat{\mathbf{V}}^{-1}\widetilde{\mathbf{\Lambda}}^{\top}\left(\mathbf{\Lambda}\mathbf{M}_{2,\cdot i}^\top+\mathbf{M}_3^\top \boldsymbol{\lambda}_i+\mathbf{M}_{4,\cdot i}\right).
	$$
	\textbf{Lemma S2} to \textbf{Lemma S4} in \cite{he2022large} and \textbf{Lemma} \ref{lm1} to \textbf{Lemma} \ref{lm2} show that
	$$
\left\|\mathbf{M}_{2,i\cdot}\right\|_2^2=O_p\left(\frac{1}{T N^2}+\frac{1}{N^3}\right), \quad\left\|\mathbf{M}_3\right\|_F^2=O_p\left(\frac{1}{T N}+\frac{1}{N^3}\right)
	$$
	while
	$$
	\left\|\mathbf{M}_{4,i\cdot}\widetilde{\mathbf{\Lambda}} \right\|_2^2=O_p\left(\frac{1}{T}+\frac{1}{N}\right)+O_p(1) \times \frac{1}{N}\|\widetilde{\mathbf{\Lambda}}-\mathbf{\Lambda} \widehat{\mathbf{H}}\|^2_F =O_p(\frac{1}{T}+\frac{1}{N}),
	$$
	where the last equation follows from \textbf{Theorem 3.1} in \cite{he2022large}.
	Therefore, by Cauchy-Schwartz inequality and triangular inequality, it's easy to prove that
	$$
	\left\|\tilde{\boldsymbol{\lambda}}_i -\widehat{\mathbf{H}}^{\top} \boldsymbol{\lambda}_i\right\|_2^2=O_p(\frac{1}{T}+\frac{1}{N}),
	$$
	then we have
	$$
	\delta_{N T}^2\left\|\tilde{\boldsymbol{\lambda}}_i-\widehat{\mathbf{H}}^{\top} \boldsymbol{\lambda}_i\right\|_2^2=O_p(1).
	$$
	To complete the proof, $ \widehat{\mathbf{H}}^{\top} \mathbf{V} \widehat{\mathbf{H}} \stackrel{p}{\rightarrow} \mathbf{I}_m $ follows directly from \textbf{Theorem 3.1} in \cite{he2022large}.{\hfill$\square$}
	\subsection{Proof of Proposition \ref{pr2}}
	Using the fact that $\max_{1\le i\le N}a^2_i\le\sum_{i=1}^{N}a^2_i$, we have$$
	\max_{1\le i\le N}\|\tilde{\boldsymbol{\lambda}}_i-\widehat{\mathbf{H}}^\top\boldsymbol{\lambda}_i\|_2^2\le \sum_{i =1}^N \|\tilde{\boldsymbol{\lambda}}_i-\widehat{\mathbf{H}}^\top\boldsymbol{\lambda}_i\|_2^2=\|\widetilde{\boldsymbol{\Lambda}}-\boldsymbol{\Lambda}\widehat{\mathbf{H}}\|_F^2.
	$$
	Then \textbf{Theorem 3.1} in \cite{he2022large} show that $$
	\max_{1\le i\le N}\|\tilde{\boldsymbol{\lambda}}_i-\widehat{\mathbf{H}}^\top\boldsymbol{\lambda}_i\|_2^2\le \|\widetilde{\boldsymbol{\Lambda}}-\boldsymbol{\Lambda}\widehat{\mathbf{H}}\|_F^2=O_p(\frac{N}{T}+\frac{1}{N}).
	$$
	As a result, $$
	\max_{1\le i\le N}\left\|\tilde{\boldsymbol{\lambda}}_i-\widehat{\mathbf{H}}^{\top} \boldsymbol{\lambda}_i\right\|_2=O_p(N^{-\frac{1}{2}})+O_p(\left(N/T\right)^{\frac{1}{2}}),
	$$
	as $N, T \rightarrow \infty$.{\hfill$\square$}
	\subsection{Proof of Theorem \ref{thm1}} To prove the theorem, we only need to show that
	\begin{equation}\label{eq1}
		P\left(\max _{1 \leq k \leq K_0} \max _{i, j \in \mathcal{G}_k} \hat{\Delta}_{i j}<\min _{1 \leq k \neq l \leq K_0} \min _{i \in \mathcal{G}_k, j \in \mathcal{G}_l} \hat{\Delta}_{i j}\right) \rightarrow 1,
	\end{equation}
	as $N, T \rightarrow \infty$. Note that for the distance between true loading vectors, we have $\Delta_{i j}^0 \equiv 0$ if $i, j \in \mathcal{G}_k$, and
	$$
	\min _{1 \leq k \neq l \leq K_0} \min _{i \in \mathcal{G}_k, j \in \mathcal{G}_l} \Delta_{i j}^0=\zeta>0 .
	$$
	Further, note that
	$$
	\Delta_{i j}^0=\frac{1}{m}\left\|\widehat{\mathbf{H}}^{\top}\left(\boldsymbol{\lambda}_i-\boldsymbol{\lambda}_j\right)\right\|_1,
	$$
	where $\widehat{\mathbf{H}}$ is a rotation matrix defined as
	$$
	\widehat{\mathbf{H}}=\mathbf{M}_1 \mathbf{\Lambda}^{\top} \widetilde{\mathbf{\Lambda}} \widehat{\mathbf{V}}^{-1}.
	$$
	To prove (\ref{eq1}), it is sufficient to show that
	\begin{equation}\label{eq2}
		\max _{1 \leq i, j \leq N}\left|\hat{\Delta}_{i j}-\Delta_{i j}^0\right|=o(\zeta)
	\end{equation}
	By the Minkowski inequality, we obtain that,
	$$
	\begin{aligned}
		\left(\hat{\Delta}_{i j}-\Delta_{i j}^0\right)^2 & \leq \frac{1}{m^2} \cdot\left(\left\|\tilde{\boldsymbol{\lambda}}_i-\widehat{\mathbf{H}}^{\top} \boldsymbol{\lambda}_i\right\|_1+\left\|\tilde{\boldsymbol{\lambda}}_j-\widehat{\mathbf{H}}^{\top} \boldsymbol{\lambda}_j\right\|_1\right)^2 \\
		& \leq \frac{2}{m^2} \cdot\left(\left\|\tilde{\boldsymbol{\lambda}}_i-\widehat{\mathbf{H}}^{\top} \boldsymbol{\lambda}_i\right\|_1^2+\left\|\tilde{\boldsymbol{\lambda}}_j-\widehat{\mathbf{H}}^{\top} \boldsymbol{\lambda}_j\right\|_1^2\right) \\
		& \leq 2\left(\left\|\tilde{\boldsymbol{\lambda}}_i-\widehat{\mathbf{H}}^{\top} \boldsymbol{\lambda}_i\right\|_2^2+\left\|\tilde{\boldsymbol{\lambda}}_j-\widehat{\mathbf{H}}^{\top} \boldsymbol{\lambda}_j\right\|_2^2\right)  .
	\end{aligned}
	$$
	Then by Proposition \ref{pr2} we have that,
	\begin{equation}\label{eq3}
		\max _{1 \leq i, j \leq N}\left|\hat{\Delta}_{i j}-\Delta_{i j}^0\right| \leq O_p\left((N / T)^{\frac{1}{2}}\right)+O_p\left(N^{-\frac{1}{2}}\right) .
	\end{equation}
	Consequently, (\ref{eq2}) follows from (\ref{eq3}) by noting that $\max \left\{(N / T)^{\frac{1}{2}}, N^{-\frac{1}{2}}\right\}=o(\zeta)$. The proof of Theorem \ref{thm1} is completed.{\hfill$\square$}
	\subsection{Proof of Theorem \ref{thm2}}
	Define $\bar{\boldsymbol{\lambda}}_i=\frac{1}{\left|\mathcal{G}_k\right|}\sum_{i\in \mathcal{G}_k}\tilde{\boldsymbol{\lambda}}_i$, and we have
	$$
	\left\|\hat{\boldsymbol{\lambda}}_i^*-\widehat{\mathbf{H}}^\top \boldsymbol{\lambda}_i\right\|^2\leq 2\left\|\hat{\boldsymbol{\lambda}}_i^*-\bar{\boldsymbol{\lambda}}_i\right\|^2+2\left\|\bar{\boldsymbol{\lambda}}_i-\widehat{\mathbf{H}}^\top \boldsymbol{\lambda}_i\right\|^2.
	$$
	Since $\mathbf{\Lambda}^\top=(\boldsymbol{\lambda}_1,\ldots,\boldsymbol{\lambda}_N)$ hold the group structure, for all $i\in\mathcal{G}_k$,
	\begin{equation}\label{thm2EQ1}
		\begin{aligned}
			\left\|\bar{\boldsymbol{\lambda}}_i-\widehat{\mathbf{H}}^\top \boldsymbol{\lambda}_i\right\|^2&=\frac{1}{\left|\mathcal{G}_k\right|^2}\left\|\sum_{i\in\mathcal{G}_k}(\tilde{\boldsymbol{\lambda}}_i-\widehat{\mathbf{H}}^\top\boldsymbol{\lambda}_i)\right\|^2\\&\lesssim \frac{1}{\left|\mathcal{G}_k\right|}\sum_{i\in\mathcal{G}_k}\left\|\tilde{\boldsymbol{\lambda}}_i-\widehat{\mathbf{H}}^\top\boldsymbol{\lambda}_i\right\|^2\\
			&\le \frac{1}{\left|\mathcal{G}_k\right|}O_p(\frac{N}{T}+\frac{1}{N}),
		\end{aligned}
	\end{equation}
	where the last equation follows from Theorem 1 in \cite{he2022large}.\\
	In order to calculate  $\left\|\hat{\boldsymbol{\lambda}}_i^*-\bar{\boldsymbol{\lambda}}_i\right\|^2$, we note that $$
	\bar{\boldsymbol{\lambda}}_i=\frac{1}{\left|\mathcal{G}_k\right|}\sum_{i\in \mathcal{G}_k}\tilde{\boldsymbol{\lambda}}_i=\frac{1}{\left|\mathcal{G}_k\right|}(\widetilde{\mathbf{F}}^\top\widetilde{\mathbf{F}})^{-1}\widetilde{\mathbf{F}}^\top\sum_{i\in \mathcal{G}_k}\widetilde{\mathbf{F}}\tilde{\boldsymbol{\lambda}}_i
	$$
	and $$
	\hat{\boldsymbol{\lambda}}_i^*=\frac{1}{\left|\mathcal{G}_k\right|}(\widetilde{\mathbf{F}}^\top\widetilde{\mathbf{F}})^{-1}\widetilde{\mathbf{F}}^\top\sum_{i\in \mathcal{G}_k}\underline{\boldsymbol{y}}_i=\frac{1}{\left|\mathcal{G}_k\right|}(\widetilde{\mathbf{F}}^\top\widetilde{\mathbf{F}})^{-1}\widetilde{\mathbf{F}}^\top\sum_{i\in \mathcal{G}_k}(\mathbf{F}\boldsymbol{\lambda}_i+\underline{\boldsymbol{\epsilon}}_i).
	$$
	Hence \begin{equation}\label{thm2eq0}
		\begin{aligned}
			\left\|\hat{\boldsymbol{\lambda}}_i^*-\bar{\boldsymbol{\lambda}}_i\right\|^2&=\frac{1}{\left|\mathcal{G}_k\right|^2}\left\|(\widetilde{\mathbf{F}}^\top\widetilde{\mathbf{F}})^{-1}\widetilde{\mathbf{F}}^\top\sum_{i\in \mathcal{G}_k}(\mathbf{F}\boldsymbol{\lambda}_i-\widetilde{\mathbf{F}}\tilde{\boldsymbol{\lambda}}_i+\underline{\boldsymbol{\epsilon}}_i)\right\|^2\\
			&\lesssim \frac{1}{\left|\mathcal{G}_k\right|}\left\|(\widetilde{\mathbf{F}}^\top\widetilde{\mathbf{F}})^{-1}\widetilde{\mathbf{F}}^\top\right\|^2\sum_{i\in \mathcal{G}_k}\left(\left\|\mathbf{F}\boldsymbol{\lambda}_i-\widetilde{\mathbf{F}}\tilde{\boldsymbol{\lambda}}_i\right\|^2+\|\underline{\boldsymbol{\epsilon}}_i\|^2\right).
		\end{aligned}
	\end{equation}
	On one hand, \textbf{Assumption} \ref{as1} and \textbf{Lemma A.2} in \cite{he2022large} show that $\|\boldsymbol{f}_t\|^2=O_p(1)$, then $$
\|\mathbf{F}^\top\|^2\le\|\mathbf{F}\|_F^2=\sum_{t=1}^{T}\|\boldsymbol{f}_t\|^2=O_p(T),$$
	similarly, $ \|\mathbf{F}\|^2=O_p(T) $. According to the \textbf{Lemma S6} in \cite{he2022large}, we have
 \begin{equation}\label{epsilon}
\|\boldsymbol{\epsilon}\|_F^2
=\sum_{t=1}^T\|\boldsymbol{\epsilon}_t\|^2=O_p(NT).
 \end{equation}
Then, we note that
    $$ \|\widetilde{\mathbf{F}}^\top\|^2=\left\|\frac{\widetilde{\mathbf{\Lambda}}^\top\mathbf{Y}^\top}{N}\right\|^2\le \frac{1}{N^2}\left\|\widetilde{\mathbf{\Lambda}}^\top{\mathbf{\Lambda}}\mathbf{F}^\top+\widetilde{\mathbf{\Lambda}}^\top\boldsymbol{\epsilon}^\top\right\|^2=O_p(T).
	$$
    Moreover, $$\begin{aligned}
	    \widehat{\mathbf{H}}\widetilde{\mathbf{F}}^\top\widetilde{\mathbf{F}}\widehat{\mathbf{H}}^\top&=(\widetilde{\mathbf{F}}\widehat{\mathbf{H}}^\top-\mathbf{F}+\mathbf{F})^\top(\widetilde{\mathbf{F}}\widehat{\mathbf{H}}^\top-\mathbf{F}+\mathbf{F})\\&=(\widetilde{\mathbf{F}}\widehat{\mathbf{H}}^\top-\mathbf{F})^\top(\widetilde{\mathbf{F}}\widehat{\mathbf{H}}^\top-\mathbf{F})+\mathbf{F}^\top(\widetilde{\mathbf{F}}\widehat{\mathbf{H}}^\top-\mathbf{F})+(\widetilde{\mathbf{F}}\widehat{\mathbf{H}}^\top-\mathbf{F})^\top\mathbf{F}+\mathbf{F}^\top\mathbf{F},
	\end{aligned}
	$$
	then by Weyl's inequality $\sigma_{min}(A+B)\le\sigma_{min}(A)+\sigma_{max}(B)$ and \textbf{Assumption} \ref{as4} and \textbf{Theorem 3.2} in \cite{he2022large},  $$
	\begin{aligned}
		&\sigma_{min}(\widehat{\mathbf{H}}\widetilde{\mathbf{F}}^\top\widetilde{\mathbf{F}}\widehat{\mathbf{H}}^\top)\\
  \geq& \sigma_{min}(\mathbf{F}^\top\mathbf{F})-\sigma_{max}\left((\widetilde{\mathbf{F}}\widehat{\mathbf{H}}^\top-\mathbf{F})^\top(\widetilde{\mathbf{F}}\widehat{\mathbf{H}}^\top-\mathbf{F})+\mathbf{F}^\top(\widetilde{\mathbf{F}}\widehat{\mathbf{H}}^\top-\mathbf{F})+(\widetilde{\mathbf{F}}\widehat{\mathbf{H}}^\top-\mathbf{F})^\top\mathbf{F}\right)\\\gtrsim& \sigma_{min}(\mathbf{F}^\top\mathbf{F})-\left(\sum_{t=1}^T\left\|\widehat{\mathbf{H}}\tilde{\boldsymbol{f}}_t-\boldsymbol{f}_t\right\|^2+2\|\mathbf{F}\|\left(\sum_{t=1}^T\left\|\widehat{\mathbf{H}}\tilde{\boldsymbol{f}}_t-\boldsymbol{f}_t\right\|^2\right)^{\frac{1}{2}}\right)\\
		\geq& CT+o_p(T) \quad \text{as}\quad T,N \to \infty,
	\end{aligned}
	$$we note that $\sigma_{max}(\mathbf{A})=\sigma_{max}(\mathbf{-A})$,
	where $\sigma_{max}$ and $\sigma_{min}$ denote the maximum and minimum singular values of a matrix, $C$ is a constant.
Thus $\widehat{\mathbf{H}}\widetilde{\mathbf{F}}^\top\widetilde{\mathbf{F}}\widehat{\mathbf{H}}^\top$ as well as $(\mathbf{V}^{\frac{1}{2}}\widehat{\mathbf{H}})\widetilde{\mathbf{F}}^\top\widetilde{\mathbf{F}}(\mathbf{V}^{\frac{1}{2}}\widehat{\mathbf{H}})^\top$ have the minimum singular value of order equal or greater than $T$. Using the fact that $(\mathbf{V}^{\frac{1}{2}}\widehat{\mathbf{H}})^\top(\mathbf{V}^{\frac{1}{2}}\widehat{\mathbf{H}})=\widehat{\mathbf{H}}^\top \mathbf{V}\widehat{\mathbf{H}}\xrightarrow{p} \mathbf{I}_m$, we obtain that $\sigma_{min}(\widetilde{\mathbf{F}}^\top\widetilde{\mathbf{F}})\ge CT+o_p(T).$\\
	Hence, we have $$\left\|\left(\widetilde{\mathbf{F}}^\top\widetilde{\mathbf{F}}\right)^{-1}\right\|=O_p(\frac{1}{T})$$.\\
	As a result,\begin{equation}\label{thm2eq1}
		\left\|(\widetilde{\mathbf{F}}^\top\widetilde{\mathbf{F}})^{-1}\widetilde{\mathbf{F}}^\top\right\|^2\le\left\|(\widetilde{\mathbf{F}}^\top\widetilde{\mathbf{F}})^{-1}\right\|^2 \left\|\widetilde{\mathbf{F}}^\top\right\|^2=O_p(\frac{1}{T}).
	\end{equation}
	On the other hand,\begin{equation}\label{thm2eq2}
		\begin{aligned}
			\frac{1}{T}\sum_{i\in \mathcal{G}_k}\left\|\mathbf{F}\boldsymbol{\lambda}_i-\widetilde{\mathbf{F}}\tilde{\boldsymbol{\lambda}}_i\right\|^2&\le \frac{1}{T}\left\|\mathbf{F}\boldsymbol{\Lambda}^{\top}-\widetilde{\mathbf{F}}\tilde{\boldsymbol{\Lambda}}^{\top}\right\|^2_{F} = \frac{1}{T}\sum_{t=1}^{T}\left\|\boldsymbol{\Lambda}\boldsymbol{f}_t-\tilde{\boldsymbol{\Lambda}}\widetilde{\boldsymbol{f}_t}\right\|^2 \\
			&=O_p(\frac{N}{T}+\frac{1}{N}),
		\end{aligned}
	\end{equation}
	where the last equation follows from \textbf{Theorem 3.3} in \cite{he2022large}.
	Furthermore, from \textbf{Assumption} \ref{as8} we have $N^{-1}\sum_{i=1}^{N}(T^{-1}\|\sum_{t=1}^T\boldsymbol{f}_t\epsilon_{it}\|^2)=O_p(1)$. Then we combine the (\ref{epsilon}) and \textbf{Theorem 3.2} in \cite{he2022large}, we note that
\begin{equation*}
\begin{aligned}
\frac{1}{N}\sum_{i=1}^N\frac{1}{T}\cdot\left\|\sum_{t=1}^T(\widetilde{\boldsymbol{f}}_t-\widehat{\mathbf{H}}^{-1}\boldsymbol{f}_t)\epsilon_{it}\right\|^2
\le& \frac{1}{N}\sum_{i=1}^N\frac{1}{T}\cdot\left\|\sum_{t=1}^T\frac{1}{N}\widetilde{\boldsymbol{\Lambda}}^{\top}(\widetilde{\boldsymbol{\Lambda}}-{\boldsymbol{\Lambda}}\widehat{\mathbf{H}})\widehat{\mathbf{H}}^{-1}\boldsymbol{f}_t
\epsilon_{it}\right\|^2\\
&+\frac{1}{N}\sum_{i=1}^N\frac{1}{T}\cdot\left\|\frac{1}{N}(\widetilde{\boldsymbol{\Lambda}}-{\boldsymbol{\Lambda}}\widehat{\mathbf{H}})^{\top}\sum_{t=1}^T\boldsymbol{\epsilon}_t
\epsilon_{it}\right\|^2\\
&+\frac{1}{N}\sum_{i=1}^N\frac{1}{T}\cdot\left\|\frac{1}{N}\widehat{\mathbf{H}}^{\top}\boldsymbol{\Lambda}^{\top}\sum_{t=1}^T\boldsymbol{\epsilon}_t
\epsilon_{it}\right\|^2
\end{aligned}
\end{equation*}
where
$$
\begin{aligned}
    \frac{1}{N}\sum_{i=1}^N\frac{1}{T}\cdot\left\|\sum_{t=1}^T\frac{1}{N}\widetilde{\boldsymbol{\Lambda}}^{\top}(\widetilde{\boldsymbol{\Lambda}}-{\boldsymbol{\Lambda}}\widehat{\mathbf{H}})\widehat{\mathbf{H}}^{-1}\boldsymbol{f}_t
\epsilon_{it}\right\|^2&\lesssim O_p(\frac{1}{N^2}+\frac{1}{T^2})\frac{1}{N}\sum_{i=1}^N\frac{1}{T}\left\|\sum_{t=1}^T\boldsymbol{f}_t\epsilon_{it}\right\|^2\\&\lesssim O_p(\frac{1}{N^2}+\frac{1}{T^2}),\\
\frac{1}{N}\sum_{i=1}^N\frac{1}{T}\cdot\left\|\frac{1}{N}(\widetilde{\boldsymbol{\Lambda}}-{\boldsymbol{\Lambda}}\widehat{\mathbf{H}})^{\top}\sum_{t=1}^T\boldsymbol{\epsilon}_t
\epsilon_{it}\right\|^2
&\lesssim O_p(\frac{1}{N^2}+\frac{1}{T})\frac{1}{N^2T}\sum_{i=1}^N\left\|\sum_{t=1}^T\boldsymbol{\epsilon}_t
\epsilon_{it}\right\|^2\\
&\lesssim O_p(\frac{1}{N^2}+\frac{1}{T})\frac{1}{N^2T}\sum_{t=1}^T\left\|\boldsymbol{\epsilon}_t\right\|^2\sum_{i=1}^N\epsilon_{it}^2\\
&\lesssim O_p(\frac{1}{N^2}+\frac{1}{T})
\end{aligned}$$

\begin{equation*}
\begin{aligned}
\frac{1}{N}\sum_{i=1}^N\frac{1}{T}\cdot\left\|\frac{1}{N}\widehat{\mathbf{H}}^{\top}\boldsymbol{\Lambda}^{\top}\sum_{t=1}^T\boldsymbol{\epsilon}_t
\epsilon_{it}\right\|^2
&\le \frac{1}{NT}\sum_{t=1}^T\left\|\frac{1}{N}\boldsymbol{\Lambda}^{\top}\boldsymbol{\epsilon}_t\right\|^2\sum_{i=1}^N\epsilon_{it}^2\\
&\lesssim O_p(\frac{1}{N})
\end{aligned}
\end{equation*}
Then we have
\begin{equation}\label{recover1}
		\begin{aligned}
\frac{1}{N}\sum_{i=1}^N\frac{1}{T}\cdot\left\|\sum_{t=1}^T\widetilde{\boldsymbol{f}}_t\epsilon_{it}\right\|^2&=\frac{1}{N}\sum_{i=1}^N\frac{1}{T}\cdot\left\|\sum_{t=1}^T(\widetilde{\boldsymbol{f}}_t-\widehat{\mathbf{H}}^{-1}\boldsymbol{f}_t)\epsilon_{it}+\widehat{\mathbf{H}}^{-1}\sum_{t=1}^T\boldsymbol{f}_t\epsilon_{it}\right\|^2\\
&\lesssim O_p(\frac{1}{T}+\frac{1}{N})+O_p(1)\\&=O_p(1).
\end{aligned}
	\end{equation}
As a result,
 \begin{equation}\label{recover2}
		\begin{aligned}
 \left\|\frac{1}{|\mathcal{G}_k|}(\widetilde{\mathbf{F}}^\top\widetilde{\mathbf{F}})^{-1}\widetilde{\mathbf{F}}^\top\sum_{i\in \mathcal{G}_k}\underline{\boldsymbol{\epsilon}_i}\right\|^2&\le \frac{1}{|\mathcal{G}_k|^2}\cdot \left\|(\widetilde{\mathbf{F}}^\top\widetilde{\mathbf{F}})^{-1}\right\|^2\cdot \left\|\sum_{i\in \mathcal{G}_k}\widetilde{\mathbf{F}}^\top\underline{\boldsymbol{\epsilon}_i}\right\|^2\\
 &\le \frac{1}{|\mathcal{G}_k|^2}\cdot O_p(\frac{1}{T^2})\cdot |\mathcal{G}_k|\sum_{i\in \mathcal{G}_k}\left\|\sum_{t=1}^T\widetilde{\boldsymbol{f}}_t\epsilon_{it}\right\|^2\\
 &=O_p(\frac{1}{T})\cdot \frac{1}{|\mathcal{G}_k|}\sum_{i\in \mathcal{G}_k}\left(\frac{1}{T}\left\|\sum_{t=1}^T\widetilde{\boldsymbol{f}}_t\epsilon_{it}\right\|^2\right)\\&
 =O_p(\frac{1}{T}).
\end{aligned}
	\end{equation}
 Therefore, we obtain that\begin{equation}\label{thm2EQ2}
		\left\|\hat{\boldsymbol{\lambda}}_i^*-\bar{\boldsymbol{\lambda}}_i\right\|^2=
\frac{1}{\left|\mathcal{G}_k\right|}O_p(\frac{N}{T}+\frac{1}{N}).
	\end{equation}
	Combining (\ref{thm2EQ1}), (\ref{thm2EQ2}) and Assumption \ref{as5}(a) completes the proof.{\hfill$\square$}
	\subsection{Proof of Theorem \ref{thm3}}
	\subsubsection{Proof of Theorem \ref{thm3}(a)} From the definition of $\hat{K}$, we only need to show that
	$$
	P\left[\mathbb{I} \mathbb{C}\left(K_0\right)=\min _{1 \leq K \leq \bar{K}} \mathbb{I} \mathbb{C}(K)\right] \rightarrow 1 .
	$$
	Consider the following two cases: (i) $1 \leq K \leq K_0-1$ and (ii) $K_0+1 \leq K \leq \bar{K}$, which correspond to under-identification and over-identification of the latent groups, respectively. Recall that $\mathcal{M}(\mathcal{G})$ denotes the event $\left\{\widehat{\mathcal{G}}_1, \widehat{\mathcal{G}}_2, \ldots, \widehat{\mathcal{G}}_{K_0}\right\}=\left\{\mathcal{G}_1, \mathcal{G}_2, \ldots, \mathcal{G}_{K_0}\right\}$. For case (i), by \textbf{Theorem} \ref{thm1} and \textbf{Lemma} \ref{lmA2}, we have
	$$
	\begin{aligned}
		& P\left[\mathbb{I} \mathbb{C}\left(K_0\right)<\mathbb{I} \mathbb{C}(K), 1 \leq K \leq K_0-1\right] \\
		= & P\left[\mathbb{I} \mathbb{C}\left(K_0\right)<\mathbb{I} \mathbb{C}(K), 1 \leq K \leq K_0-1, \mathcal{M}(\mathcal{G})\right]\\
  &+P\left[\mathbb{I} \mathbb{C}\left(K_0\right) <\mathbb{I} \mathbb{C}(K), 1 \leq K \leq K_0-1, \mathcal{M}^c(\mathcal{G})\right] \\
		= & 1+o(1) .
	\end{aligned}
	$$
	On the other hand, for case (ii), by \textbf{Theorem} \ref{thm1} and \textbf{Lemma} \ref{lmA3}, we have
	$$
	\begin{aligned}
		& P\left[\mathbb{I} \mathbb{C}\left(K_0\right)<\mathbb{I} \mathbb{C}(K), K_0+1 \leq K \leq \bar{K}\right] \\
		= & P\left[\mathbb{I} \mathbb{C}\left(K_0\right)<\mathbb{I} \mathbb{C}(K), K_0+1 \leq K \leq \bar{K}, \mathcal{M}(\mathcal{G})\right]\\
  &+P\left[\mathbb{I} \mathbb{C}\left(K_0\right)<\mathbb{I} \mathbb{C}(K), K_0+1 \leq K \leq \bar{K}, \mathcal{M}^c(\mathcal{G})\right] \\
		= & 1+o(1) .
	\end{aligned}
	$$
	Combining the above results completes the proof of Theorem \ref{thm3} (a).{\hfill$\square$}
	\subsubsection{Proof of Theorem \ref{thm3}(b)}
	By \textbf{Theorem} \ref{thm1} and \textbf{Theorem} \ref{thm3} (a), we have
	$$
	\begin{aligned}
		P\left(\hat{\boldsymbol{\lambda}}_i=\hat{\boldsymbol{\lambda}}_i^*\right)  =&P\left[\mathbb{I} \mathbb{C}\left(K_0\right)=\min _{1 \leq K \leq \bar{K}} \mathbb{I} \mathbb{C}(K), \mathcal{M}(\mathcal{G})\right]\\
  &+P\left[\mathbb{I} \mathbb{C}\left(K_0\right)=\min _{1 \leq K \leq \bar{K}} \mathbb{I} \mathbb{C}(K), \mathcal{M}^c(\mathcal{G})\right] \\
		=&1+o(1),
	\end{aligned}
	$$
	where $\bar{K}$ is a pre-specified finite positive integer which is larger than $K_0$. The proof of Theorem \ref{thm3} (b) is completed.{\hfill$\square$}
	\section{Additional proofs}\label{appC}
	\subsection{Proof of Lemma \ref{lm1}}
	Assume $T$ is even and $\bar{T}=T / 2$, otherwise we can delete the last observation. Given a permutation of $\{1, \ldots, T\}$, denoted as $\sigma$, let $\boldsymbol{f}_t^\sigma, \boldsymbol{\epsilon}_t^\sigma$ and $\boldsymbol{y}_t^\sigma$ be the rearranged factors, errors and observations, further define
	$$
	\mathbf{M}_2^\sigma=\frac{1}{\bar{T}} \sum_{s=1}^{\bar{T}} \frac{\left(\boldsymbol{\epsilon}_{2 s-1}^\sigma-\boldsymbol{\epsilon}_{2 s}^\sigma\right)\left(\boldsymbol{f}_{2 s-1}^\sigma-\boldsymbol{f}_{2 s}^\sigma\right)^{\top}}{\left\|\boldsymbol{y}_{2 s-1}^\sigma-\boldsymbol{y}_{2 s}^\sigma\right\|^2}
	$$
	Denote $\mathcal{S}_T$ as the set containing all the permutations of $\{1, \ldots, T\}$, then it's not hard to prove that
	$$
	\sum_{\sigma \in \mathcal{S}_T} \frac{T}{2} \mathbf{M}_2^\sigma=T \times(T-2) ! \times \frac{T(T-1)}{2} \mathbf{M}_2
	$$
	That is,
	$$
	\mathbf{M}_{2,i\cdot}=\frac{1}{T !} \sum_{\sigma \in \mathcal{S}_T} \mathbf{M}_{2,i\cdot}^\sigma
	$$
	$$
	\Rightarrow \mathbb{E}\left\|\mathbf{M}_{2,i\cdot}\right\|_2\leq \frac{1}{T !} \sum_{\sigma \in \mathcal{S}_T} \mathbb{E}\left\|\mathbf{M}_{2,i\cdot}^\sigma\right\|_2=\mathbb{E}\left\|\mathbf{M}_{2,i\cdot}^\sigma\right\|_2 \leq \sqrt{\mathbb{E}\left\|\mathbf{M}_{2,i\cdot}^\sigma\right\|_2^2} \text { for any } \sigma .
	$$
	Now take $\sigma$ as given, i.e., $\sigma=\{1, \ldots, T\}$ which is the original order. By the property of elliptical distribution, for any $s=1, \ldots, \bar{T}$,
	$$
	\left(\begin{array}{c}
		\boldsymbol{f}_{2 s-1}-\boldsymbol{f}_{2 s} \\
		\boldsymbol{\epsilon}_{2 s-1}-\boldsymbol{\epsilon}_{2 s}
	\end{array}\right) \stackrel{d}{=} \xi_1\left(\begin{array}{cc}
		\mathbf{I}_m & \mathbf{0} \\
		\mathbf{0} & \mathbf{A}
	\end{array}\right) \frac{\boldsymbol{g}}{\|\boldsymbol{g}\|}
	$$
	where $\xi_1$ is determined by $\xi$. $\boldsymbol{g} \sim \mathcal{N}_{m+N}(\mathbf{0}, \mathbf{I})$ and $\boldsymbol{g}$ is independent of $\xi_1$. $ \mathbf{A A}^{\top}=\boldsymbol{\Sigma}_\epsilon$. Hence,
	$$
	\mathbf{X}_s:=\frac{\left(\boldsymbol{\epsilon}_{2 s-1}^\sigma-\boldsymbol{\epsilon}_{2 s}^\sigma\right)\left(\boldsymbol{f}_{2 s-1}^\sigma-\boldsymbol{f}_{2 s}^\sigma\right)^{\top}}{\left\|\boldsymbol{y}_{2 s-1}^\sigma-\boldsymbol{y}_{2 s}^\sigma\right\|^2} \stackrel{d}{=} \frac{\mathbf{A} \boldsymbol{g}_2 \boldsymbol{g}_1^{\top}}{\left\|\mathbf{\Lambda} \boldsymbol{g}_1+\mathbf{A} \boldsymbol{g}_2\right\|^2}
	$$
	where $\boldsymbol{g}_1$ is composed of the first $m$ entries of $\boldsymbol{g}$ and $\boldsymbol{g}_2$ is composed of the left ones. Note that $\mathbf{X}_s$ and $\mathbf{X}_t$ are independently and identically distributed when $s \neq t$, so
	$$
	\mathbb{E}\left\|\mathbf{M}_{2,i\cdot}^\sigma\right\|_2^2=\mathbb{E}\left\|\frac{1}{\bar{T}} \sum_{s=1}^{\bar{T}}\mathbf{X}_{s,i\cdot}\right\|_2^2=\frac{1}{\bar{T}} \mathbb{E}\left\|\mathbf{X}_{1,i\cdot}\right\|_2^2+\frac{\bar{T}(\bar{T}-1)}{\bar{T}^2}\left\|\mathbb{E} \mathbf{X}_{1,i\cdot}\right\|_2^2
	$$
	We first focus on $\mathbb{E} \mathbf{X}_{1,i\cdot}$, where $$\mathbf{X}_{1,i\cdot}=\frac{-\boldsymbol{\lambda}_i^{\top}\boldsymbol{u}_1\boldsymbol{u}_1^{\top}-\boldsymbol{\lambda}_i^{\top}\boldsymbol{u}_1\mathbf{\Lambda}^{\top}\boldsymbol{\Sigma}_y^{-1}\boldsymbol{u}_2+\mathbf{A}_{i\cdot} \mathbf{A}^{\top} \boldsymbol{\Sigma}_y^{-1} \boldsymbol{u}_2\boldsymbol{u}_2^{\top}\boldsymbol{\Sigma}_y^{-1}\mathbf{\Lambda}}
	{\left\|\boldsymbol{u}_2\right\|^2}.$$
	and then we define
	$$
	\left(\begin{array}{l}
		\boldsymbol{u}_1 \\
		\boldsymbol{u}_2
	\end{array}\right)=\left(\begin{array}{cc}
		\mathbf{I}_m & -\mathbf{\Lambda}^{\top} \boldsymbol{\Sigma}_y^{-1} \\
		\mathbf{0} & \mathbf{I}_N
	\end{array}\right)\left(\begin{array}{cc}
		\mathbf{I}_m & \mathbf{0} \\
		\mathbf{\Lambda} & \mathbf{A}
	\end{array}\right)\left(\begin{array}{l}
		\boldsymbol{g}_1 \\
		\boldsymbol{g}_2
	\end{array}\right) \sim \mathcal{N}\left(\mathbf{0},\left(\begin{array}{cc}
		\boldsymbol{\Sigma}_{\boldsymbol{u}_1} & \mathbf{0} \\
		\mathbf{0} & \boldsymbol{\Sigma}_y
	\end{array}\right)\right),
	$$
	where $\boldsymbol{\Sigma}_{\boldsymbol{u}_1}=\mathbf{I}_m-\mathbf{\Lambda}^{\top} \boldsymbol{\Sigma}_y^{-1} \mathbf{\Lambda}$, then $\boldsymbol{u}_1$ and $\boldsymbol{u}_2$ are independent and
	$$
 \begin{aligned}
	\left(\begin{array}{l}
		\boldsymbol{g}_1 \\
		\boldsymbol{g}_2
	\end{array}\right)&=\left(\begin{array}{cc}
		\mathbf{I}_m & \mathbf{0} \\
		-\mathbf{A}^{-1} \mathbf{\Lambda} & \mathbf{A}^{-1}
	\end{array}\right)\left(\begin{array}{cc}
		\mathbf{I}_m & \mathbf{\Lambda}^{\top} \boldsymbol{\Sigma}_y^{-1} \\
		\mathbf{0} & \mathbf{I}_N
	\end{array}\right)\left(\begin{array}{l}
		\boldsymbol{u}_1 \\
		\boldsymbol{u}_2
	\end{array}\right) \\
 &=\left(\begin{array}{cc}
		\mathbf{I}_m & \mathbf{\Lambda}^{\top} \boldsymbol{\Sigma}_y^{-1} \\
		-\mathbf{A}^{-1} \mathbf{\Lambda} & \mathbf{A}^{\top} \boldsymbol{\Sigma}_y^{-1}
	\end{array}\right)\left(\begin{array}{l}
		\boldsymbol{u}_1 \\
		\boldsymbol{u}_2
	\end{array}\right) .
 \end{aligned}
	$$
	As a result,
	$$
	\frac{\mathbf{A} \boldsymbol{g}_2 \boldsymbol{g}_1^{\top}}{\left\|\mathbf{\Lambda} \boldsymbol{g}_1+\mathbf{A} \boldsymbol{g}_2\right\|^2}=\frac{\left(-\mathbf{\Lambda} \boldsymbol{u}_1+\mathbf{A} \mathbf{A}^{\top} \boldsymbol{\Sigma}_y^{-1} \boldsymbol{u}_2\right)\left(\boldsymbol{u}_1+\mathbf{\Lambda}^{\top} \boldsymbol{\Sigma}_y^{-1} \boldsymbol{u}_2\right)^{\top}}{\left\|\boldsymbol{u}_2\right\|^2}
	$$
	Because $\boldsymbol{u}_1$ and $\boldsymbol{u}_2$ are zero-mean independent Gaussian vectors, we have
	$$
	\mathbb{E} \frac{\boldsymbol{u}_2 \boldsymbol{u}_1^{\top}}{\left\|\boldsymbol{u}_2\right\|^2}=\mathbf{0}, \quad \mathbb{E}\left\|\boldsymbol{u}_2\right\|^{-2} \leq \frac{1}{\lambda_N\left(\boldsymbol{\Sigma}_y\right)} \mathbb{E} \frac{1}{\chi_N^2} \asymp N^{-1},
	$$
	where $\chi^2_N$ is a Chi-square random variable with degree $N$. Hence,
	$$
	\mathbb{E} \mathbf{X}_1=-\mathbb{E}\left\|\boldsymbol{u}_2\right\|^{-2} \mathbf{\Lambda} \boldsymbol{\Sigma}_{\boldsymbol{u}_1}+\mathbf{A A}^{\top} \boldsymbol{\Sigma}_y^{-1} \mathbb{E} \frac{\boldsymbol{u}_2 \boldsymbol{u}_2^{\top}}{\left\|\boldsymbol{u}_2\right\|^2} \boldsymbol{\Sigma}_y^{-1} \mathbf{\Lambda}
	$$
	then
	$$
	\mathbb{E} \mathbf{X}_{1,i\cdot}=-\mathbb{E}\left\|\boldsymbol{u}_2\right\|^{-2} \boldsymbol{\Sigma}_{\boldsymbol{u}_1}\boldsymbol{\lambda}_i+\boldsymbol{\Sigma}_{\epsilon,i\cdot} \boldsymbol{\Sigma}_y^{-1} \mathbb{E} \frac{\boldsymbol{u}_2 \boldsymbol{u}_2^{\top}}{\left\|\boldsymbol{u}_2\right\|^2} \boldsymbol{\Sigma}_y^{-1} \mathbf{\Lambda}
	$$
	so we have
	$$
 \begin{aligned}
     \left\|\mathbb{E} \mathbf{X}_{1,i\cdot}\right\|_2^2 &\leq \left\|\boldsymbol{\Sigma}_{\boldsymbol{u}_1}\right\|_F^2\left(\mathbb{E}\left\|\boldsymbol{u}_2\right\|^{-2}\right)^2\|\boldsymbol{\lambda}_i\|^2+\left\|\boldsymbol{\Sigma}_{\epsilon,i\cdot}\right\|^2\left\|\boldsymbol{\Sigma}_y^{-1} \mathbb{E} \frac{\boldsymbol{u}_2 \boldsymbol{u}_2^{\top}}{\left\|\boldsymbol{u}_2\right\|^2} \boldsymbol{\Sigma}_y^{-1} \mathbf{\Lambda}\right\|_F^2 \\&=O(N^{-3}).
 \end{aligned}
	$$
	Now we move to the calculation of $\mathbb{E}\left\|\mathbf{X}_{1,i\cdot}\right\|_2^2$. As \cite{he2022large} noted that
	$$
\begin{aligned}
&\mathbb{E}\left\|\boldsymbol{u}_1\right\|^2 \lesssim N^{-1}, \quad \mathbb{E}\left\|\boldsymbol{u}_1\right\|^4 \leq 2^m \times 3 \mathbb{E}\left\|\boldsymbol{u}_1\right\|^2 \lesssim N^{-1}, \\
&\mathbb{E}\left\|\boldsymbol{u}_2\right\|^{-4} \leq \frac{1}{\lambda_N^2\left(\boldsymbol{\Sigma}_y\right)} \mathbb{E} \chi_N^{-2} \asymp N^{-2}.
\end{aligned}
 $$
	Let $\mathbf{D}=(d_{ij})_{N\times N}:=\boldsymbol{\Sigma}_\epsilon \boldsymbol{\Sigma}_y^{-\frac{1}{2}}$, and $\boldsymbol{\Sigma}_y^{-\frac{1}{2}}\boldsymbol{u}_2:=(\boldsymbol{e}_1,\ldots,\boldsymbol{e}_N)^\top \sim \mathcal{N}\left(\mathbf{0},\mathbf{I}_N\right) $, we have
	$$
	\begin{aligned}
\mathbb{E}\left(\left|\boldsymbol{\Sigma}_{\epsilon,i\cdot}  \boldsymbol{\Sigma}_y^{-1} \boldsymbol{u}_2\right|^4\right)&=\mathbb{E}\left((\Sigma_{j=1}^{N}d_{ij}\boldsymbol{e}_j)^4\right)\\
  &=6\Sigma_{1\le j\neq k\le N} \mathbb{E}(d_{ij}^2 \boldsymbol{e}_j^2 d_{ik}^2 \boldsymbol{e}_k^2)+\Sigma_{j=1}^N\mathbb{E}(d_{ij}^4 \boldsymbol{e}_j^4) \\
		&\le 6(\Sigma_{j=1}^N d_{ij}^2)^2\leq 6(\left\|\mathbf{D}\right\|^2)^2=O(1).
	\end{aligned}
	$$
	Then by Cauchy-Schwartz inequality and H{\"o}lder inequality,
	$$
	\begin{aligned}
&\mathbb{E}\left\|\mathbf{X}_{1,i\cdot}\right\|_2^2  \\&\lesssim \mathbb{E} \frac{\left(\left|\boldsymbol{\lambda}_i^\top \boldsymbol{u}_1\right|^2+\left|\boldsymbol{\Sigma}_{\epsilon,i\cdot}  \boldsymbol{\Sigma}_y^{-1} \boldsymbol{u}_2\right|^2\right)\left(\left\|\boldsymbol{u}_1\right\|^2+\left\|\mathbf{\Lambda}^{\top} \boldsymbol{\Sigma}_y^{-1} \boldsymbol{u}_2\right\|^2\right)}{\left\|\boldsymbol{u}_2\right\|^4} \\
		& \lesssim \mathbb{E}\left(\frac{\|\boldsymbol{\lambda}_i\|^2\left\|\boldsymbol{u}_1\right\|^4}{\left\|\boldsymbol{u}_2\right\|^4}+\frac{\|\boldsymbol{\lambda}_i\|^2\left\|\mathbf{\Lambda}^{\top} \boldsymbol{\Sigma}_y^{-1}\right\|^2\left\|\boldsymbol{u}_1\right\|^2}{\left\|\boldsymbol{u}_2\right\|^2}+\frac{\left\|\boldsymbol{\Sigma}_{\epsilon,i\cdot}  \boldsymbol{\Sigma}_y^{-1}\right\|^2\left\|\boldsymbol{u}_1\right\|^2}{\left\|\boldsymbol{u}_2\right\|^2}\right)\\
&\quad+\mathbb{E}\left(|\boldsymbol{\Sigma}_{\epsilon,i\cdot}  \boldsymbol{\Sigma}_y^{-1} \boldsymbol{u}_2|^2\frac{\left\|\mathbf{\Lambda}^{\top} \boldsymbol{\Sigma}_y^{-1}\right\|^2}{\left\|\boldsymbol{u}_2\right\|^2}\right)\\
		&\lesssim O(N^{-2})+\left(\mathbb{E}\left(\left|\boldsymbol{\Sigma}_{\epsilon,i\cdot}  \boldsymbol{\Sigma}_y^{-1} \boldsymbol{u}_2\right|^4\right)\right)^{\frac{1}{2}}\left(\mathbb{E}(\left\|\boldsymbol{u}_2\right\|^{-4})\right)^{\frac{1}{2}}\left\|\mathbf{\Lambda}^{\top} \boldsymbol{\Sigma}_y^{-1}\right\|^2 \\
  &=O\left(N^{-2}\right)
	\end{aligned}
	$$
	As a result,
	$$
	\mathbb{E}\left\|\mathbf{M}_{2,i\cdot}^\sigma\right\|_2 \lesssim \sqrt{(T N^2)^{-1}+N^{-3}} \Rightarrow\left\|\mathbf{M}_{2,i\cdot}\right\|_2^2=O_p\left(\frac{1}{T N^2}+\frac{1}{N^3}\right)
	$$
	which concludes the lemma.{\hfill$\square$}
	\subsection{Proof of Lemma \ref{lm2}}
	By the decomposition that $\widetilde{\mathbf{\Lambda}}=\widetilde{\mathbf{\Lambda}}-\mathbf{\Lambda} \widehat{\mathbf{H}}+\mathbf{\Lambda} \widehat{\mathbf{H}}$, we have
	$$
	\left\|\mathbf{M}_{4,i\cdot} \widetilde{\mathbf{\Lambda}}\right\|_2^2 \lesssim \left\|\mathbf{M}_{4,i\cdot} \mathbf{\Lambda}\right\|_2^2\|\widehat{\mathbf{H}}\|_F^2+N\left\|\mathbf{M}_{4,i\cdot}\right\|_2^2 \times \frac{1}{N}\|\widehat{\mathbf{\Lambda}}-\mathbf{\Lambda} \widehat{\mathbf{H}}\|_F^2 ,
	$$
	where $\|\widehat{\mathbf{H}}\|_F^2=O_p(1)$ from \textbf{Lemma S3} in \cite{he2022large}.
	
	We start with $\mathbf{M}_{4,i\cdot} \mathbf{\Lambda}$. Similarly to the proof of \textbf{Lemma} \ref{lm1}, we have $$\mathbb{E}\left\|\mathbf{M}_{4,i\cdot} \mathbf{\Lambda}\right\|_2 \leq \sqrt{\frac{1}{\bar{T}} \mathbb{E}\|\mathbf{X}_{i\cdot}\|_2^2+\frac{\bar{T}(\bar{T}-1)}{\bar{T}^2}\|\mathbb{E} \mathbf{X}_{i\cdot}\|_2^2},$$
	where $\mathbf{X} \stackrel{d}{=} \frac{\left(-\mathbf{\Lambda} \boldsymbol{u}_1+\mathbf{A} \mathbf{A}^{\top} \boldsymbol{\Sigma}_y^{-1} \boldsymbol{u}_2\right)\left(-\mathbf{\Lambda} \boldsymbol{u}_1+\mathbf{A} \mathbf{A}^{\top} \boldsymbol{\Sigma}_y^{-1} \boldsymbol{u}_2\right)^{\top} \mathbf{\Lambda}}{\left\|\boldsymbol{u}_2\right\|^2}$, and $\boldsymbol{u}_1$ and $\boldsymbol{u}_2$ are the same as in the proof of \textbf{Lemma} \ref{lm1}. Therefore,
	$$
 \begin{aligned}
	\|\mathbb{E} \mathbf{X}_{i\cdot}\|_2^2 &\lesssim\|\boldsymbol{\lambda}_i\|_2^2\|\mathbf{\Lambda}^\top\mathbf{\Lambda}\|_F^2\left\|\boldsymbol{\Sigma}_{\boldsymbol{u}_1}\right\|_F^2\left(\mathbb{E}\left\|\boldsymbol{u}_2\right\|^{-2}\right)^2+\left\|\boldsymbol{\Sigma}_y^{-1} \mathbb{E} \frac{\boldsymbol{u}_2 \boldsymbol{u}_2^{\top}}{\left\|\boldsymbol{u}_2\right\|^2} \boldsymbol{\Sigma}_y^{-1}\right\|^2\|\mathbf{\Lambda}\|_F^2\\
 &=O\left(N^{-1}\right) .
 \end{aligned}
	$$
	On the other hand,
	$$
	\begin{aligned}
		\mathbb{E}\|\mathbf{X}_{i\cdot}\|_2^2 \lesssim & \|\boldsymbol{\lambda}_i\|_2^2 \mathbb{E}\left\|\boldsymbol{u}_1\right\|^4\left\|\mathbf{\Lambda}^{\top} \mathbf{\Lambda}\right\|_F^2 \mathbb{E}\left\|\boldsymbol{u}_2\right\|^{-4}\\
  &+\left\|\boldsymbol{\Sigma}_y^{-1}\right\|^2 \mathbb{E}\left\|\boldsymbol{u}_1\right\|^2\left\|\mathbf{\Lambda}^{\top} \mathbf{\Lambda}\right\|_F^2 \mathbb{E}\left\|\boldsymbol{u}_2\right\|^{-2} \\
		& +\|\boldsymbol{\lambda}_i\|_2^2 \mathbb{E}\left\|\boldsymbol{u}_1\right\|^2\left\|\boldsymbol{\Sigma}_y^{-1} \mathbf{A} \mathbf{A}^{\top} \mathbf{\Lambda}\right\|_F^2 \mathbb{E}\left\|\boldsymbol{u}_2\right\|^{-2}\\
		&+\left(\mathbb{E}\left(\left|\boldsymbol{\Sigma}_{\epsilon,i\cdot}  \boldsymbol{\Sigma}_y^{-1} \boldsymbol{u}_2\right|^4\right)\right)^{\frac{1}{2}}\left(\mathbb{E}(\left\|\boldsymbol{u}_2\right\|^{-4})\right)^{\frac{1}{2}}\left\|\boldsymbol{\Sigma}_\epsilon \boldsymbol{\Sigma}_y^{-1}\mathbf{\Lambda}\right\|^2\\
  =&O(1) .
	\end{aligned}
	$$
	As a result,
	$$
	\left\|\mathbf{M}_{4,i\cdot} \mathbf{\Lambda}\right\|_2^2=O_p\left(\frac{1}{T}+\frac{1}{N}\right) .
	$$
	It remains to show that $N\left\|\mathbf{M}_{4,i\cdot}\right\|_2^2=O_p(1)$.
	Similarly to the proof of \textbf{Lemma} \ref{lm1}, we have
	$$
	\mathbb{E}\left\|\mathbf{M}_{4,i\cdot}\right\|_2 \leq \sqrt{\frac{1}{\bar{T}} \mathbb{E}\|\mathbf{Z}_{i\cdot}\|_2^2+\frac{\bar{T}(\bar{T}-1)}{\bar{T}^2}\|\mathbb{E} \mathbf{Z}_{i\cdot}\|_2^2},$$
	where
	$$\mathbf{Z} \stackrel{d}{=} \frac{\left(-\mathbf{\Lambda} \boldsymbol{u}_1+\mathbf{A} \mathbf{A}^{\top} \boldsymbol{\Sigma}_y^{-1} \boldsymbol{u}_2\right)\left(\mathbf{\Lambda} \boldsymbol{u}_1+\mathbf{A} \mathbf{A}^{\top} \boldsymbol{\Sigma}_y^{-1} \boldsymbol{u}_2\right)^{\top}}{\left\|\boldsymbol{u}_2\right\|^2} .
	$$
	On one hand,
	$$
	\|\mathbb{E} \mathbf{Z}_{i\cdot}\|_2^2 \lesssim\|\boldsymbol{\lambda}_i\|_2^2\left\|\boldsymbol{\Sigma}_{\boldsymbol{u}_1}\right\|^2\|\mathbf{\Lambda}\|^2\left(\mathbb{E}\left\|\boldsymbol{u}_2\right\|^{-2}\right)^2+\left\|\boldsymbol{\Sigma}_y^{-1} \mathbb{E} \frac{\boldsymbol{u}_2 \boldsymbol{u}_2^{\top}}{\left\|\boldsymbol{u}_2\right\|^2} \boldsymbol{\Sigma}_y^{-1}\right\|^2=O\left(N^{-1}\right)
	$$
	On the other hand,
	$$
	\begin{aligned}
		\mathbb{E}\|\mathbf{Z}_{i\cdot}\|_F^2 \lesssim&\|\boldsymbol{\lambda}_i\|^2\|\mathbf{\Lambda}\|^2 \mathbb{E}\left\|\boldsymbol{u}_1\right\|^4 \mathbb{E}\left\|\boldsymbol{u}_2\right\|^{-4}+\left\|\boldsymbol{\Sigma}_y^{-1}\right\|^2\|\mathbf{\Lambda}\|^2 \mathbb{E}\left\|\boldsymbol{u}_1\right\|^2\mathbb{E}\left\|\boldsymbol{u}_2\right\|^{-2}\\
		&+\left(\mathbb{E}\left(\left|\boldsymbol{\Sigma}_{\epsilon,i\cdot}  \boldsymbol{\Sigma}_y^{-1} \boldsymbol{u}_2\right|^4\right)\right)^{\frac{1}{2}}\left(\mathbb{E}(\left\|\boldsymbol{u}_2\right\|^{-4})\right)^{\frac{1}{2}}\left\|\boldsymbol{\Sigma}_\epsilon \boldsymbol{\Sigma}_y^{-1}\right\|^2 =O(N^{-1}) .
	\end{aligned}
	$$
	Hence, $
	N\left\|\mathbf{M}_{4,i\cdot} \right\|_2^2=NO_p\left(\frac{1}{TN}+\frac{1}{N}\right)=O_p(1).
	${\hfill$\square$}
	\subsection{Proof of Lemma \ref{lmA2}} Without loss of generality, we only consider the case where $K=K_{0}-1$ as the other cases can be dealt with in the same manner. Since
	$$
	\mathbb{I} \mathbb{C}\left(K_{0}-1\right)-\mathbb{I} \mathbb{C}\left(K_{0}\right)=\ln \left[S\left(K_{0}-1\right) / S\left(K_{0}\right)\right]-\rho
	$$
	it is sufficient to prove
	\begin{equation}\label{C2}
		P\left[S\left(K_{0}-1\right)-S\left(K_{0}\right)<\rho, \mathcal{M}(\mathcal{G})\right] \rightarrow 0
	\end{equation}
	as $N, T \rightarrow \infty$ and the rest is completely analogous to the proof of Corollary 1 in \cite{bai2002determining}.\\
	Conditional on the event $\mathcal{M}(\mathcal{G})$, two of the groups among $\left\{\mathcal{G}_{1}, \mathcal{G}_{2}, \ldots, \mathcal{G}_{K_{0}}\right\}$ are falsely merged when the AHC algorithm stops at $K=K_{0}-1$. Without loss of generality, we assume that $\mathcal{G}_{K_{0}-1}$ and $\mathcal{G}_{K_{0}}$ are falsely merged. Denote $\mathcal{G}_{K_{0}-1 \mid K_{0}-1}=\mathcal{G}_{K_{0}-1} \bigcup \mathcal{G}_{K_{0}}$ and
	$\overline{\mathcal{G}}=\left\{\mathcal{G}_{1}, \mathcal{G}_{2}, \ldots, \mathcal{G}_{K_{0}-2}, \mathcal{G}_{K_{0}-1 \mid K_{0}-1}\right\}$. To evaluate $S\left(K_{0}-1\right)-S\left(K_{0}\right)$, we first prove that for $k=1, \ldots, K_{0}-2$,
	\begin{equation}\label{C3}
		\hat{\boldsymbol{\lambda}}_{i \mid \mathcal{G}}=\hat{\boldsymbol{\lambda}}_{i \mid \overline{\mathcal{G}}} \text {, if } i \in \mathcal{G}_{k},
	\end{equation}
	where $\hat{\boldsymbol{\lambda}}_{i \mid \mathcal{G}}$ and $\hat{\boldsymbol{\lambda}}_{i \mid \overline{\mathcal{G}}}$ are the group-specific loading estimates given in (\ref{groupedlambda}) in the special cases where $\widetilde{\mathcal{G}}(K)$ is replaced by $\mathcal{G}$ and $\overline{\mathcal{G}}$, respectively. Note that by (\ref{groupedlambda}), $\text { if } i \in \mathcal{G}_{k}$, we have
	\begin{equation}\label{C4}
		\begin{aligned}
			& \hat{\boldsymbol{\lambda}}_{i \mid \mathcal{G}}=\frac{1}{\left|\mathcal{G}_{k}\right|}\left(\widetilde{\mathbf{F}}^{\top} \widetilde{\mathbf{F}}\right)^{-1} \widetilde{\mathbf{F}}^{\top} \sum_{i \in \mathcal{G}_{k}} \underline{\boldsymbol{y}}_{i}\quad \quad k=1, \ldots, K_{0} \\
			& \hat{\boldsymbol{\lambda}}_{i \mid \overline{\mathcal{G}}}=\left\{\begin{array}{lll}
				\frac{1}{\left|\mathcal{G}_{k}\right|}\left(\widetilde{\mathbf{F}}^{\top} \widetilde{\mathbf{F}}\right)^{-1} \widetilde{\mathbf{F}}^{\top} \sum_{i \in \mathcal{G}_{k}} \underline{\boldsymbol{y}}_{i}, & k=1, \ldots, K_{0}-2,\\
				\frac{1}{\left|\mathcal{G}_{K_{0}-1}\right|+\left|\mathcal{G}_{K_{0}}\right|}\left(\widetilde{\mathbf{F}}^{\top} \widetilde{\mathbf{F}}\right)^{-1} \widetilde{\mathbf{F}}^{\top} \sum_{i \in \mathcal{G}_{K_{0}-1 \mid K_{0}-1}} \underline{\boldsymbol{y}}_{i},&k=K_{0}-1, K_{0},
			\end{array}\right.
		\end{aligned}
	\end{equation}
	where $\widetilde{\mathbf{F}}=\mathbf{Y} \widetilde{\mathbf{\Lambda}} / N$ is the RTS estimate of the factor matrix. Then (\ref{C4}) immediately implies (\ref{C3}).
	With the definition of $S(K)$ in (\ref{SK}), (\ref{C3}) then indicates that\begin{equation}\label{C5}
		\begin{aligned}
			S\left(K_{0}-1\right)-S\left(K_{0}\right)  =&(N T)^{-1} \sum_{k=K_{0}-1}^{K_{0}} \sum_{i \in \mathcal{G}_{k}} \sum_{t=1}^{T}\left(y_{i t}-\hat{\boldsymbol{\lambda}}_{i| \bar{\mathcal{G}}}^\top \tilde{\boldsymbol{f}}_{t}\right)^{2} \\
			& -(N T)^{-1} \sum_{k=K_{0}-1}^{K_{0}} \sum_{i \in \mathcal{G}_{k}} \sum_{t=1}^{T}\left(y_{i t}-\hat{\boldsymbol{\lambda}}_{i \mid \mathcal{G}}^\top \tilde{\boldsymbol{f}}_{t}\right)^{2} \\
   :=&\frac{\left|\mathcal{G}_{K_{0}-1}\right|+\left|\mathcal{G}_{K_{0}}\right|}{N} \cdot\left[\left.S(1)\right|_{\mathcal{G}_{K_{0}-1 \mid K_{0}-1}}-\left.S(2)\right|_{\mathcal{G}_{K_{0}-1 \mid K_{0}-1}}\right] .
		\end{aligned}
	\end{equation}
	Here $\left.S(K)\right|_{\mathcal{G}_{K_{0}-1 \mid K_{0}-1}}$ denotes the restricted objective function value $S(K)$ on the set $\mathcal{G}_{K_{0}-1 \mid K_{0}-1}:=\mathcal{G}_{K_{0}-1} \bigcup \mathcal{G}_{K_{0}}$, and conditional on $\mathcal{M}(\mathcal{G})$, we have $\left.S(2)\right|_{\mathcal{G}_{K_{0}-1 \mid K_{0}-1}}=\left.S(1)\right|_{\mathcal{G}_{K_{0}-1}}+$ $\left.S(1)\right|_{\mathcal{G}_{K_{0}}}$. To prove the asymptotic properties of $\left.S(K)\right|_{\mathcal{G}_{K_{0}-1 \mid K_{0}-1}}$, let$$\left.V(\widetilde{\mathbf{F}})\right|_{\mathcal{G}_{K_{0}-1 \mid K_{0}-1}}=\min _{\boldsymbol{\Lambda}}\left[\left|\mathcal{G}_{K_{0}-1 \mid K_{0}-1}\right| \cdot T\right]^{-1} \times \sum_{i \in \mathcal{G}_{K_{0}-1 \mid K_{0}-1}} \sum_{t=1}^{T}\left(y_{i t}-\boldsymbol{\lambda}_{i}^\top \tilde{\boldsymbol{f}}_{t}\right)^{2}
	$$
	denote the sum of squared residuals in the unconstrained model within $\mathcal{G}_{K_{0}-1 \mid K_{0}-1}$ (divided by $\left.\left|\mathcal{G}_{K_{0}-1 \mid K_{0}-1}\right| \cdot T\right)$ when $\widetilde{\mathbf{F}}$ is estimated via RTS. Then, we can rewrite the last expression in (\ref{C5}) as
	\begin{equation}\label{C7}
		\begin{aligned}
  &\frac{\left|\mathcal{G}_{K_{0}-1}\right|+\left|\mathcal{G}_{K_{0}}\right|}{N} \cdot\left[\left.S(1)\right|_{\mathcal{G}_{K_{0}-1 \mid K_{0}-1}}-\left.S(2)\right|_{\mathcal{G}_{K_{0}-1 \mid K_{0}-1}}\right] \\
   := & \left[\left.\frac{\left|\mathcal{G}_{K_{0}-1}\right|+\left|\mathcal{G}_{K_{0}}\right|}{N} \cdot V(\widetilde{\mathbf{F}})\right|_{\mathcal{G}_{K_{0}-1 \mid K_{0}-1}}+\omega_{1}\right] \\
   &- \left[\left.\frac{\left|\mathcal{G}_{K_{0}-1}\right|+\left|\mathcal{G}_{K_{0}}\right|}{N} \cdot V(\widetilde{\mathbf{F}})\right|_{\mathcal{G}_{K_{0}-1 \mid K_{0}-1}}+\omega_{2}\right] \\
			= & \omega_{1}-\omega_{2},
		\end{aligned}
	\end{equation}
	where
	$$
	\begin{aligned}
		\omega_{1} & :=\frac{\left|\mathcal{G}_{K_{0}-1}\right|+\left|\mathcal{G}_{K_{0}}\right|}{N} \cdot\left[\left.S(1)\right|_{\mathcal{G}_{K_{0}-1 \mid K_{0}-1}}-\left.V(\widetilde{\mathbf{F}})\right|_{\mathcal{G}_{K_{0}-1 \mid K_{0}-1}}\right] \\
		& =(N T)^{-1} \sum_{k=K_{0}-1}^{K_{0}} \sum_{i \in \mathcal{G}_{k}}\left(\tilde{\boldsymbol{\lambda}}_{i}-\hat{\boldsymbol{\lambda}}_{i \mid \overline{\mathcal{G}}}\right)^{\top} \widetilde{\mathbf{F}}^{\top} \underline{\boldsymbol{y}}_{i}, \\
		\omega_{2} & :=\frac{\left|\mathcal{G}_{K_{0}-1}\right|+\left|\mathcal{G}_{K_{0}}\right|}{N} \cdot\left[\left.S(2)\right|_{\mathcal{G}_{K_{0}-1 \mid K_{0}-1}}-\left.V(\widetilde{\mathbf{F}})\right|_{\mathcal{G}_{K_{0}-1 \mid K_{0}-1}}\right] \\
		& =(N T)^{-1} \sum_{i \in \mathcal{G}_{K_{0}-1}}\left(\tilde{\boldsymbol{\lambda}}_{i}-\hat{\boldsymbol{\lambda}}_{i \mid \mathcal{G}}\right)^{\top} \widetilde{\mathbf{F}}^{\top} \underline{\boldsymbol{y}}_{i}+(N T)^{-1} \sum_{i \in \mathcal{G}_{K_{0}}}\left(\tilde{\boldsymbol{\lambda}}_{i}-\hat{\boldsymbol{\lambda}}_{i \mid \mathcal{G}}\right)^\top \widetilde{\mathbf{F}}^{\top} \underline{\boldsymbol{y}}_{i} .
	\end{aligned}
	$$
	Then, (\ref{C2}) holds under (\ref{C5}) and (\ref{C7}), if we can prove\begin{equation}\label{C8}
		P\left[\omega_{1}-\omega_{2}<\rho, \mathcal{M}(\mathcal{G})\right] \rightarrow 0
	\end{equation}
	To prove (\ref{C8}), we define $\hat{\boldsymbol{\lambda}}_{K_{0}-1 \mid K_{0}}:=\hat{\boldsymbol{\lambda}}_{i \mid \mathcal{G}}$ where $i \in \mathcal{G}_{K_{0}-1}$ and $\hat{\boldsymbol{\lambda}}_{K_{0} \mid K_{0}}:=\hat{\boldsymbol{\lambda}}_{i \mid \mathcal{G}}$ where $i \in \mathcal{G}_{K_{0}}$. Note that conditional on $\mathcal{M}(\mathcal{G})$, for $i \in \mathcal{G}_{K_{0}-1 \mid K_{0}-1}$,
	$$
	\begin{aligned}
		\hat{\boldsymbol{\lambda}}_{i \mid \overline{\mathcal{G}}} & =\frac{1}{\left|\mathcal{G}_{K_{0}-1}\right|+\left|\mathcal{G}_{K_{0}}\right|}\left(\widetilde{\mathbf{F}}^{\top} \widetilde{\mathbf{F}}\right)^{-1} \widetilde{\mathbf{F}}^{\top} \sum_{i \in \mathcal{G}_{K_{0}-1} \cup \mathcal{G}_{K_{0}}} \underline{\boldsymbol{y}}_{i} \\
		& =\left(\left|\mathcal{G}_{K_{0}-1}\right|+\left|\mathcal{G}_{K_{0}}\right|\right)^{-1}\left|\mathcal{G}_{K_{0}-1}\right| \hat{\boldsymbol{\lambda}}_{K_{0}-1 \mid K_{0}}+\left(\left|\mathcal{G}_{K_{0}-1}\right|+\left|\mathcal{G}_{K_{0}}\right|\right)^{-1}\left|\mathcal{G}_{K_{0}}\right| \hat{\boldsymbol{\lambda}}_{K_{0} \mid K_{0}} .
	\end{aligned}
	$$
 Then, $\omega_{1}-\omega_{2}$ can be rewritten as
 \setlength\abovedisplayskip{-0.5pt}
	\begin{equation}\label{C9}
		\begin{aligned}
			\omega_{1}-\omega_{2} =& (N T)^{-1} \sum_{i \in \mathcal{G}_{K_{0}-1}} \frac{\left|\mathcal{G}_{K_{0}}\right|}{\left|\mathcal{G}_{K_{0}-1}\right|+\left|\mathcal{G}_{K_{0}}\right|}\left(\hat{\boldsymbol{\lambda}}_{K_{0}-1 \mid K_{0}}-\tilde{\boldsymbol{\lambda}}_{i}\right)^{\top} \widetilde{\mathbf{F}}^{\top} \underline{\boldsymbol{y}}_{i} \\
			& +(N T)^{-1} \sum_{i \in \mathcal{G}_{K_{0}-1}} \frac{\left|\mathcal{G}_{K_{0}}\right|}{\left|\mathcal{G}_{K_{0}-1}\right|+\left|\mathcal{G}_{K_{0}}\right|}\left(\tilde{\boldsymbol{\lambda}}_{i}-\hat{\boldsymbol{\lambda}}_{K_{0} \mid K_{0}}\right)^{\top} \widetilde{\mathbf{F}}^{\top} \underline{\boldsymbol{y}}_{i} \\
			& +(N T)^{-1} \sum_{i \in \mathcal{G}_{K_{0}}} \frac{\left|\mathcal{G}_{K_{0}-1}\right|}{\left|\mathcal{G}_{K_{0}-1}\right|+\left|\mathcal{G}_{K_{0}}\right|}\left(\tilde{\boldsymbol{\lambda}}_{i}-\hat{\boldsymbol{\lambda}}_{K_{0}-1 \mid K_{0}}\right)^{\top} \widetilde{\mathbf{F}}^{\top} \underline{\boldsymbol{y}}_{i} \\
			& +(N T)^{-1} \sum_{i \in \mathcal{G}_{K_{0}}} \frac{\left|\mathcal{G}_{K_{0}-1}\right|}{\left|\mathcal{G}_{K_{0}-1}\right|+\left|\mathcal{G}_{K_{0}}\right|}\left(\hat{\boldsymbol{\lambda}}_{K_{0} \mid K_{0}}-\tilde{\boldsymbol{\lambda}}_{i}\right)^{\top} \widetilde{\mathbf{F}}^{\top} \underline{\boldsymbol{y}}_{i} \\
			 =&(N T)^{-1} \sum_{i \in \mathcal{G}_{K_{0}-1}} \frac{\left|\mathcal{G}_{K_{0}}\right|}{\left|\mathcal{G}_{K_{0}-1}\right|+\left|\mathcal{G}_{K_{0}}\right|}\left(\hat{\boldsymbol{\lambda}}_{K_{0}-1 \mid K_{0}}-\hat{\boldsymbol{\lambda}}_{K_{0} \mid K_{0}}\right)^{\top} \widetilde{\mathbf{F}}^{\top} \underline{\boldsymbol{y}}_{i} \\
			& +(N T)^{-1} \sum_{i \in \mathcal{G}_{K_{0}}} \frac{\left|\mathcal{G}_{K_{0}-1}\right|}{\left|\mathcal{G}_{K_{0}-1}\right|+\left|\mathcal{G}_{K_{0}}\right|}\left(\hat{\boldsymbol{\lambda}}_{K_{0} \mid K_{0}}-\hat{\boldsymbol{\lambda}}_{K_{0}-1 \mid K_{0}}\right)^{\top} \widetilde{\mathbf{F}}^{\top} \underline{\boldsymbol{y}}_{i} .
		\end{aligned}
	\end{equation}
	Additionally, from $\underline{\boldsymbol{y}}_{i}=\mathbf{F} \boldsymbol{\lambda}_{i}+\underline{\boldsymbol{\epsilon}}_{i}$ and (\ref{Kendall estimator}), we can show that
\setlength\abovedisplayskip{-0.5pt}
	\begin{equation}\label{C10}
		\begin{aligned}
		&\hat{\boldsymbol{\lambda}}_{K_{0}-1 \mid K_{0}}-\hat{\boldsymbol{\lambda}}_{K_{0} \mid K_{0}}\\ =& \frac{1}{\left|\mathcal{G}_{K_{0}-1}\right|}\left(\widetilde{\mathbf{F}}^{\top} \widetilde{\mathbf{F}}\right)^{-1} \widetilde{\mathbf{F}}^{\top} \sum_{i \in \mathcal{G}_{K_{0}-1}} \underline{\boldsymbol{y}}_{i}-\frac{1}{\left|\mathcal{G}_{K_{0}}\right|}\left(\widetilde{\mathbf{F}}^{\top} \widetilde{\mathbf{F}}\right)^{-1} \widetilde{\mathbf{F}}^{\top} \sum_{i \in \mathcal{G}_{K_{0}}} \underline{\boldsymbol{y}}_{i} \\
			=& (\widetilde{\mathbf{F}}^{\top}\widetilde{\mathbf{F}})^{-1} \widetilde{\mathbf{F}}^{\top}\left[\mathbf{F}\left(\boldsymbol{\lambda}_{\mathcal{G}, K_{0}-1}-\boldsymbol{\lambda}_{\mathcal{G}, K_{0}}\right)+\left|\mathcal{G}_{K_{0}-1}\right|^{-1} \sum_{l \in \mathcal{G}_{K_{0}-1}} \underline{\boldsymbol{\epsilon}}_{l}-\left|\mathcal{G}_{K_{0}}\right|^{-1} \sum_{q \in \mathcal{G}_{K_{0}}} \underline{\boldsymbol{\epsilon}}_{q}\right].
		\end{aligned}
	\end{equation}
	Combining (\ref{C9}) and (\ref{C10}), we have
 \begin{equation}\label{C11}
		\begin{aligned}
			&\omega_{1}-\omega_{2}\\
   =&(N T)^{-1} \sum_{i \in \mathcal{G}_{K_{0}-1}} \frac{\left|\mathcal{G}_{K_{0}}\right|}{\left|\mathcal{G}_{K_{0}-1}\right|+\left|\mathcal{G}_{K_{0}}\right|}\left(\boldsymbol{\lambda}_{\mathcal{G}, K_{0}-1}-\boldsymbol{\lambda}_{\mathcal{G}, K_{0}}\right)^{\top} \\&\quad\cdot\mathbf{F}^{\top} \widetilde{\mathbf{F}}\left(\widetilde{\mathbf{F}}^{\top} \widetilde{\mathbf{F}}\right)^{-1} \widetilde{\mathbf{F}}^{\top} \mathbf{F} \boldsymbol{\lambda}_{\mathcal{G}, K_{0}-1} \\
			& +(N T)^{-1} \sum_{i \in \mathcal{G}_{K_{0}}} \frac{\left|\mathcal{G}_{K_{0}-1}\right|}{\left|\mathcal{G}_{K_{0}-1}\right|+\left|\mathcal{G}_{K_{0}}\right|}\left(\boldsymbol{\lambda}_{\mathcal{G}, K_{0}}-\boldsymbol{\lambda}_{\mathcal{G}, K_{0}-1}\right)^{\top} \\&\quad\cdot\mathbf{F}^{\top} \widetilde{\mathbf{F}}\left(\widetilde{\mathbf{F}}^{\top} \widetilde{\mathbf{F}}\right)^{-1} \widetilde{\mathbf{F}}^{\top} \mathbf{F} \boldsymbol{\lambda}_{\mathcal{G}, K_{0}} \\
			& +(N T)^{-1} \sum_{i \in \mathcal{G}_{K_{0}-1}} \frac{\left|\mathcal{G}_{K_{0}}\right|}{\left|\mathcal{G}_{K_{0}-1}\right|+\left|\mathcal{G}_{K_{0}}\right|}\left(\boldsymbol{\lambda}_{\mathcal{G}, K_{0}-1}-\boldsymbol{\lambda}_{\mathcal{G}, K_{0}}\right)^{\top}
   \\&\quad\cdot\mathbf{F}^{\top} \widetilde{\mathbf{F}}\left(\widetilde{\mathbf{F}}^{\top} \widetilde{\mathbf{F}}\right)^{-1} \widetilde{\mathbf{F}}^{\top} \underline{\boldsymbol{\epsilon}}_{i} \\
			& +(N T)^{-1} \sum_{i \in \mathcal{G}_{K_{0}-1}} \frac{\left|\mathcal{G}_{K_{0}}\right|}{\left|\mathcal{G}_{K_{0}-1}\right|+\left|\mathcal{G}_{K_{0}}\right|}\left(\left|\mathcal{G}_{K_{0}-1}\right|^{-1} \sum_{l \in \mathcal{G}_{K_{0}-1}} \underline{\boldsymbol{\epsilon}}_{l}-\left|\mathcal{G}_{K_{0}}\right|^{-1} \sum_{q \in \mathcal{G}_{K_{0}}} \underline{\boldsymbol{\epsilon}}_{q}\right)^{\top} \\&\quad\cdot\widetilde{\mathbf{F}}(\widetilde{\mathbf{F}}^\top\widetilde{\mathbf{F}})^{-1} \widetilde{\mathbf{F}}^{\top} \mathbf{F} \boldsymbol{\lambda}_{\mathcal{G}, K_{0}-1} \\
			& +(N T)^{-1} \sum_{i \in \mathcal{G}_{K_{0}-1}} \frac{\left|\mathcal{G}_{K_{0}}\right|}{\left|\mathcal{G}_{K_{0}-1}\right|+\left|\mathcal{G}_{K_{0}}\right|}\left(\left|\mathcal{G}_{K_{0}-1}\right|^{-1} \sum_{l \in \mathcal{G}_{K_{0}-1}} \underline{\boldsymbol{\epsilon}}_{l}-\left|\mathcal{G}_{K_{0}}\right|^{-1} \sum_{q \in \mathcal{G}_{K_{0}}} \underline{\boldsymbol{\epsilon}}_{q}\right)^{\top} \\&\quad\cdot\widetilde{\mathbf{F}}(\widetilde{\mathbf{F}}^\top\widetilde{\mathbf{F}})^{-1} \widetilde{\mathbf{F}}^{\top} \underline{\boldsymbol{\epsilon}}_{i} \\
			& +(N T)^{-1} \sum_{i \in \mathcal{G}_{K_{0}}} \frac{\left|\mathcal{G}_{K_{0}-1}\right|}{\left|\mathcal{G}_{K_{0}-1}\right|+\left|\mathcal{G}_{K_{0}}\right|}\left(\boldsymbol{\lambda}_{\mathcal{G}, K_{0}}-\boldsymbol{\lambda}_{\mathcal{G}, K_{0}-1}\right)^{\top} \mathbf{F}^{\top} \\&\quad\cdot\widetilde{\mathbf{F}}(\widetilde{\mathbf{F}}^\top\widetilde{\mathbf{F}})^{-1} \widetilde{\mathbf{F}}^{\top} \underline{\boldsymbol{\epsilon}}_{i} \\
			& +(N T)^{-1} \sum_{i \in \mathcal{G}_{K_{0}}} \frac{\left|\mathcal{G}_{K_{0}-1}\right|}{\left|\mathcal{G}_{K_{0}-1}\right|+\left|\mathcal{G}_{K_{0}}\right|}\left(\left|\mathcal{G}_{K_{0}}\right|^{-1} \sum_{q \in \mathcal{G}_{K_{0}}} \underline{\boldsymbol{\epsilon}}_{q}-\left|\mathcal{G}_{K_{0}-1}\right|^{-1} \sum_{l \in \mathcal{G}_{K_{0}-1}} \underline{\boldsymbol{\epsilon}}_{l}\right)^{\top} \\&\quad\cdot\widetilde{\mathbf{F}}(\widetilde{\mathbf{F}}^\top\widetilde{\mathbf{F}})^{-1} \widetilde{\mathbf{F}}^{\top} \mathbf{F} \boldsymbol{\lambda}_{\mathcal{G}, K_{0}} \\
			& +(N T)^{-1} \sum_{i \in \mathcal{G}_{K_{0}}} \frac{\left|\mathcal{G}_{K_{0}-1}\right|}{\left|\mathcal{G}_{K_{0}-1}\right|+\left|\mathcal{G}_{K_{0}}\right|}\left(\left|\mathcal{G}_{K_{0}}\right|^{-1} \sum_{q \in \mathcal{G}_{K_{0}}} \underline{\boldsymbol{\epsilon}}_{q}-\left|\mathcal{G}_{K_{0}-1}\right|^{-1} \sum_{l \in \mathcal{G}_{K_{0}-1}} \underline{\boldsymbol{\epsilon}}_{l}\right)^{\top} \\&\quad\cdot\widetilde{\mathbf{F}}(\widetilde{\mathbf{F}}^\top\widetilde{\mathbf{F}})^{-1} \widetilde{\mathbf{F}}^{\top} \underline{\boldsymbol{\epsilon}}_{i} \\
			 \equiv& I+I I+I I I+I V+V+V I+V I I+V I I I .
		\end{aligned}
	\end{equation}
	First, we consider $I+I I$. Applying Assumption \ref{as5}(a), we have

	$$
	\begin{aligned}
		&I+I I \\
  \geq& (N T)^{-1} \sum_{i \in \mathcal{G}_{K_{0}-1}} \tau_{1} \cdot\left(\boldsymbol{\lambda}_{\mathcal{G}, K_{0}-1}-\boldsymbol{\lambda}_{\mathcal{G}, K_{0}}\right)^{\top} \mathbf{F}^{\top} \widetilde{\mathbf{F}}\left(\widetilde{\mathbf{F}}^{\top} \widetilde{\mathbf{F}}\right)^{-1} \widetilde{\mathbf{F}}^{\top} \mathbf{F} \boldsymbol{\lambda}_{\mathcal{G}, K_{0}-1} \\
		& -(N T)^{-1} \sum_{i \in \mathcal{G}_{K_{0}}} \tau_{1}\left(\boldsymbol{\lambda}_{\mathcal{G}, K_{0}-1}-\boldsymbol{\lambda}_{\mathcal{G}, K_{0}}\right)^{\top} \mathbf{F}^{\top} \widetilde{\mathbf{F}}\left(\widetilde{\mathbf{F}}^{\top} \widetilde{\mathbf{F}}\right)^{-1} \widetilde{\mathbf{F}}^{\top} \mathbf{F} \boldsymbol{\lambda}_{\mathcal{G}, K_{0}} \\
		\geq&  \tau_{1}^{2}\left(\boldsymbol{\lambda}_{\mathcal{G}, K_{0}-1}-\boldsymbol{\lambda}_{\mathcal{G}, K_{0}}\right)^{\top}\left(\frac{\mathbf{F}^{\top} \widetilde{\mathbf{F}}}{T}\right)\left(\widetilde{\mathbf{F}}^{\top} \widetilde{\mathbf{F}} / T\right)^{-1}\left(\frac{\widetilde{\mathbf{F}}^{\top} \mathbf{F}}{T}\right)\left(\boldsymbol{\lambda}_{\mathcal{G}, K_{0}-1}-\boldsymbol{\lambda}_{\mathcal{G}, K_{0}}\right) \\=&\tau_{1}^{2}\left(\boldsymbol{\lambda}_{\mathcal{G}, K_{0}-1}-\boldsymbol{\lambda}_{\mathcal{G}, K_{0}}\right)^{\top} \cdot \mathbf{M} \mathbf{V}^{-1} \mathbf{M}^{\top} \cdot\left(\boldsymbol{\lambda}_{\mathcal{G}, K_{0}-1}-\boldsymbol{\lambda}_{\mathcal{G}, K_{0}}\right)\\
  &+o_{p}(1) \cdot \tau_{1}^{2}\left\|\boldsymbol{\lambda}_{\mathcal{G}, k_{1}}-\boldsymbol{\lambda}_{\mathcal{G}, k_{2}}\right\|_{2}^{2},
	\end{aligned}
	$$
	where $\mathbf{M}$ and $\mathbf{V}$ are limits of $\mathbf{F}^{\top} \widetilde{\mathbf{F}}/T$ and $\widetilde{\mathbf{V}}=\left(\widetilde{\mathbf{F}}^{\top} \widetilde{\mathbf{F}} / T\right)$, respectively. $\mathbf{F}^{\top} \widetilde{\mathbf{F}}/T=\mathbf{F}^\top(\widetilde{\mathbf{F}}\widehat{\mathbf{H}}^\top-\mathbf{F})(\widehat{\mathbf{H}}^\top)^{-1}/T+\mathbf{F}^{\top} \mathbf{F}(\widehat{\mathbf{H}}^\top)^{-1}/T$, where the first term is $o_p(1)$ and the second term is invertible, so $\mathbf{M}$ is invertible. Note that $\mathbf{M} \mathbf{V}^{-1} \mathbf{M}^{\top}$ is positive definite because $\mathbf{M}$ is invertible and $\mathbf{V}$ is positive definite by the proof of Theorem \ref{thm2}. In addition, by the definition of $\zeta$ in Theorem \ref{thm1}, we have $\min _{1 \leq k_{1} \neq k_{2} \leq K_{0}} \| \boldsymbol{\lambda}_{\mathcal{G}, k_{1}}-$ $\boldsymbol{\lambda}_{\mathcal{G}, k_{2}} \|_{2}^{2}=O\left(\zeta^{2}\right)$. Thus it then follows that $I+I I \geq O_{p}\left(\zeta^{2}\right)$. In the following, we shall show that the remaining terms in (\ref{C11}) are bounded by $O_{p}\left(\frac{1}{\sqrt{N}}\right)$, which converges to zero at a faster rate than $O_{p}\left(\zeta^{2}\right)$ by Assumption \ref{as5}(b).\\
	For $I I I$, we rewrite it as
	$$
	\begin{aligned}
		I I I  =&(N T)^{-1} \sum_{i \in \mathcal{G}_{K_{0}-1}} \frac{\left|\mathcal{G}_{K_{0}}\right|}{\left|\mathcal{G}_{K_{0}-1}\right|+\left|\mathcal{G}_{K_{0}}\right|} \boldsymbol{\lambda}_{\mathcal{G}, K_{0}-1}^{\top} \mathbf{F}^{\top} \widetilde{\mathbf{F}}\left(\widetilde{\mathbf{F}}^{\top} \widetilde{\mathbf{F}}\right)^{-1} \widetilde{\mathbf{F}}^{\top} \underline{\boldsymbol{\epsilon}}_{i} \\
		& -(N T)^{-1} \sum_{i \in \mathcal{G}_{K_{0}-1}} \frac{\left|\mathcal{G}_{K_{0}}\right|}{\left|\mathcal{G}_{K_{0}-1}\right|+\left|\mathcal{G}_{K_{0}}\right|} \boldsymbol{\lambda}_{\mathcal{G}, K_{0}}^{\top} \mathbf{F}^{\top} \widetilde{\mathbf{F}}\left(\widetilde{\mathbf{F}}^{\top} \widetilde{\mathbf{F}}\right)^{-1} \widetilde{\mathbf{F}}^{\top} \underline{\boldsymbol{\epsilon}}_{i} \\
		 =&I I I_{1}-I I I_{2} .
	\end{aligned}
	$$
	For the first term,
$$
	\begin{aligned}
		I I I_{1} &\leq(N T)^{-1} \cdot \boldsymbol{\lambda}_{\mathcal{G}, K_{0}-1}^{\top} \frac{\mathbf{F}^{\top} \widetilde{\mathbf{F}}}{T} \cdot \widetilde{\mathbf{V}}^{-1} \widetilde{\mathbf{F}}^{\top}\cdot\sum_{i \in \mathcal{G}_{K_{0}-1}} \underline{\boldsymbol{\epsilon}}_{i} \\
		\left\|I I I_{1}\right\| &\leq\max _{k}\left\|\boldsymbol{\lambda}_{\mathcal{G}, k}\right\|\left\|\frac{\mathbf{F}^{\top} \mathbf{F}}{T}\right\|^{\frac{1}{2}}\left\|\widetilde{\mathbf{V}}^{-1}\right\|\left\|\frac{\widetilde{\mathbf{F}}^{\top} \widetilde{\mathbf{F}}}{T}\right\| \\&\qquad\qquad\cdot\left(\frac{1}{N T\left|\mathcal{G}_{K_{0}-1}\right|} \sum_{t=1}^{T} \sum_{i \in \mathcal{G}_{K_{0}-1}} \sum_{l \in \mathcal{G}_{K_{0}-1}} \epsilon_{i t} \epsilon_{l t}\right)^{\frac{1}{2}} .
	\end{aligned}
	$$
	The first expression is $O_{p}(1)$ by Assumptions \ref{as2} and \ref{as4} and $\|\widetilde{\mathbf{V}}\|=O_{p}(1)$. For the second
	expression, since\begin{equation}\label{C13}
		\begin{aligned}
			\left|\frac{1}{T\left|\mathcal{G}_{K_{0}-1}\right|} \sum_{t=1}^{T} \sum_{i \in \mathcal{G}_{K_{0}-1}} \sum_{l \in \mathcal{G}_{K_{0}-1}} \epsilon_{i t} \epsilon_{l t}\right| & \leq \frac{1}{T\left|\mathcal{G}_{K_{0}-1}\right|} \sum_{t=1}^{T} \sum_{i \in \mathcal{G}_{K_{0}-1}} \sum_{l \in \mathcal{G}_{K_{0}-1}}\left|\sigma_{i l}\right| \\
			& \leq \frac{1}{T\left|\mathcal{G}_{K_{0}-1}\right|} \sum_{t=1}^{T} \sum_{i=1}^{N} \sum_{l=1}^{N}\left|\sigma_{i l}\right|=O_{p}(1),
		\end{aligned}
	\end{equation}
	by Assumption \ref{as4-1}(b), we can conclude that $I I I_{1} \leq O_{p}\left(\frac{1}{\sqrt{N}}\right)$. Analogously, we can also prove that the second term $I I I_{2} \leq O_{p}\left(\frac{1}{\sqrt{N}}\right)$ and thus $I I I \leq O_{p}\left(\frac{1}{\sqrt{N}}\right)$. By the same token, we can show that $V I \leq O_{p}\left(\frac{1}{\sqrt{N}}\right)$. \\
 Then we consider
 \begin{small}
     \begin{equation*}
     \begin{aligned}
&I V\\
  =& (N T)^{-1} \sum_{i \in \mathcal{G}_{K_{0}-1}} \frac{\left|\mathcal{G}_{K_{0}}\right|}{\left|\mathcal{G}_{K_{0}-1}\right|+\left|\mathcal{G}_{K_{0}}\right|}\left(\left|\mathcal{G}_{K_{0}-1}\right|^{-1} \sum_{l \in \mathcal{G}_{K_{0}-1}} \underline{\boldsymbol{\epsilon}}_{l}\right)^{\top} \widetilde{\mathbf{F}}\left(\widetilde{\mathbf{F}}^{\top} \widetilde{\mathbf{F}}\right)^{-1} \widetilde{\mathbf{F}}^{\top} \mathbf{F} \boldsymbol{\lambda}_{\mathcal{G}, K_{0}-1} \\
		& -(N T)^{-1} \sum_{i \in \mathcal{G}_{K_{0}-1}} \frac{\left|\mathcal{G}_{K_{0}}\right|}{\left|\mathcal{G}_{K_{0}-1}\right|+\left|\mathcal{G}_{K_{0}}\right|}\left(\left|\mathcal{G}_{K_{0}}\right|^{-1} \sum_{q \in \mathcal{G}_{K_{0}}} \underline{\boldsymbol{\epsilon}}_{q}\right)^{\top} \widetilde{\mathbf{F}}\left(\widetilde{\mathbf{F}}^{\top} \widetilde{\mathbf{F}}\right)^{-1} \widetilde{\mathbf{F}}^{\top} \mathbf{F} \boldsymbol{\lambda}_{\mathcal{G}, K_{0}-1} \\
		=& I V_{1}-I V_{2} .
	\end{aligned}
 \end{equation*}
 \end{small}
	By arguments analogous to those for $I I I_{1}$, the first term

	$$
  \begin{small}
	\begin{aligned}
		&I V_{1} \\\leq&\left(\left|\mathcal{G}_{K_{0}-1}\right|\right)^{-\frac{1}{2}} \cdot \max _{k}\left\|\boldsymbol{\lambda}_{\mathcal{G}, k}\right\|\left\|\frac{\mathbf{F}^{\top} \mathbf{F}}{T}\right\|^{\frac{1}{2}}\left\|\frac{\widetilde{\mathbf{F}}^{\top} \widetilde{\mathbf{F}}}{T}\right\|\cdot\left(\frac{1}{T\left|\mathcal{G}_{K_{0}-1}\right|} \sum_{t=1}^{T} \sum_{l \in \mathcal{G}_{K_{0}-1}} \sum_{q \in \mathcal{G}_{K_{0}-1}} \epsilon_{l t} \epsilon_{q t}\right)^{\frac{1}{2}} \\
		=&O_{p}\left(\frac{1}{\sqrt{\left|\mathcal{G}_{K_{0}-1}\right|}}\right) .
	\end{aligned}
 \end{small}
	$$
	By Assumption \ref{as4-1}(b) and the fact that $\left|\mathcal{G}_{1}\right|, \ldots,\left|\mathcal{G}_{K_{0}}\right|$ and $N$ are of the same order of magnitude, $I V_{1}$ is bounded by $O_{p}\left(\frac{1}{\sqrt{N}}\right)$. In the same way, we can show that the second term $I V_{2} \leq O_{p}\left(\frac{1}{\sqrt{N}}\right)$ and thus $I V \leq O_{p}\left(\frac{1}{\sqrt{N}}\right)$. Using arguments analogous to those used in the proof of $I V \leq O_{p}\left(\frac{1}{\sqrt{N}}\right)$, we can show that $V I I \leq O_{p}\left(\frac{1}{\sqrt{N}}\right)$.
	Finally we consider the terms $V$ and $V I I I$. Note that
	$$
	\begin{aligned}
		V =& (N T)^{-1} \sum_{i \in \mathcal{G}_{K_{0}-1}} \frac{\left|\mathcal{G}_{K_{0}}\right|}{\left|\mathcal{G}_{K_{0}-1}\right|+\left|\mathcal{G}_{K_{0}}\right|}\left(\left|\mathcal{G}_{K_{0}-1}\right|^{-1} \sum_{l \in \mathcal{G}_{K_{0}-1}} \underline{\boldsymbol{\epsilon}}_{l}\right)^{\top} \widetilde{\mathbf{F}}\left(\widetilde{\mathbf{F}}^{\top} \widetilde{\mathbf{F}}\right)^{-1} \widetilde{\mathbf{F}}^{\top} \underline{\boldsymbol{\epsilon}}_{i} \\
		& -(N T)^{-1} \sum_{i \in \mathcal{G}_{K_{0}-1}} \frac{\left|\mathcal{G}_{K_{0}}\right|}{\left|\mathcal{G}_{K_{0}-1}\right|+\left|\mathcal{G}_{K_{0}}\right|}\left(\left|\mathcal{G}_{K_{0}}\right|^{-1} \sum_{q \in \mathcal{G}_{K_{0}}} \underline{\boldsymbol{\epsilon}}_{q}\right)^{\top} \widetilde{\mathbf{F}}\left(\widetilde{\mathbf{F}}^{\top} \widetilde{\mathbf{F}}\right)^{-1} \widetilde{\mathbf{F}}^{\top} \underline{\boldsymbol{\epsilon}}_{i} \\
		=& V_{1}-V_{2} .
	\end{aligned}
	$$
	For $V_{1}$, we have
	$$
	\begin{aligned}
		V_{1} =& (N T)^{-1} \frac{\left|\mathcal{G}_{K_{0}}\right|}{\left|\mathcal{G}_{K_{0}-1}\right|+\left|\mathcal{G}_{K_{0}}\right|}\left(\left|\mathcal{G}_{K_{0}-1}\right|^{-1} \sum_{l \in \mathcal{G}_{K_{0}-1}} \underline{\boldsymbol{\epsilon}}_{l}\right)^{\top} \widetilde{\mathbf{F}}\left(\widetilde{\mathbf{F}}^{\top} \widetilde{\mathbf{F}}\right)^{-1} \widetilde{\mathbf{F}}^{\top} \sum_{i \in \mathcal{G}_{K_{0}-1}} \underline{\boldsymbol{\epsilon}}_{i} \\
		 \leq&(N T)^{-1}\left(\left|\mathcal{G}_{K_{0}-1}\right|^{-1} \sum_{l \in \mathcal{G}_{K_{0}-1}} \underline{\boldsymbol{\epsilon}}_{l}\right)^{\top} \widetilde{\mathbf{F}}\left(\widetilde{\mathbf{F}}^{\top} \widetilde{\mathbf{F}}\right)^{-1} \widetilde{\mathbf{F}}^{\top} \sum_{i \in \mathcal{G}_{K_{0}-1}} \underline{\boldsymbol{\epsilon}}_{i} \\
		 \leq&\left(\frac{1}{N} \cdot\left\|\left(\frac{\widetilde{\mathbf{F}}^{\top} \widetilde{\mathbf{F}}}{T}\right)^{-1}\right\|\left\|\frac{\widetilde{\mathbf{F}}^{\top} \widetilde{\mathbf{F}}}{T}\right\|\right) \cdot\left(\frac{1}{T\left|\mathcal{G}_{K_{0}-1}\right|} \sum_{t=1}^{T} \sum_{l \in \mathcal{G}_{K_{0}-1}} \sum_{q \in \mathcal{G}_{K_{0}-1}} \epsilon_{l t} \epsilon_{q t}\right)^{\frac{1}{2}} \\
		& \cdot\left(\frac{1}{T\left|\mathcal{G}_{K_{0}-1}\right|} \sum_{t=1}^{T} \sum_{i \in \mathcal{G}_{K_{0}-1}} \sum_{j \in \mathcal{G}_{K_{0}-1}} \epsilon_{i t} \epsilon_{j t}\right)^{\frac{1}{2}} .
	\end{aligned}
	$$
	We next show that each term on the right hand side of the last expression can be bounded. The first term is bounded by $O_{p}\left(\frac{1}{N}\right)$ since $\widetilde{\mathbf{F}}^{\top} \widetilde{\mathbf{F}} / T=\widetilde{\mathbf{V}}$ and $\|\widetilde{\mathbf{V}}\|=O_{p}(1)$; the second and last term are both bounded by $O_{p}(1)$ by (\ref{C13}). Hence $V_{1} \leq O_{p}\left(\frac{1}{N}\right)$. Similarly, we can also prove that $V_{2} \leq O_{p}\left(\frac{1}{N}\right)$ and thus $V \leq O_{p}\left(\frac{1}{N}\right)$. Using arguments as used in the analysis of $V$, we can also show that $V I I I \leq O_{p}\left(\frac{1}{N}\right)$. Thus we have proved that as long as $\zeta^{2} \cdot \sqrt{N} \rightarrow \infty, I+I I$ will be the dominant term. In sum, we readily have
	$$
	\omega_{1}-\omega_{2} \geq O_{p}\left(\zeta^{2}\right),
	$$
	by Assumption \ref{as5}(b). Moreover, as the tuning parameter $\rho$ satisfies $\rho / \zeta^{2} \rightarrow 0$ by Assumption \ref{as5}(b), $\omega_{1}-\omega_{2}<\rho$ with probability approaching zero. This implies (\ref{C8}) and thus (\ref{C2}) holds. The proof of Lemma \ref{lmA2} has been completed.{\hfill$\square$}
	\subsection{Proof of Lemma \ref{lmA3}}
	Without loss of generality, we only consider the case where $K=K_{0}+1$ and prove that
	\begin{equation}\label{C14}
		P\left[\mathbb{I} \mathbb{C}\left(K_{0}\right)<\mathbb{I} \mathbb{C}\left(K_{0}+1\right), \mathcal{M}(\mathcal{G})\right] \rightarrow 1 .
	\end{equation}
	By analogous arguments as used in the proof of Corollary 1 in \cite{bai2002determining}, we only need to show that
	\begin{equation}\label{C15}
		P\left[S\left(K_{0}\right)-S\left(K_{0}+1\right)<\rho, \mathcal{M}(\mathcal{G})\right] \rightarrow 1 .
	\end{equation}
	Conditional on the event $\mathcal{M}(\mathcal{G})$, one of the clusters $\left\{\mathcal{G}_{1}, \mathcal{G}_{2}, \ldots, \mathcal{G}_{K_{0}}\right\}$ is split into two sub-clusters when the AHC algorithm stops at $K=K_{0}+1$. Without loss of generality, we assume that $\mathcal{G}_{K_{0}}$ is divided into two sub-clusters and denote the resulting $K_{0}+1$ clusters as $\mathcal{G}^{*}=\left\{\mathcal{G}_{1}^{*}, \mathcal{G}_{2}^{*}, \ldots, \mathcal{G}_{K_{0}}^{*}, \mathcal{G}_{K_{0}+1}^{*}\right\}$ with $\mathcal{G}_{k}^{*}=\mathcal{G}_{k}$ for $k=1, \ldots, K_{0}-1$ and $\mathcal{G}_{K_{0}}^{*} \cup \mathcal{G}_{K_{0}+1}^{*}=\mathcal{G}_{K_{0}}$.
	For $k=1, \ldots, K_{0}-1$, similarly following the proof of (\ref{C3}) in Lemma \ref{lmA2}, we can show that
	\begin{equation}\label{C16}
		\hat{\boldsymbol{\lambda}}_{i \mid \mathcal{G}}=\hat{\boldsymbol{\lambda}}_{i \mid \mathcal{G}^{*}},  \quad i \in \mathcal{G}_{k} ,
	\end{equation}
	where $\hat{\boldsymbol{\lambda}}_{i \mid \mathcal{G}}$ is given in (\ref{C4}) and
	$$
	\hat{\boldsymbol{\lambda}}_{i \mid \mathcal{G}^{*}}= \begin{cases}\frac{1}{\left|\mathcal{G}_{k}\right|}\left(\widetilde{\mathbf{F}}^{\top} \widetilde{\mathbf{F}}\right)^{-1} \widetilde{\mathbf{F}}^{\top} \sum_{i \in \mathcal{G}_{k}} \underline{\boldsymbol{y}}_{i}, & \text { if } i \in \mathcal{G}_{k}, \quad k=1, \ldots, K_{0}-1, \\ \frac{1}{\left|\mathcal{G}_{K_{0}}^{*}\right|}\left(\widetilde{\mathbf{F}}^{\top} \widetilde{\mathbf{F}}\right)^{-1} \widetilde{\mathbf{F}}^{\top} \sum_{i \in \mathcal{G}_{K_{0}}^{*}} \underline{\boldsymbol{y}}_{i}, & \text { if } i \in \mathcal{G}_{K_{0}}^{*}, \\ \frac{1}{\left|\mathcal{G}_{K_{0}+1}^{*}\right|}\left(\widetilde{\mathbf{F}}^{\top} \widetilde{\mathbf{F}}\right)^{-1} \widetilde{\mathbf{F}}^{\top} \sum_{i \in \mathcal{G}_{K_{0}+1}^{*}}^{*} \underline{\boldsymbol{y}}_{i}, & \text { if } i \in \mathcal{G}_{K_{0}+1}^{*} .\end{cases}
	$$
	By (\ref{C16}), we have
	\begin{equation}\label{C17}
		\begin{aligned}
			S\left(K_{0}\right)-S\left(K_{0}+1\right)  =&(NT)^{-1} \sum_{i \in \mathcal{G}_{K_{0}}} \sum_{t=1}^{T}\left(y_{i t}-\hat{\boldsymbol{\lambda}}_{i \mid \mathcal{G}}^{\top} \tilde{\boldsymbol{f}}_{t}\right)^{2} \\
			& -(NT)^{-1} \sum_{k=K_{0}}^{K_{0}+1} \sum_{i \in \mathcal{G}_{k}^{*}} \sum_{t=1}^{T}\left(y_{i t}-\hat{\boldsymbol{\lambda}}_{i \mid \mathcal{G}^{*}}^{\top} \tilde{\boldsymbol{f}}_{t}\right)^{2} \\
			 :=&\frac{\left|\mathcal{G}_{K_{0}}\right|}{N}\left[\left.S(1)\right|_{\mathcal{G}_{K_{0}}}-\left.S(2)\right|_{\mathcal{G}_{K_{0}}}\right] .
		\end{aligned}
	\end{equation}
	Similarly, here $\left.S(\cdot)\right|_{\mathcal{G}_{K_{0}}}$ denotes the restricted objective function value $S(K)$ on the set $\mathcal{G}_{K_{0}}$, and conditional on $\mathcal{M}(\mathcal{G})$, we have $\left.S(2)\right|_{\mathcal{G}_{K_{0}}}=\left.S(1)\right|_{\mathcal{G}_{K_{0}}^{*}}+\left.S(1)\right|_{\mathcal{G}_{K_{0}+1}^{*}}$. Let $\mathbf{M}_{F}=\mathbf{I}-\mathbf{P}_{F}$ denote the idempotent matrix spanned by null space of $\mathbf{F}$ and define
	\begin{equation}\label{C18}
\left.V(\mathbf{F})\right|_{\mathcal{G}_{K_{0}}}=\left(\left|\mathcal{G}_{K_{0}}\right| \cdot T\right)^{-1} \sum_{i \in \mathcal{G}_{K_{0}}} \underline{\boldsymbol{y}}_{i}^{\top} \mathbf{M}_{F} \underline{\boldsymbol{y}}_{i} .
	\end{equation}
	By the triangle inequality, we have
	\begin{equation}\label{C19}
		\begin{aligned}
			|S(1)|_{\mathcal{G}_{K_{0}}}-S(2)|_{\mathcal{G}_{K_{0}}}| & \leq|S(1)|_{\mathcal{G}_{K_{0}}}-V(\mathbf{F})|_{\mathcal{G}_{K_{0}}}|+| V(\mathbf{F})|_{\mathcal{G}_{K_{0}}}-S(2)|_{\mathcal{G}_{K_{0}}}|  \\
			& \leq 2 \max _{1 \leq K \leq 2}|S(K)|_{\mathcal{G}_{K_{0}}}-V(\mathbf{F})|_{\mathcal{G}_{K_{0}}}|.
		\end{aligned}
	\end{equation}
	By (\ref{C17}), (\ref{C19}) and Assumption \ref{as5}(b), (\ref{C15}) would follow if we can show that for each $K=1,2$,
	\begin{equation}\label{C20}
		\left.S(K)\right|_{\mathcal{G}_{K_{0}}}-\left.V(\mathbf{F})\right|_{\mathcal{G}_{K_{0}}}=O_{p}\left(C_{NT}^{-2}\right),
	\end{equation}
	where $C_{NT}$ is given in Assumption \ref{as5}(b). To study the asymptotic properties of $\left.S(K)\right|_{\mathcal{G}_{K_{0}}}-\left.V(\mathbf{F})\right|_{\mathcal{G}_{K_{0}}}$, we let $\omega_{K}^{*}$ denote the difference in the sum of squared errors in the constrained
	and unconstrained models varying in $K$ on $\mathcal{G}_{K_{0}}$ :
	\begin{equation}\label{C21}
		\begin{aligned}
			\omega_{1}^{*} & :=\left.S(1)\right|_{\mathcal{G}_{K_{0}}}-\left.V(\widetilde{\mathbf{F}})\right|_{\mathcal{G}_{K_{0}}} \\
			& =\left(\left|\mathcal{G}_{K_{0}}\right| \cdot T\right)^{-1} \sum_{i \in \mathcal{G}_{K_{0}}}\left(\tilde{\boldsymbol{\lambda}}_{i}-\hat{\boldsymbol{\lambda}}_{i \mid \mathcal{G}}\right)^{\top} \widetilde{\mathbf{F}}^{\top} \underline{\boldsymbol{y}}_{i}, \\
			\omega_{2}^{*} & :=\left.S(2)\right|_{\mathcal{G}_{K_{0}}}-\left.V(\widetilde{\mathbf{F}})\right|_{\mathcal{G}_{K_{0}}} \\
			& =\left(\left|\mathcal{G}_{K_{0}}\right| \cdot T\right)^{-1} \sum_{i \in \mathcal{G}_{K_{0}}^{*}}\left(\tilde{\boldsymbol{\lambda}}_{i}-\hat{\boldsymbol{\lambda}}_{i \mid \mathcal{G}^{*}}\right)^{\top} \widetilde{\mathbf{F}}^{\top} \underline{\boldsymbol{y}}_{i}\\&\quad+\left(\left|\mathcal{G}_{K_{0}}\right| \cdot T\right)^{-1} \sum_{i \in \mathcal{G}_{K_{0}+1}^{*}}\left(\tilde{\boldsymbol{\lambda}}_{i}-\hat{\boldsymbol{\lambda}}_{i \mid \mathcal{G}^{*}}\right)^{\top} \widetilde{\mathbf{F}}^{\top} \underline{\boldsymbol{y}}_{i},
		\end{aligned}
	\end{equation}
	Using (\ref{C21}), we can decompose $\left.S(K)\right|_{\mathcal{G}_{K_{0}}}-\left.V(\mathbf{F})\right|_{\mathcal{G}_{K_{0}}}$ as follows:
	\begin{equation}\label{C22}
		\begin{aligned}
			\left.S(K)\right|_{\mathcal{G}_{K_{0}}}-\left.V(\mathbf{F})\right|_{\mathcal{G}_{K_{0}}} & =\left[\left.V(\widetilde{\mathbf{F}})\right|_{\mathcal{G}_{K_{0}}}+\omega_{K}^{*}\right]-\left.V(\mathbf{F})\right|_{\mathcal{G}_{K_{0}}} \\
			& =\left[\left.V(\widetilde{\mathbf{F}})\right|_{\mathcal{G}_{K_{0}}}-\left.V(\mathbf{F})\right|_{\mathcal{G}_{K_{0}}}\right]+\omega_{K}^{*} \\
			& \equiv a+\omega_{K}^{*}.
		\end{aligned}
	\end{equation}
	By Lemma \ref{bai2002lm4}, we readily have
	\begin{equation}\label{C23}
a=\left.V(\widetilde{\mathbf{F}})\right|_{\mathcal{G}_{K_{0}}}-\left.V(\mathbf{F})\right|_{\mathcal{G}_{K_{0}}} \leq O_{p}\left(C_{NT}^{-2}\right).
	\end{equation}
	Thus, to prove (\ref{C20}), it is sufficient to show that both $\omega_{1}^{*} \leq O_{p}\left(C_{NT}^{-2}\right)$ and $\omega_{2}^{*} \leq O_{p}\left(C_{NT}^{-2}\right)$ hold.\\
	First, we verify $\omega_{1}^{*} \leq O_{p}\left(C_{NT}^{-2}\right)$. Combining Theorem 1 in \cite{he2022large} and Theorem \ref{thm2}, we can show that conditional on $\mathcal{M}(\mathcal{G})$ and for $i \in \mathcal{G}_{K_{0}}$,
	\begin{equation}\label{C24}
 \begin{aligned}
     \frac{1}{\left|\mathcal{G}_{K_{0}}\right|}\sum_{i\in \mathcal{G}_{K_{0}}}\left\|\tilde{\boldsymbol{\lambda}}_{i}-\hat{\boldsymbol{\lambda}}_{i \mid \mathcal{G}}\right\|^{2} &\leq(\frac{1}{\left|\mathcal{G}_{K_{0}}\right|}\sum_{i\in \mathcal{G}_{K_{0}}}\left\|\tilde{\boldsymbol{\lambda}}_{i}-\widetilde{\mathbf{H}}^{\top} \boldsymbol{\lambda}_{\mathcal{G}, K_{0}}\right\|^{2})\\&\quad+\left\|\hat{\boldsymbol{\lambda}}_{K_{0} \mid K_{0}}-\widetilde{\mathbf{H}}^{\top} \boldsymbol{\lambda}_{\mathcal{G}, K_{0}}\right\|^{2}\\&=O_{p}\left(\delta_{NT}^{-2}\right),
 \end{aligned}
	\end{equation}
	where $\delta_{NT}$ is defined in Proposition \ref{pr1}. \\
 Using \(
	\omega_{1}^{*} =\left(\left|\mathcal{G}_{K_{0}}\right| \cdot T\right)^{-1} \sum_{i \in \mathcal{G}_{K_{0}}}\left(\tilde{\boldsymbol{\lambda}}_{i}-\hat{\boldsymbol{\lambda}}_{i \mid \mathcal{G}}\right)^{\top}\widetilde{\mathbf{F}}^{\top} \underline{\boldsymbol{y}}_{i}\), we can further bound $\omega_{1}^{*}$ as follows by Assumption \ref{as4-1}(b), which shows that $\max_{1\le i\le N}\|\underline{\boldsymbol{\epsilon}}_i\|^2=O_p(T)$:
	\begin{equation}\label{C25}
		\begin{aligned}
			|\omega^*_1|^2&\leq \frac{1}{T^2|\mathcal{G}_{K_0}|^2}\sum_{i \in \mathcal{G}_{K_0}} ||\tilde{\boldsymbol{\lambda}}_i-\hat{\boldsymbol{\lambda}}_{i|\mathcal{G}}||^2||\widetilde{\mathbf{F}}^{\top}\mathbf{F}\boldsymbol{\lambda}_i+\widetilde{\mathbf{F}}^{\top}\underline{\boldsymbol{\epsilon}}_i||^2 \\
			&
			\lesssim \frac{1}{|\mathcal{G}_{K_0}|^2}\sum_{i \in \mathcal{G}_{K_0}} ||\tilde{\boldsymbol{\lambda}}_i-\hat{\boldsymbol{\lambda}}_{i|\mathcal{G}}||^2,
		\end{aligned}
	\end{equation}
	Now by Assumption \ref{as5}(a),
	\begin{equation}\label{C26}
		|\omega^*_1|  \lesssim \frac{1}{\sqrt{|\mathcal{G}_{K_0}|}} \cdot O_p(\delta^{-1}_{NT} ) \leq \frac{1}{\tau_1\sqrt{N}}\cdot O_p(\delta^{-1}_{NT} )  ,
	\end{equation}
	Consequently, $\omega_{1}^{*}  \leq O_{p}\left(C_{NT}^{-2}\right)$.\\
	Next, $\omega_{2}^{*} \leq O_{p}\left(C_{NT}^{-2}\right)$ can be proved by using analogous arguments as used for (\ref{C25}) and (\ref{C26}). It follows that (\ref{C20}) holds and thus
	\begin{equation}\label{C27}
		S\left(K_{0}\right)-S\left(K_{0}+1\right)=O_{p}\left(C_{NT}^{-2}\right),
	\end{equation}
	in view of (\ref{C17}), (\ref{C19}) and (\ref{C20}). This, in conjunction with the fact that $\rho$ converges to zero at a slower rate than $C_{NT}^{-2}$ under Assumption \ref{as5}(b), implies that $P\left[S\left(K_{0}\right)-S\left(K_{0}+1\right)<\right.$ $\rho, \mathcal{M}(\mathcal{G})] \rightarrow 1$ as $N, T \rightarrow \infty$. It follows that $P\left[\mathbb{I} \mathbb{C}\left(K_{0}\right)<\mathbb{I} \mathbb{C}\left(K_{0}+1\right), \mathcal{M}(\mathcal{G})\right] \rightarrow 1$.\\ This completes the proof of Lemma \ref{lmA3}.{\hfill$\square$}
\subsection{Proof of Lemma \ref{bai2002lm4}}
	Since $\widehat{\mathbf{H}}^\top \mathbf{V} \widehat{\mathbf{H}} \xrightarrow{p} \mathbf{I}_m$, we have $\|\widehat{\mathbf{H}}\|_F =O_p(1)$ is invertible with probability approaching to 1. From $\underline{\boldsymbol{y}}_i=\mathbf{F}\boldsymbol{\lambda}_i+\underline{\boldsymbol{\epsilon}}_i$, we have $\underline{\boldsymbol{y}}_i=\mathbf{F} (\widehat{\mathbf{H}}^\top)^{-1} \widehat{\mathbf{H}}^{\top} \boldsymbol{\lambda}_i+\underline{\boldsymbol{\epsilon}}_i$. This implies
	$$
	\begin{aligned}
		\underline{\boldsymbol{y}}_i & =\widetilde{\mathbf{F}} \widehat{\mathbf{H}}^{\top} \boldsymbol{\lambda}_i+\underline{\boldsymbol{\epsilon}}_i-\left(\widetilde{\mathbf{F}}-\mathbf{F} (\widehat{\mathbf{H}}^\top)^{-1}\right) \widehat{\mathbf{H}}^\top \boldsymbol{\lambda}_i \\
		& =\widetilde{\mathbf{F}} \widehat{\mathbf{H}}^{\top} \boldsymbol{\lambda}_i+\underline{\mathbf{u}}_i,
	\end{aligned}
	$$
	where $\underline{\mathbf{u}}_i=\underline{\boldsymbol{\epsilon}}_i-\left(\widetilde{\mathbf{F}}\widehat{\mathbf{H}}^\top-\mathbf{F} \right)  \boldsymbol{\lambda}_i$. Note that
	$$
	\begin{aligned}
		V(\mathbf{F})= & N^{-1} T^{-1} \sum_{i=1}^N \underline{\boldsymbol{\epsilon}}_i^{\top} \mathbf{M}_F \underline{\boldsymbol{\epsilon}}_i, \\
		V(\widetilde{\mathbf{F}})= & N^{-1} T^{-1} \sum_{i=1}^N \underline{\mathbf{u}}_i^{\top} \mathbf{M}_{\widetilde{F}} \underline{\mathbf{u}}_i\\= & N^{-1} T^{-1} \sum_{i=1}^N\left(\underline{\boldsymbol{\epsilon}}_i-\left(\widetilde{\mathbf{F}}\widehat{\mathbf{H}}^\top-\mathbf{F} \right)  \boldsymbol{\lambda}_i\right)^{\top} \mathbf{M}_{\widetilde{F}}\left(\underline{\boldsymbol{\epsilon}}_i-\left(\widetilde{\mathbf{F}}\widehat{\mathbf{H}}^\top-\mathbf{F} \right)  \boldsymbol{\lambda}_i\right) \\
		= & N^{-1} T^{-1} \sum_{i=1}^N \underline{\boldsymbol{\epsilon}}_i^{\top} \mathbf{M}_{\widetilde{F}} \underline{\boldsymbol{\epsilon}}_i-2 N^{-1} T^{-1} \sum_{i=1}^N \boldsymbol{\lambda}_i^\top \left(\widetilde{\mathbf{F}}\widehat{\mathbf{H}}^\top-\mathbf{F} \right)^{\top} \mathbf{M}_{\widetilde{F}} \underline{\boldsymbol{\epsilon}}_i \\
		+&N^{-1} T^{-1} \sum_{i=1}^N \boldsymbol{\lambda}_i^\top \left(\widetilde{\mathbf{F}}\widehat{\mathbf{H}}^\top-\mathbf{F} \right)^{\top} \mathbf{M}_{\widetilde{F}}\left(\widetilde{\mathbf{F}}\widehat{\mathbf{H}}^\top-\mathbf{F} \right)\boldsymbol{\lambda}_i \\
		= & a+b+c .
	\end{aligned}
	$$
	Because $\mathbf{I}-\mathbf{M}_{\widetilde{F}}$ is positive semi-definite, $\boldsymbol{x}^{\top} \mathbf{M}_{\widetilde{F}} \boldsymbol{x} \leq \boldsymbol{x}^{\top} \boldsymbol{x}$. Thus,
	$$
	\begin{aligned}
		c & \leq N^{-1} T^{-1} \sum_{i=1}^N \boldsymbol{\lambda}_i^\top \left(\widetilde{\mathbf{F}}\widehat{\mathbf{H}}^\top-\mathbf{F} \right)^{\top} \left(\widetilde{\mathbf{F}}\widehat{\mathbf{H}}^\top-\mathbf{F} \right)  \boldsymbol{\lambda}_i \\
		& \leq T^{-1} \sum_{t=1}^T\left\|\widehat{\mathbf{H}}\tilde{\boldsymbol{f}}_t-\boldsymbol{f}_t\right\|^2 \cdot\left(N^{-1} \sum_{i=1}^N\left\|\boldsymbol{\lambda}_i\right\|^2\right) \\
		& =O_p\left(C_{N T}^{-2}\right) \cdot O_p(1)
	\end{aligned}
	$$
	by Theorem 3.2 in \cite{he2022large} and Assumption \ref{as2}.
	For term $b$, we use the fact that $|tr(\mathbf{A})| \leq m\|\mathbf{A}\|$ for any $m \times m$ matrix $\mathbf{A}$, and we have$$
	\mathbb{E}\left\|N^{-1 / 2} \sum_{i=1}^N \epsilon_{i t} \boldsymbol{\lambda}_i\right\|^2=\frac{1}{N} \sum_{i=1}^N \sum_{j=1}^N \mathbb{E}\left[\epsilon_{i t} \epsilon_{j t}\right] \boldsymbol{\lambda}_i^{\top} \boldsymbol{\lambda}_j \leq \overline{\boldsymbol{\lambda}}^2 \frac{1}{N} \sum_{i=1}^N \sum_{j=1}^N\left|\tau_{i j}\right| \leq \overline{\boldsymbol{\lambda}}^2 M,
	$$
	Thus
	$$
	\begin{aligned}
		b & =2 T^{-1} tr\left(\left(\widetilde{\mathbf{F}}\widehat{\mathbf{H}}^\top-\mathbf{F} \right)^{\top} \mathbf{M}_{\widetilde{F}}\left(N^{-1} \sum_{i=1}^N \underline{\boldsymbol{\epsilon}}_i \boldsymbol{\lambda}_i^\top\right)\right) \\
		& \leq 2m \left\|\frac{\widetilde{\mathbf{F}}\widehat{\mathbf{H}}^\top-\mathbf{F} }{\sqrt{T}}\right\| \cdot\left\|\frac{1}{\sqrt{T} N} \sum_{i=1}^N \underline{\boldsymbol{\epsilon}}_i \boldsymbol{\lambda}_i^\top\right\| \\
		& \leq 2 m\left(T^{-1} \sum_{i=1}^T\left\|\widehat{\mathbf{H}}\tilde{\boldsymbol{f}}_t-\boldsymbol{f}_t\right\|^2\right)^{1 / 2} \cdot \frac{1}{\sqrt{N}}\left(\frac{1}{T} \sum_{i=1}^T\left\|\frac{1}{\sqrt{N}} \sum_{i=1}^N \epsilon_{i t} \boldsymbol{\lambda}_i\right\|^2\right)^{1 / 2} \\
		& =O_p\left(C_{N T}^{-1}\right) \cdot \frac{1}{\sqrt{N}}=O_p\left(C_{N T}^{-2}\right)
	\end{aligned}
	$$
	by Theorem 3.2 in \cite{he2022large}. Therefore,
	$$
	V(\widetilde{\mathbf{F}})=N^{-1} T^{-1} \sum_{i=1}^N \underline{\boldsymbol{\epsilon}}_i^{\top} \mathbf{M}_{\widetilde{F}} \underline{\boldsymbol{\epsilon}}_i+O_p\left(C_{N T}^{-2}\right) .
	$$
	Using the fact that $V(\widetilde{\mathbf{F}})-V(\mathbf{F}) \geq 0$,
	$$ 0 \geq V(\mathbf{F})-V(\widetilde{\mathbf{F}})=\frac{1}{N T} \sum_{i=1}^N \underline{\boldsymbol{\epsilon}}_i^{\top} \mathbf{P}_{\widetilde{F}} \underline{\boldsymbol{\epsilon}}_i-\frac{1}{N T} \sum_{i=1}^N \underline{\boldsymbol{\epsilon}}_i^{\top} \mathbf{P}_{F} \underline{\boldsymbol{\epsilon}}_i+O_p\left(C_{N T}^{-2}\right).$$
	
	Note that
	$$
	\begin{aligned}
		\frac{1}{N T} \sum_{i=1}^N \underline{\boldsymbol{\epsilon}}_i^{\top} \mathbf{P}_{F} \underline{\boldsymbol{\epsilon}}_i & \leq\left\|\left(\mathbf{F}^{\top} \mathbf{F} / T\right)^{-1}\right\| \cdot N^{-1} T^{-2} \sum_{i=1}^N \underline{\boldsymbol{\epsilon}}_i^{\top} \mathbf{F} \mathbf{F}^{\top} \underline{\boldsymbol{\epsilon}}_i \\
		& =O_p(1) T^{-1} N^{-1} \sum_{i=1}^N\left\|T^{-1 / 2} \sum_{i=1}^N \boldsymbol{f}_t \epsilon_{i t}\right\|^2=O_p\left(T^{-1}\right) \leq O_p\left(C_{N T}^{-2}\right)
	\end{aligned}
	$$
	by Assumption \ref{as8}. Thus
	$$
	0 \geq \frac{1}{N T} \sum_{i=1}^N \underline{\boldsymbol{\epsilon}}_i^{\top} \mathbf{P}_{\widetilde{F}} \underline{\boldsymbol{\epsilon}}_i+O_p\left(C_{N T}^{-2}\right) .
	$$
	This implies that $0 \leq N^{-1} T^{-1} \sum_{i=1}^N \underline{\boldsymbol{\epsilon}}_i^{\top} \mathbf{P}_{\widetilde{F}} \underline{\boldsymbol{\epsilon}}_i=O_p\left(C_{N T}^{-2}\right)$. In summary
	$$
	V(\widetilde{\mathbf{F}})-V(\mathbf{F})=O_p\left(C_{N T}^{-2}\right) .
	$$
  The proof is complete.{\hfill$\square$}

\end{appendix}
\section*{Funding}
The authors gratefully acknowledge National Key R\&D Program of China (No. 2023YFA1008701), National Science Foundation of China (12171282),  Qilu Young Scholars Program of Shandong University.

\bibliographystyle{model2-names}
\bibliography{Ref}

\end{document}